\documentclass[a4paper,11pt]{article}
\pdfoutput=1 

\usepackage{jinstpub} 

\usepackage{color}
\usepackage{enumitem}
\usepackage{threeparttable}
\usepackage[normalem]{ulem}
\usepackage{lineno}
\usepackage{graphicx}
\usepackage{subcaption}
\usepackage{mwe}
\usepackage{amsmath}

\usepackage{amssymb}

\usepackage{siunitx}
\usepackage{newtxmath,newtxtext}

\usepackage{overpic}
\usepackage{longtable}
\usepackage{multirow}

\newlength{\figwidth}
\newlength{\fighalfwidth}
\title{\boldmath \center \LARGE Neutrino Event Selection in the MicroBooNE Liquid Argon Time Projection Chamber using Wire-Cell 3-D Imaging, Clustering, and Charge-Light Matching}

\author[jj]{P.~Abratenko}
\author[o]{M.~Alrashed}
\author[n]{R.~An}
\author[d]{J.~Anthony}
\author[ii]{J.~Asaadi}
\author[s]{A.~Ashkenazi}
\author[mm]{S.~Balasubramanian}
\author[k]{B.~Baller}
\author[t]{C.~Barnes}
\author[x]{G.~Barr}
% ...

% ...
\affiliation[d]{University of Cambridge, Cambridge CB3 0HE, United Kingdom}
% ...
\affiliation[k]{Fermi National Accelerator Laboratory (FNAL), Batavia, IL 60510, USA}
% ...
\affiliation[n]{Illinois Institute of Technology (IIT), Chicago, IL 60616, USA}
\affiliation[o]{Kansas State University (KSU), Manhattan, KS, 66506, USA}
% ...
\affiliation[s]{Massachusetts Institute of Technology (MIT), Cambridge, MA, 02139, USA}
\affiliation[t]{University of Michigan, Ann Arbor, MI, 48109, USA}
% ...
\affiliation[x]{University of Oxford, Oxford OX1 3RH, United Kingdom}
% ...
\affiliation[ii]{University of Texas, Arlington, TX, 76019, USA}
\affiliation[jj]{Tufts University, Medford, MA, 02155, USA}
% ...
\affiliation[mm]{Wright Laboratory, Department of Physics, Yale University, New Haven, CT, 06520, USA}

  \emailAdd{microboone\_info@fnal.gov}
\date{}

\abstract{An accurate and efficient event reconstruction is required to realize the full scientific capability of
liquid argon time projection chambers (LArTPCs). The current and future neutrino experiments that rely on massive
LArTPCs create a need for new ideas and reconstruction approaches.  Wire-Cell, proposed in recent years, is a novel
tomographic event reconstruction method for LArTPCs.
The Wire-Cell 3D imaging approach capitalizes on charge, sparsity, time, and geometry information to reconstruct a topology-agnostic 3D image of the ionization electrons prior to pattern recognition.
A second novel method, the many-to-many charge-light matching, then pairs the TPC charge activity to the detected
scintillation light signal, thus enabling a powerful rejection of cosmic-ray muons in the MicroBooNE detector.
A robust processing of the scintillation light signal and an appropriate clustering of the reconstructed 3D image are
fundamental to this technique. In this paper, we describe the principles and algorithms of these techniques and
their successful application in the MicroBooNE experiment. A quantitative evaluation of the performance of these
techniques is presented. Using these techniques, a 95\% efficient pre-selection of neutrino charged-current events
is achieved with a 30-fold reduction of non-beam-coincident cosmic-ray muons, and about 80\% of the selected neutrino
charged-current events are reconstructed with at least 70\% completeness and 80\% purity. 
}


\begin{document}
\maketitle
\flushbottom

\section{Introduction}\label{sec:intro}

The Liquid Argon Time Projection Chamber (LArTPC)~\cite{rubbia77,Chen:1976pp,willis74,Nygren:1976fe} is a novel detector technology 
under rapid development. 
It is a fully active calorimeter with excellent 3D tracking capability,
which can enable particle identification (PID) of unprecedented power in neutrino detection.
This detector technology has been utilized in many current accelerator neutrino experiments, such as MicroBooNE~\cite{Acciarri:2016smi} and the Short Baseline Neutrino (SBN) program~\cite{Machado:2019oxb}, and it will be used in the future massive LArTPC experiments, such as DUNE~\cite{Abi:2020wmh}.

Event reconstruction is one of the most challenging tasks in analyzing the data from current and future large-scale LArTPCs. A high-performance event reconstruction is vital to take full advantage of the capability of LArTPCs for physics 
measurements. Multiple reconstruction approaches are being developed in MicroBooNE, including the Pandora multi-algorithm pattern 
recognition~\cite{uboone-pandora} and deep learning with convolutional neural networks~\cite{uboone-dl, uboone-dl2}. Another
novel event reconstruction method, Wire-Cell, has also been under rapid development for MicroBooNE. 
The Wire-Cell 3D imaging~\cite{wc_imaging} capitalizes on the most fundamental LArTPC detector information -- time, charge, and 
geometry -- to tomographically reconstruct a topology-agnostic three-dimensional image of the ionization electrons prior to any 
pattern recognition step. 
The early construction of the 3D image without the involvement of pattern recognition is the primary distinction between Wire-Cell and other reconstruction paradigms~\cite{uboone-pandora,uboone-dl, uboone-dl2}. This is beneficial because in 3D the particle activities are more separated than in 2D, which reduces the difficulties in clustering and other pattern recognition tasks.
Enabled by the high-performance ionization 
electron signal processing procedure in MicroBooNE~\cite{noise_filter_paper, SP1_paper, SP2_paper}, the Wire-Cell 3D imaging reduces 
the degeneracies -- integrated charge measured along each wire other than pixelated measurement of charge -- inherent in the LArTPC wire readouts as used by MicroBooNE and numerous other experiments.

Detector defects such as nonfunctional channels (10\% of all wire readouts in MicroBooNE) and the numerous cosmic-ray muons (20--30 per TPC 
readout window) in the MicroBooNE detector pose additional challenges to the overall success of the event reconstruction. 
We address the first problem by allowing for the reconstruction in regions where two out of three channels, one from each wire plane, are functional. For these regions, an analysis that also relies on information from nearby fully functional regions is performed. Our method significantly reduces the extent of unusable 
regions by a factor of ten. To deal with the high rate of cosmic rays, we developed a many-to-many 
TPC-charge and PMT-light (charge-light) 
matching method, to distinguish the candidate neutrino activity, which is in coincidence with the beam spill, from the numerous cosmic rays 
spanning the entire MicroBooNE detector and the TPC readout window. TPC activity hereafter refers to the energy deposition in LArTPC by ionization. It originates from either a cosmic-ray muon or a neutrino interaction. This method relies on the Wire-Cell 3D imaging and 
emphasizes the interplay between the scintillation light and the ionization charge 
signals created by charged particles traversing the LAr. A robust processing of the scintillation light signals from the photon detector system 
and an appropriate clustering, which groups the TPC activities that represent signals initiated by an individual primary particle, are fundamental to this technique.

In this paper, we describe the principles, algorithms, and performance evaluation of the Wire-Cell 3D imaging and the many-to-many charge-light matching, 
including the light signal processing and the 3D clustering. These techniques provide a solid foundation to reject coincident 
in-beam cosmic-ray muons~\cite{Wire-Cell-Generic-PRD} with downstream reconstruction techniques (e.g.~track trajectory fitting and 
pattern recognition). The outcome of these tools, e.g.~the Wire-Cell 2D and 3D images of the neutrino activities with the surrounding cosmic-ray
activities removed, can also improve the performance of other reconstruction paradigms~\cite{uboone-dl, uboone-dl2,uboone-pandora}.  
The principle and implementation of the Wire-Cell 3D imaging is presented in section~\ref{sec:imaging}.
The many-to-many charge-light matching to pair the TPC activities to the reconstructed PMT activities is described in section~\ref{sec:clustering_matching} as the final step to select the candidate neutrino activities. 
Evaluations of the quality of the Wire-Cell 3D imaging and the efficacy of the many-to-many charge-light matching are demonstrated in section~\ref{sec:evaluation}.
A summary of the performance and discussion is presented in section~\ref{sec:summary}.

\section{The MicroBooNE detector}\label{sec:uboone}

\begin{figure}[htbp!]
   \includegraphics[width=0.9\textwidth]{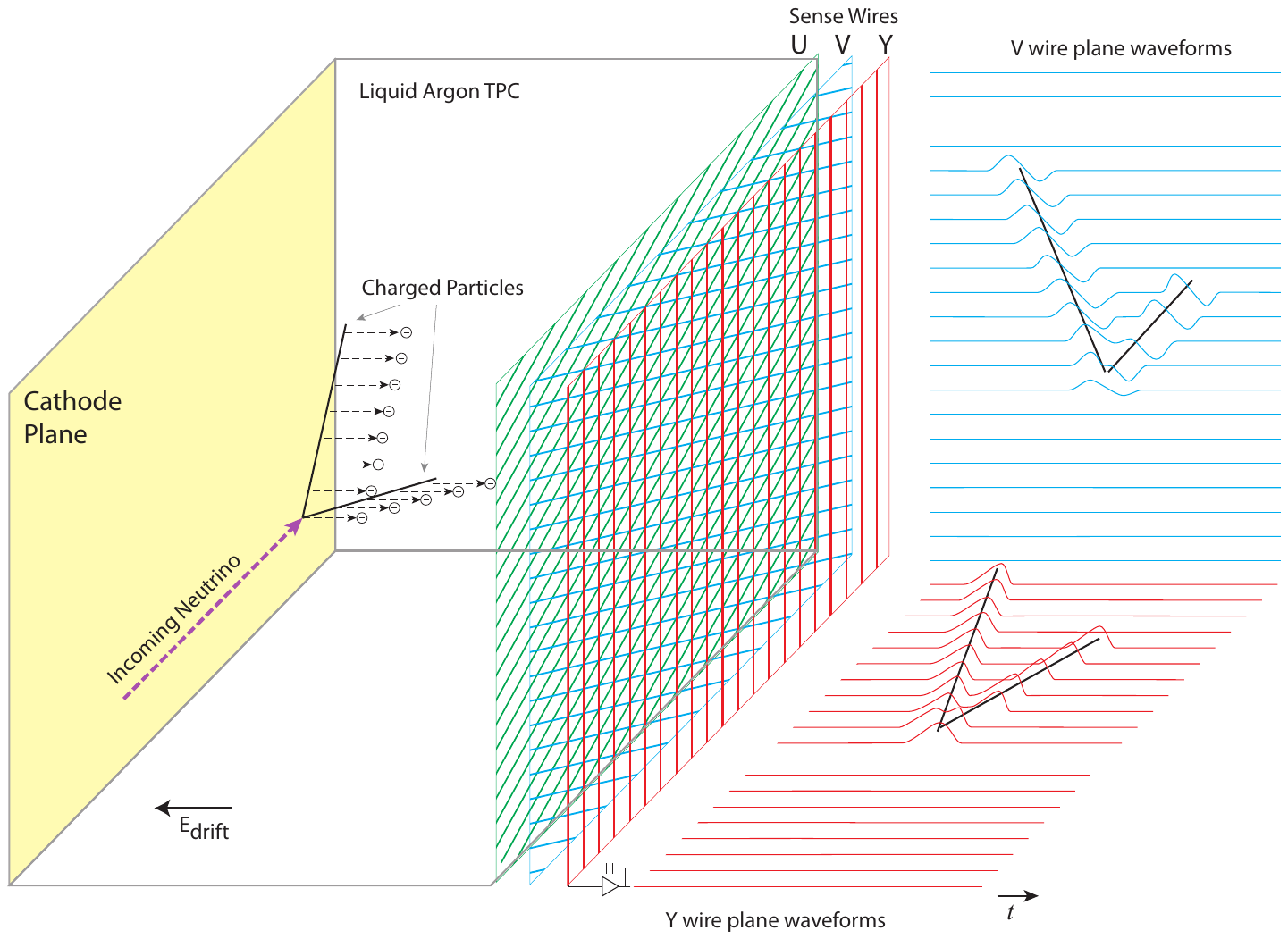}
   \caption{Illustration of single-phase LArTPCs~\cite{Acciarri:2016smi}. Each wire plane provides a 2D image of the ionization electrons with respect to a specific wire orientation.}
  \label{fig:tpccartoon}
\end{figure}

The MicroBooNE detector is the first LArTPC in the SBN program to measure neutrino interactions from the on-axis Booster neutrino beam (BNB)~\cite{bnb} at Fermi National Accelerator Laboratory in Batavia, IL.
MicroBooNE uses a single-phase (i.e.~liquid phase only) LArTPC with a rectangular active volume of the following dimensions: 2.6 m (width, along the drift direction), 2.3 m (height, vertical), and 10.4 m (length, along the beam direction), as illustrated in figure~\ref{fig:tpccartoon}. The TPC has an active mass of 85 metric tonnes and is immersed in a single-walled and cylindrical shaped cryostat with a 170 tonne liquid argon capacity.

A high voltage of \SI{-70}{\kV} applied on the cathode plane provides a drift field of \SI{273}{\V/\cm}.
The electrons ionized by any energy deposition from traversing charged particles drift towards the anode wire planes along the electric field at a nominal speed of about \SI{1.10}{\mm/\us}.
In this paper, we use the X-axis to represent the direction away from the readout wire plane and opposite to the ionization charge drift, Y-axis to represent the vertical-up direction, and Z-axis to represent the BNB beam direction.
There are three parallel wire readout planes~\cite{Acciarri:2017wp} on the anode side with different wire orientations.
The first wire plane facing the cathode is labeled ``U'', and the second and third plane are labeled ``V'' and ``Y'', respectively.
The 3456 wires in the Y plane are oriented vertically and the 2400 wires in the U (V) plane are oriented +($-$)60$^\circ$ with respect to the vertical direction.
The spacing between adjacent wires and adjacent wire planes are both \SI{3}{\mm}. 
Different bias voltages, \SI{-110}{\V}, \SI{0}{\V}, and \SI{+230}{\V}, are applied to the U, V, and Y wire planes, respectively, to ensure all ionization electrons drift through the U and V planes before being collected by the Y plane.
The U and V planes are commonly referred to as the induction planes and the ionization electrons induce bipolar electrical signals as they pass through the planes; the Y plane is referred to as the collection plane and sees unipolar electrical pulses.

The TPC readout is defined with respect to the event trigger and includes three 1.6 ms frames, spanning -1.6 ms to +3.2 ms relative to the trigger time, with a sampling rate of 2 MHz (0.5 $\mu$s per time tick). 
Therefore, each wire plane records a 2D image (time versus wire) of the ionization electrons within the full 4.8 ms TPC readout.

Behind the wire planes and external to the TPC, there is an array of thirty-two 8'' photomultiplier tubes (PMTs)~\cite{Conrad:2015xta} to detect scintillation light for triggering, timing, and other purposes.
The PMT readout includes four 1.6 ms frames with the beam gate window (1.6 $\mu$s beam-spill) contained within the second 1.6 ms frame. The sampling rate is 64 MHz (15.625 ns per sample) for each PMT and the signal is recorded in a dynamic-range-based, paired form for each channel -- a high gain (x10) signal and a low gain (x1) signal.
The 32 PMTs promptly (in a few nanoseconds) detect the scintillation light and provide the intensity and position information of the photo-electrons originating from either a cosmic-ray muon or a neutrino interaction. The TPC and PMT readouts cover the full time range of the beam neutrino activities as well as cosmic-ray activities that enter the beam spill frame during the relatively slow drift of ionization electrons, which has a maximum drift time of 2.3~ms.

\section{Wire-Cell 3D Imaging}\label{sec:imaging}

Analysis of the single-phase LArTPC with a wire readout scheme is a natural application of the
tomography technique, which the Wire-Cell 3D imaging strictly follows.
Ref.~\cite{wc_imaging} introduces the basic concepts and the key mathematics of the Wire-Cell 3D imaging. 
In this section, we focus more on the realistic issues when applying the Wire-Cell 3D imaging to MicroBooNE data.

The fundamental information provided by a LArTPC is as follows:
\begin{itemize}[noitemsep, nolistsep]
	\item[(i)] \textbf{Time} - when the ionization electrons arrive at the anode wire plane\footnote{The absolute starting time of each cosmic-ray muon needs to be corrected by using the light signal information with the charge-light matching technique described in section~\ref{sec:matching_alg}}.
	\item[(ii)] \textbf{Geometry} - the positions of the wires from each plane that have signals from the ionization electrons, i.e.~hit wires.
	\item[(iii)] \textbf{Charge} - the number of ionization electrons measured by the hit wires from each wire plane.
\end{itemize}

The time and charge information comes from the time distribution of the deconvolved charge, which is obtained via advanced signal processing techniques. 
In particular, the 2D deconvolution technique~\cite{SP1_paper, SP2_paper} significantly improves the signal processing for the induction planes and makes the deconvolved charge consistent across the multiple wire planes.
The geometry information is the wire position, along the wire pitch direction (perpendicular to the wire orientation).
Since the wire planes have different wire orientations, signals on each wire are taken as a 1D projection of the charge depositions with the summation of the charge available in the proximity of each wire. The position of each individual charge deposition along the wire itself can only be provided by other wire planes.

The Wire-Cell 3D imaging uses two major steps to reconstruct the 3D image of the ionization electrons arriving at the anode plane:
1) Reconstruct the 2D image of the ionization electrons on the anode plane in a given time slice, e.g.~2 $\mu$s (4 ticks in the TPC readout) considering the intrinsic time smearing of about 1.5 $\mu$s after signal processing~\cite{SP1_paper}. The integrated charge within the time slice on each hit wire is used; 
2) Concatenate the 2D images from the previous step in the sequence of time slices to form the 3D image. From three wire readout planes, at most three 1D projection views are available within one time slice, in contrast with the dozens or even hundreds of 1D projection views available in common tomography applications, such as those for medical imaging.
Compared to a pixelated readout with n$^2$ pixels, the $\mathcal{O}$(n) wires (3$\times$n for three wire planes) afforded by a wire readout scheme reduces the heat loads and the cost of the readout system, but result in a considerable loss of information.
To recover from the loss of information, additional constraints are used:
\begin{itemize}[noitemsep, nolistsep]
	\item[(iv)] \textbf{Sparsity} - the distribution of ionization electrons in space is expected to be sparse, typically occupying less than 10\% of the local bounding volume that contains the activities, for any physical signals.
	\item[(v)] \textbf{Proximity} - the ionization electrons are read out by consecutive wires because a charged particle ionizes argon atoms continuously in the fully active LArTPC volume. 
	\item[(vi)] \textbf{Positivity} - the number of the drifted ionization electrons can only be positive.
\end{itemize}

The actual procedure we use to incorporate the above information is divided into two processes: \emph{tiling} and \emph{solving}, as described in section~\ref{sec:tiling} and section~\ref{sec:solving}, respectively.
In the implementation of Wire-Cell 3D imaging in MicroBooNE, the nonfunctional wires~\cite{noise_filter_paper}
aggravate the wire readout ambiguity, and introduce a large number of \emph{ghost} energy depositions.
A dedicated de-ghosting algorithm, discussed in section~\ref{sec:deghosting}, is developed to mitigate this effect.

\subsection{Tiling}\label{sec:tiling}

\begin{figure}[thpb!]
  \centering
            \begin{overpic}[width=0.8\textwidth, height=0.5\textwidth]{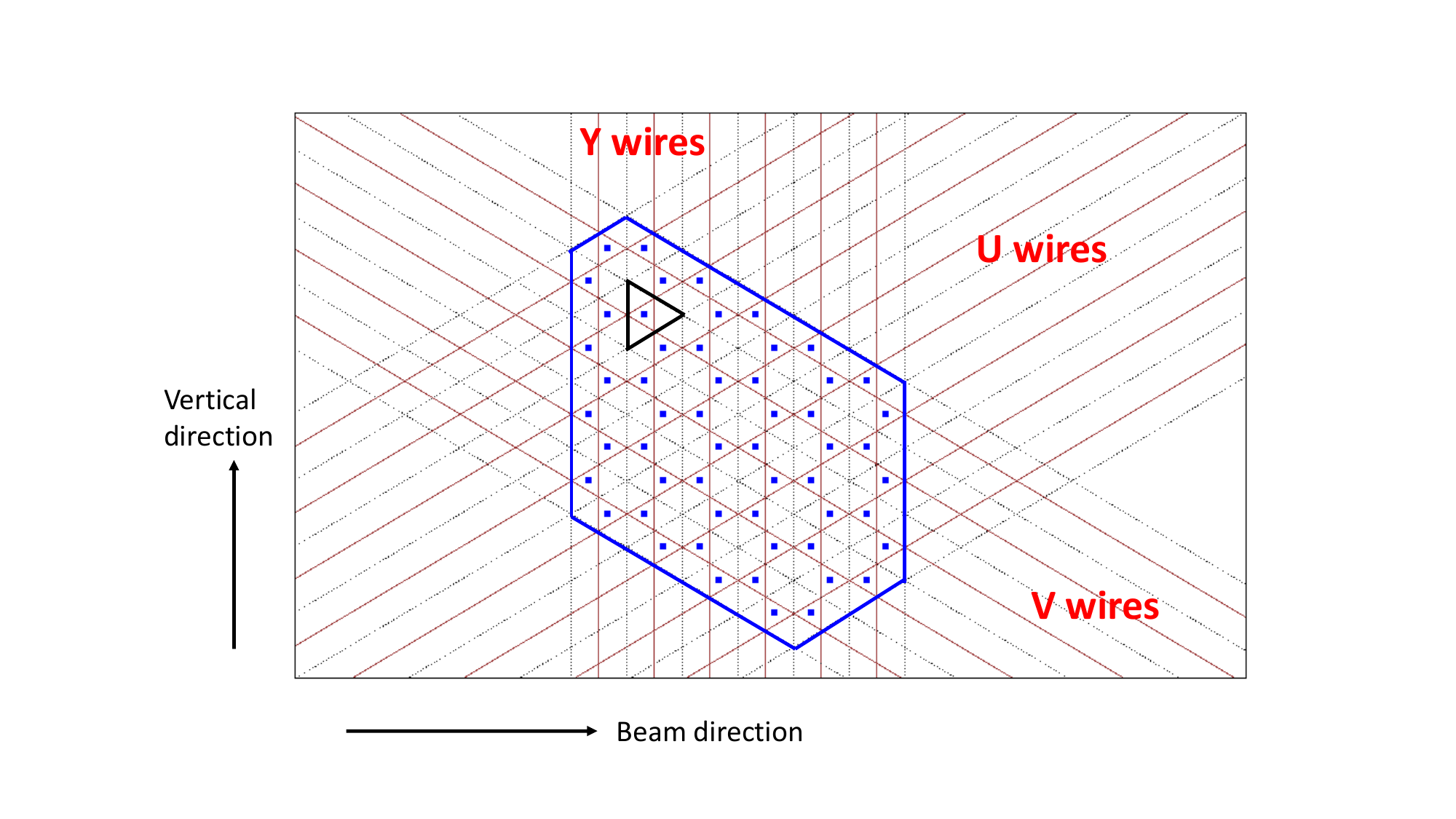}
		                      \put(68,55){\text{\small MicroBooNE}}
				                  \end{overpic}
    \caption{An example of the hit cells and blob constructed by the hit wires with the MicroBooNE detector geometry. Each wire is represented by a solid red line and the wire (pitch) boundaries are represented by dashed black lines. All hit cells have equilateral triangular shapes and are marked with blue dots at their centers. An example cell is marked by the black triangle. A blob is formed by the contiguous hit cells and marked by solid blue lines.}
  \label{fig:blob}
\end{figure}

The 2D image of the ionization electrons in a time slice consists of {\it cells}, which are the smallest geometric units formed by wires from three planes.
Figure~\ref{fig:blob} shows tens of cells, for example the black triangle, which is the overlapping area of three wires from the three wire planes. Each wire represents a 2D region centered around the wire location with its width equal to the wire pitch.
All cells have equilateral triangular shapes because of the MicroBooNE wire orientations and positioning.

The smallest time unit in the Wire-Cell imaging is a time slice, whose 2 $\mu$s width contains four sampling ticks from the TPC readout.
The width of the time slice introduces negligible information loss because the software filtering in the signal processing has a cut-off frequency at about 0.25 MHz to optimize the signal-to-noise ratio, which in turn smears the time resolution.
Geometry is used to determine all possible hit cells within a time slice by finding the intersections of the hit wires. 
In figure~\ref{fig:blob}, there are 8 hit U wires (2.4 cm wide), 5 hit V wires (1.5 cm wide), and 6 hit Y wires (1.8 cm wide),  leading to 55 possible hit cells.
The fact that there are fewer knowns (19 hit wires) than unknowns (55 cells) indicates an ambiguity is introduced by the wire readout.
Meanwhile, the amount of integrated charge in a time slice and the identities of active wires in that time slice are affected by diffusion and long-range induction effects (especially for induction planes) as charge drifts in the TPC, as well as the action of software filters applied to the waveforms~\cite{SP1_paper, SP2_paper}.
To mitigate the impact from wire ambiguity and charge smearing, 
a procedure to merge the consecutive hit cells is developed, called {\it tiling}.
The groups of hit cells after tiling are called {\it blobs}.
The blob in figure~\ref{fig:blob} is marked by solid blue lines.
The connected hit wires are merged as {\it wire bundles} in the tiling procedure, and a blob is the overlapping area of three wire bundles from each wire plane as shown in figure~\ref{fig:blob}.
Note that a cell or a blob can be taken as a 3D object and its length along the drift direction is the width of the time slice, i.e.~2 $\mu$s or about 2.2 mm. In the following sections, 3D space points will be used to describe the algorithms and a ``space point'' is equivalent to a ``cell'' hereafter, which represents a 3D voxel of the space with a finite size. Its charge is deduced by the total charge of the blob that contains it, divided by the number of space points within the blob.

There are three advantages to the tiling. Firstly, it completely collects the reconstructed charge smeared to the adjacent wires, resulting in more consistent charge values across the wire planes. Secondly, it greatly reduces the number of unknowns in the later stage of {\it solving}. Thirdly, it significantly reduces the computational cost.
The charge smearing is different for different wire planes. Obtaining consistent charge measurements across multiple wire planes by the tiling is fundamental to construct and solve the Wire-Cell 3D imaging equation as described in section~\ref{sec:solving}.  
    
\begin{figure}[htpb!]
  \centering
   \begin{overpic}[width=0.8\figwidth]{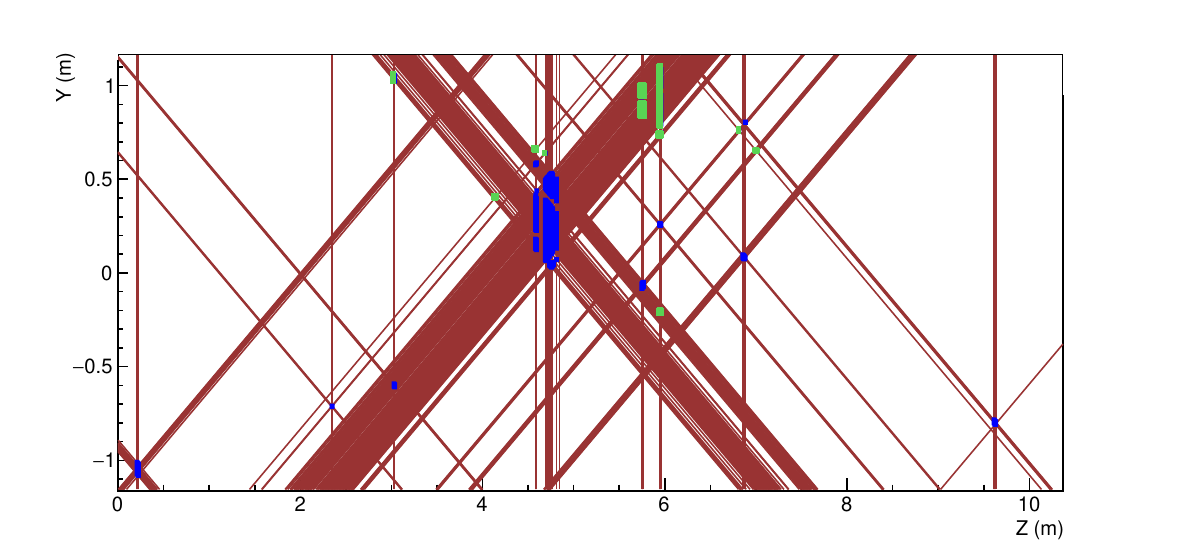}
	        \put(65,43){\text{\small MicroBooNE data}}
		   \end{overpic}
    \caption{ An example event with hit wires and blobs after applying the deghosting algorithm (see section~\ref{sec:deghosting}).
    Blobs are marked in blue or green.
    Blue blobs correspond to 3-plane tiling requiring all three wire planes to be functional. 
    Green blobs correspond to the additional blobs created in 2-plane tiling requiring at least two wire planes to be functional.
    Hit wires are represented by solid red lines.}
  \label{fig:tiling_comparison}
\end{figure}

Figure~\ref{fig:blob} corresponds to a single track traversing the time slice in a local area.
In reality, there could be multiple tracks from cosmic-ray muons or a neutrino interaction traversing the time slice (a fixed {\it x} position) at various Y-Z locations as shown in figure~\ref{fig:tiling_comparison}. 
The solid red lines represent the hit wires from each wire plane. The resulting blobs are marked in blue or green.
One may notice that in figure~\ref{fig:tiling_comparison}, the green blobs only have two corresponding wire bundles from two wire planes. 
This is because the hit wires in the third wire plane are not able to provide reasonable signals if they are nonfunctional or too noisy. Note that figure~\ref{fig:tiling_comparison} is the result after applying the de-ghosting algorithm as introduced in section~\ref{sec:deghosting}, so some blobs are determined to be fake and removed.

\begin{figure}[thbp!]     
  \centering
  \begin{overpic}[width=0.9\figwidth]{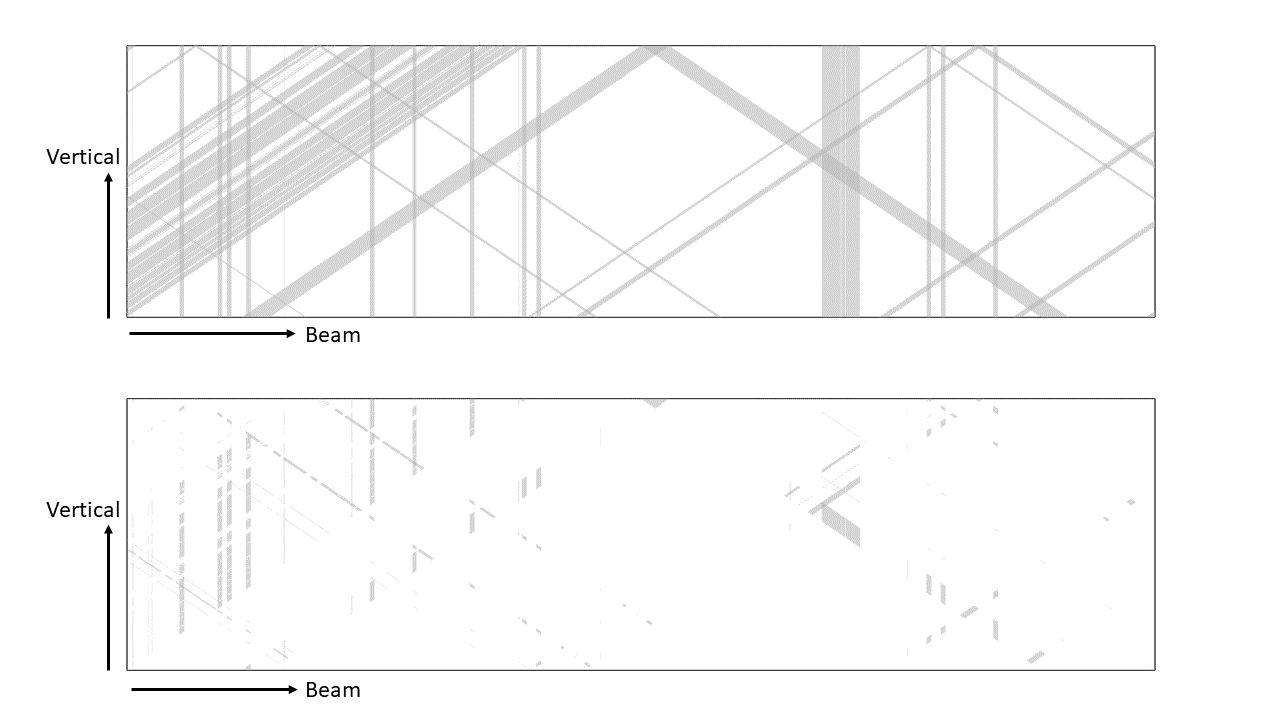}
    \put(15,54){\text{Active detector if three live wires are required prior to tiling}}
    \put(20,26){\text{Active detector if two live wires are required to tile}}
    \put(72,5){\text{MicroBooNE}}
  \end{overpic}
    \caption{Impact of the nonfunctional wires (gray) on the anode plane. The borders of the two figures correspond to the boundaries of the LArTPC active volume.
      Top: the gray area that has at least one wire nonfunctional is 30\%.
      Bottom: the gray area that has at least two wires nonfunctional is 3\%.
    }
  \label{fig:dead_region}
\end{figure}

Generally, a 3-plane tiling approach requires the wires from all three wire planes to be functional.
Given that about 10\% of channels are nonfunctional in MicroBooNE for various reasons~\cite{noise_filter_paper}, this requirement introduces 30\% inactive regions on the 2D anode plane as illustrated in the top panel of figure~\ref{fig:dead_region}. 
To address this issue, we allow for a 2-plane tiling procedure in areas where at least two planes have functional wires.
This means that only the area having two or three nonfunctional wires is regarded as the nonfunctional region.
This drastically reduces the nonfunctional volume from 30\% to 3\% as shown in the bottom panel of figure~\ref{fig:dead_region}, and an increase of the number of blobs (green blobs) can be seen in figure~\ref{fig:tiling_comparison}.
Outside this 3\% nonfunctional region, the 2-plane tiling procedure assumes all the nonfunctional wires are assumed to be hit all the time.

While the missing 3-plane blobs are recovered with 2-plane tiling, a number of fake blobs, or ``ghosts'', are created in areas where two functional hit wires cross a third nonfunctional wire, where no true physical charge is responsible for the corresponding wires' measurements.
Some ghosts could still appear when all three wire planes are functional because of the intrinsic ambiguity of the wire-readout scheme, but the number of ghosts is significantly increased when 2-plane tiling is allowed, given the sizable number of nonfunctional wires.

\begin{figure}[!htpb] 
  \centering
    \begin{overpic}[width=0.8\figwidth]{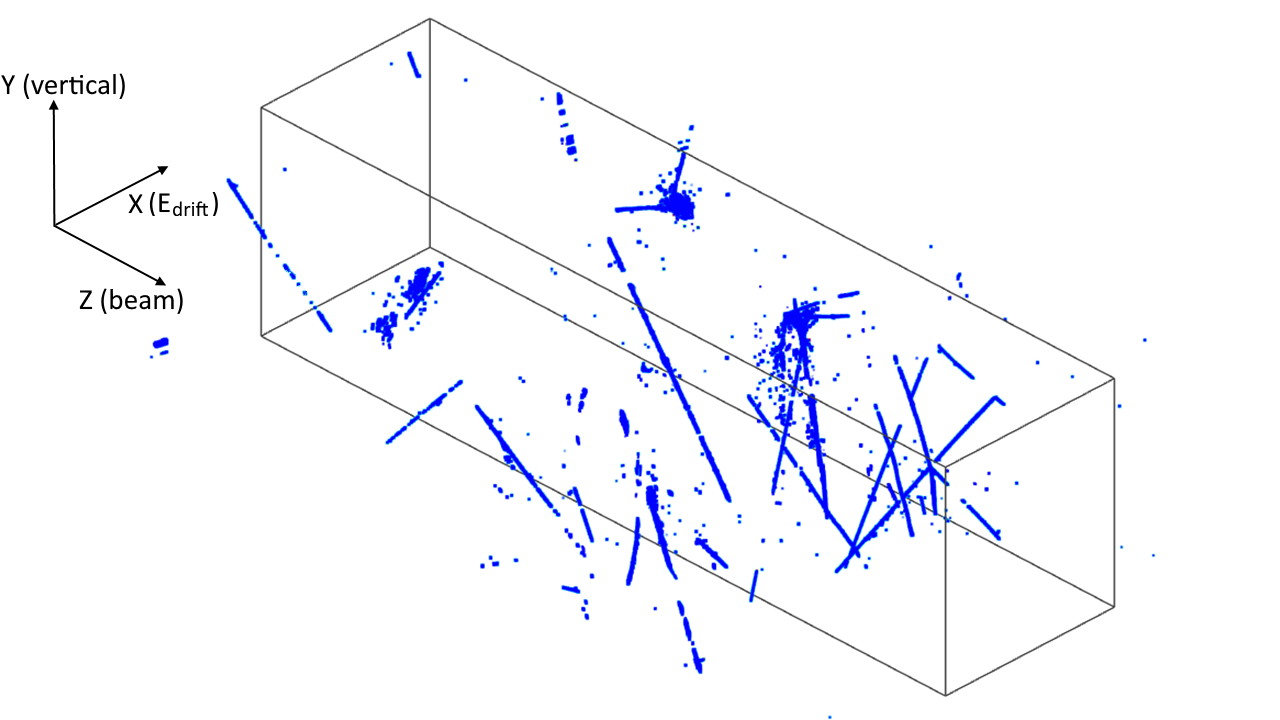}
        \put(10,10){\textbf{\small MicroBooNE Data}}
    \end{overpic}
    \begin{overpic}[width=0.85\figwidth]{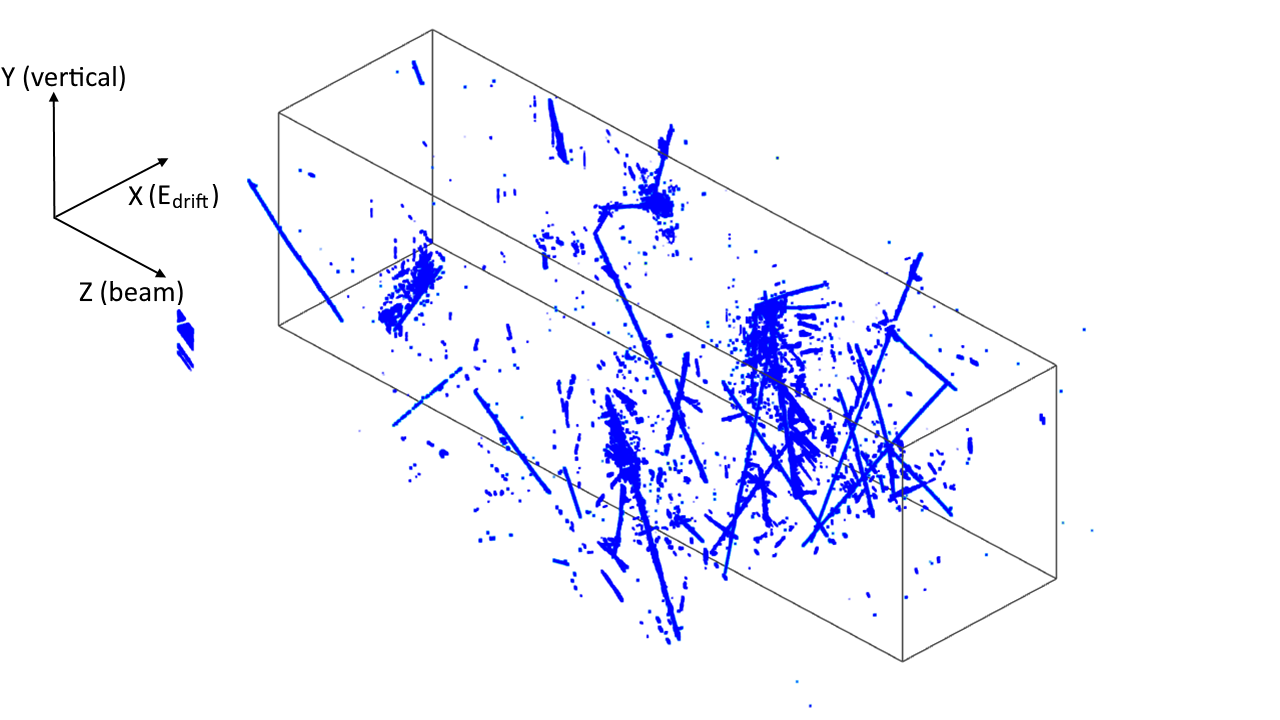}
        \put(10,10){\textbf{\small MicroBooNE Data}}
    \end{overpic}
    \begin{overpic}[width=0.8\figwidth]{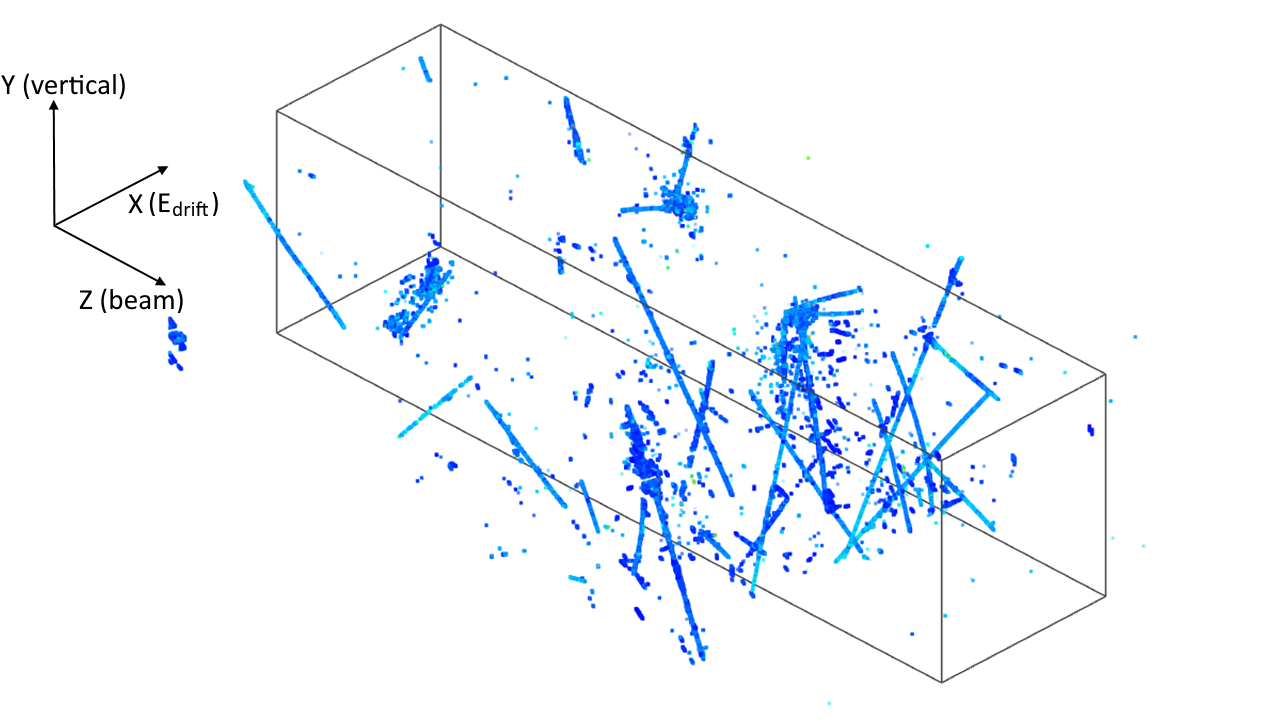}
        \put(10,10){\textbf{\small MicroBooNE Data}}
    \end{overpic}
    \caption{Comparison of the tiling results and the charge solving result from MicroBooNE data (event 41075, run 3493). The solid black box represents the LArTPC active volume with an X-position (converted from the readout time) relative to the neutrino interaction time.
    Only time and geometry information are used in the tiling. 
    Sparsity, positivity, and proximity information are incorporated in the charge solving as described in section~\ref{sec:solving}.
    Top: 3-plane tiling with 70\% active volume. 
    Middle: 2-plane tiling with 97\% active volume.
    Bottom: 2-plane tiling result after the charge solving. 
    The color scale represents the resulting charge values in the charge solving.
  }
  \label{fig:3plane_2plane_tiling}
\end{figure}

Using the time and geometry information, concatenating the 2D blobs in each time slice from tiling provides a 3D image of all the possible charge depositions, as shown in the example in figure~\ref{fig:3plane_2plane_tiling}.
The top panel corresponds to the 3-plane tiling, yielding a 70\% functional volume.
The middle panel corresponds to the 2-plane tiling, providing a 97\% functional volume.
Since it is essential to limit the nonfunctional volume in physics measurements to increase the charge collection efficiency and improve the later reconstruction performance, the next task is to remove the ghosts, which originate from the wire readout ambiguity and worsened by nonfunctional wires in the 2-plane tiling procedure.

\subsection{Charge solving}\label{sec:solving}

Charge is one of the most fundamental bases on which to remove the ghosts.
A system of linear equations can be constructed by relating the measured charge of a hit wire to the unknown charges of the possible hit cells along this wire. 
In practice, after the tiling step, blobs and wire bundles are considered here rather than cells and wires.
The equation can be expressed as follows:
\begin{equation}
    y = Ax,
\label{eq:charge}
\end{equation}
where $y$ is a vector of the integrated measured charges for the hit wire bundles, $x$ is a vector of the unknown charges of all blobs, 
and $A$ is a matrix with its element $A_{ij}=1(0)$ if the blob corresponding to $x_j$ is (not) on the wire bundle corresponding to $y_i$.
We call eq.~\eqref{eq:charge} the imaging equation of the first principle.
In an ideal solution of eq.~\eqref{eq:charge}, the true hit blob will have charges equal to their truth values, and the fake blobs will have zero charge. 
However, even if the charges are measured completely and accurately, eq.~\eqref{eq:charge} generally has no unique solution.
The problem is the result of the fact that there are generally more unknowns than knowns in this system, and this under-determined linear system stems from the wire readout ambiguity.
As a consequence, the matrix $A^TA$ usually does not have full rank and it is not invertible, and the general solution of eq.~\eqref{eq:charge}, $x=(A^T\cdot A)^{-1}\cdot A^T\cdot y$, cannot be used.

As elaborated in ref.~\cite{wc_imaging}, one can find an optimized solution to eq.~\eqref{eq:charge} by making it an optimization problem after applying additional constraints,
\begin{equation}
	\text{minimize} \, ||x||_p, \quad \text{subject to:} \,\, y = Ax,
	\label{eq:lp}
\end{equation}
where $||x||_p = (\sum_i|x_i|^p)^{1/p}$ is the $\ell_p$-norm of a vector $x$.
Since the physics activities in LArTPCs are generally sparse, i.e.~most of the elements of $x$ are zero, the $\ell_0$-norm (a count of the nonzero elements) can be used to seek the most sparse or the simplest solution that explains the measurements.
The minimization of $||x||_0$ can be achieved by removing the unknowns until the linear equation is solvable.  
For example in figure~\ref{fig:tiling_comparison}, there are 25 blobs, while only about 10 hits are true.
One can remove 15 unknowns\footnote{The number of unknowns to be removed is the number of zero eigenvalues of matrix $A^TA$.} out of the 25 to solve the equation and find the ``best'' one satisfying the optimization condition.
However, in this case there are $C_{25}^{10} \approx 3.3 \times 10^{6}$ combinatorial ways to remove the unknowns and in general this optimization is an NP-hard problem that is extremely expensive in computation.
Mathematicians~\cite{compressed_sensing} have discovered that an alternative constraint, the $\ell_1$-norm, can well approximate the $\ell_0$-norm result with a much faster minimization.
This $\ell_1$ technique, also known as {\it compressed sensing}, is widely applied in many other fields for signal processing and computational photography.
As shown in section~\ref{sec:matching_alg}, the compressed sensing technique is also used to perform the many-to-many charge-light matching.

In practice, a chi-square function is constructed to take into account the uncertainties of the measured charge from signal processing~\cite{SP2_paper}, and the compressed sensing technique is implemented by adding an $\ell_1$-regularization term to the chi-square function:
\begin{equation}
    \chi^2 = ||y'-A'x||_2^2 + \lambda ||x||_1,
    \label{eq:chi2_solv_l1_raw}
\end{equation}
where the vector $y$ and $x$ are pre-normalized through $V^{-1} = Q^TQ$ (Cholesky decomposition), 
$y' = Q \cdot y$, $A' = Q \cdot A$, and $\lambda$ regulates the strength of $||x||_1$. 
The matrix $V$ is the real symmetric covariance matrix of the charge measurement uncertainties.
The $\ell_1$-regularized chi-square function is convex with a unique global minimum, enabling fast minimization algorithms such as coordinate descent~\cite{coordesc}. 
An implementation of the coordinate descent method can be found in a Wire-Cell git repository~\cite{wc_coordesc}.
Another constraint, {\it positivity} of the charge (number of the ionization electrons), is added in the coordinate descent method to help remove the ghosts.

So far we have shown the incorporation of charge, sparsity, and positivity to seek the unique solution to the imaging equation of the first principle.
To further improve the robustness of the $\ell_1$-regularization result, {\it proximity} information is incorporated given the fact that the LArTPC is a fully active detector, therefore the measured activities from charged particles are spatially continuous, in contrast to other sampling detectors. 
For those adjacent blobs over different time slices, the regularization strength $\lambda$ is applied with an additional scaling factor of $a^{n}$ to lower the chance of removing the corresponding element in $x$ during the $\ell_1$ minimization.
$n$ represents the number of the adjacent blobs that are connected to the target blob, and $a$ is a predefined scaling factor.  
The final chi-square function in the Wire-Cell imaging is transformed to be:
\begin{equation}
    \chi^2 = ||y'-A'x||_2^2 + \lambda ||\omega \cdot x||_1,
    \label{eq:chi2_solv_l1}
\end{equation}
where $\lambda$ is an overall regularization strength parameter, and $\omega_i=a^{n_i}$ is the weight for $x_i$ as described in the text. The two hyper-parameters $\lambda$ and $a$ are tuned by data events.
Note that $y$ is a vector of the integrated measured charge for each wire bundle in the tiling and $x$ is a vector of the charge to be solved for each blob. 

The bottom panel of figure~\ref{fig:3plane_2plane_tiling} shows the result after applying the charge solving procedure to the 2-plane tiling result in the middle panel. It is clear that the ghosts are further reduced and the 3D voxels are now associated with different charge values, which correspond to the solution $x$ of the imaging equation of the first principle. 
As elaborated in section~\ref{sec:matching_alg}, such 3D charge solving is critical to predict the scintillation light signals for each PMT, allowing for comparison to and matching with the observed light information.

\subsection{De-ghosting}\label{sec:deghosting}

The amount of ghosts is considerably reduced after the charge solving but the result is still unsatisfactory.
The sparsity combined with the proximity is already incorporated in the charge solving to resolve the wire readout ambiguity; however, this procedure is performed in a ``local'' manner restricted within each time slice or over adjacent time slices.
Within the 3D imaging, all connected blobs in 3D space are grouped together as {\it proto-clusters}. 
A proto-cluster does not necessarily group all related TPC activities from a cosmic-ray muon or a neutrino interaction, since there might be true or artificial gaps in the 3D image.
The principle of the sparsity of the LArTPC physics activities will be further used in a ``global'' manner to reconstruct the sparsest 3D images of the TPC activities by removing the less prominent proto-clusters that are redundant to explain the observed 2D-projection measurements from wire planes.
Following this philosophy, a dedicated algorithm, {\it deghosting}, is developed to remove the residual ghosts based on their two main characteristics.
\begin{itemize}[noitemsep, nolistsep, leftmargin=*]
    \item[] \textbf{Position} - the ghosts are mainly present in areas where one wire plane is nonfunctional.
    \item[] \textbf{Projection} - the ghost proto-clusters, mostly track-like, are generally redundant in all three 2D projection views of wire-versus-time.
\end{itemize}
The area with one nonfunctional wire plane provides significantly less constraints in the tiling and charge solving. 
This introduces a large ambiguity in the wire readout and a high probability of the presence of ghosts.
As indicated in eq.~\eqref{eq:chi2_solv_l1}, the 3D space points are reconstructed by matching the charge for all the functional wire planes, and the charge that forms a ghost proto-cluster must come from the original measurement from a genuine track.
Generally speaking, in one of the wire-versus-time views, the ghost tracks are in the nonfunctional region, and match or coincide with genuine tracks in the other two views.
So an effective way to identify ghosts is to check each individual wire-versus-time view to test if a proto-cluster is present as redundant pieces or missing pieces of another more prominent proto-cluster.

\begin{figure}[!htpb] 
  \centering
\begin{overpic}[width=0.9\figwidth]{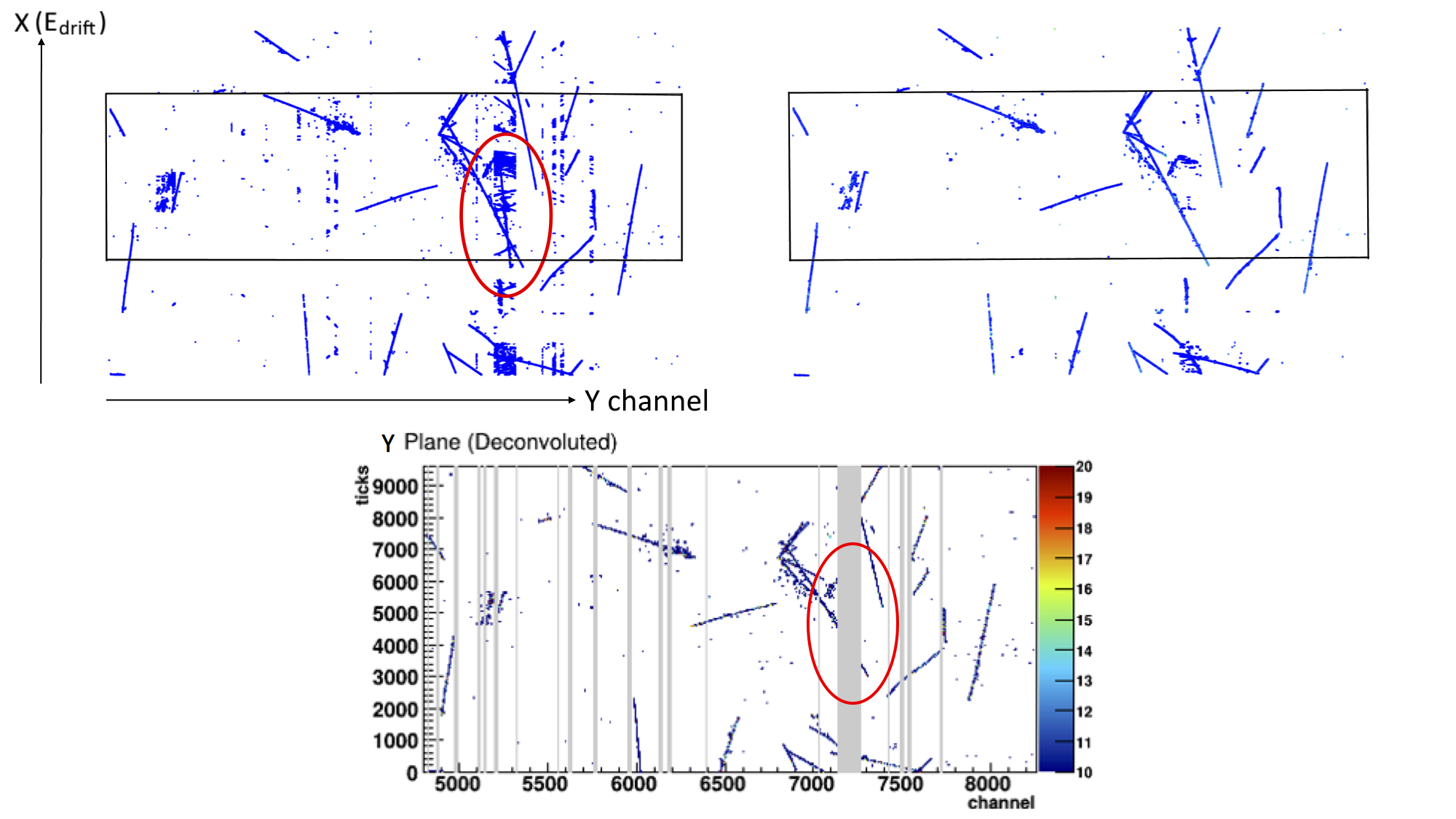}
  \put(52,25){\scriptsize MicroBooNE Data}
  \put(7.5,52){\scriptsize MicroBooNE Data}
  \put(54,52){\scriptsize MicroBooNE Data}
  \end{overpic}
  \caption{
    Top left: 2D projection to the Y plane's wire-versus-time view of the reconstructed
    3D image without the de-ghosting algorithm. The black box represents the full detector length in Y and the full cathode-to-anode drift distance in X. The red circle corresponds to the same volume in the TPC as in figure~\ref{fig:uplane} and figure~\ref{fig:vplane}.
    Top right: after the de-ghosting algorithm. 
	Bottom: Original charge measurement. The vertical axis bin width (time) is 4 ticks (2 microseconds), and the color scale represents the number of ionization electrons scaled by a factor 1/500 (comparable to ADC counts from raw waveforms). The nonfunctional wires are marked in gray.
  }
  \label{fig:wplane}
\end{figure}

\begin{figure}[!htbp]    
  \centering
\begin{overpic}[width=0.9\figwidth]{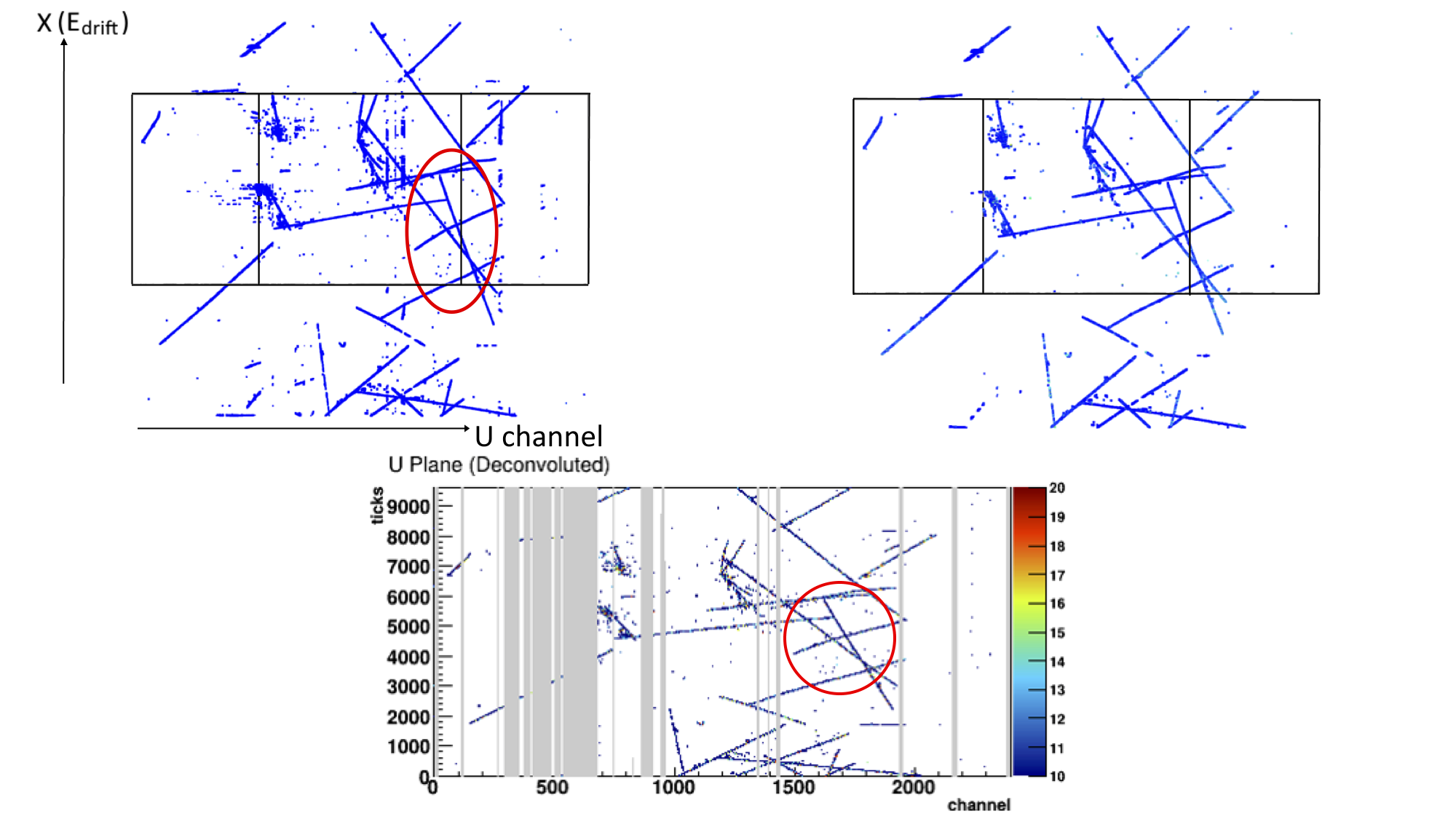}
  \put(49,23){\scriptsize MicroBooNE Data}
  \put(10,52.25){\scriptsize MicroBooNE Data}
  \put(59,52){\scriptsize MicroBooNE Data}
\end{overpic}
  \caption{
    Top left: 2D projection to the U plane's wire-versus-time view of the reconstructed
    3D image without the de-ghosting algorithm. The black box represents the full array of U channels and the full cathode-to-anode drift distance in X. The red circle corresponds to the same volume in the TPC as in figure~\ref{fig:wplane} and figure~\ref{fig:vplane}.
    Top right: after the de-ghosting algorithm. 
	Bottom: Original charge measurement. Y-axis bin width (time) is ticks (2 microseconds), and Z-axis value represents the number of ionization electrons scaled by a factor 1/500 (comparable to ADC counts from raw waveforms). The nonfunctional wires are marked in gray.
  }
  \label{fig:uplane}
\end{figure}

\begin{figure}[!htpb]
  \centering
\begin{overpic}[width=0.9\figwidth]{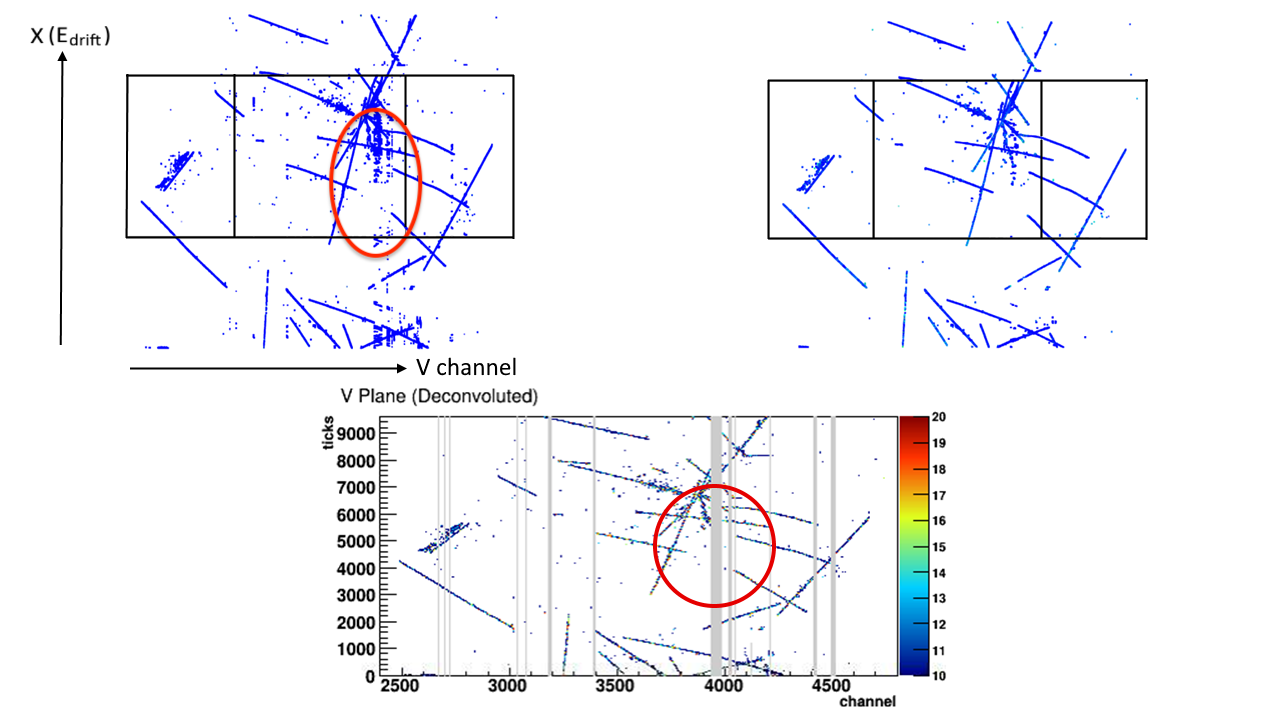}
  \put(50,24){\scriptsize MicroBooNE Data} 
  \put(10,51){\scriptsize MicroBooNE Data}
  \put(60,50.5){\scriptsize MicroBooNE Data}
\end{overpic}
  \caption{
    Top left: 2D projection to the V plane's wire-versus-time view of the reconstructed
    3D image without the de-ghosting algorithm. The black box represents the full array of V channels and the full cathode-to-anode drift distance in X. The red circle corresponds to the same volume in the TPC as in figure~\ref{fig:wplane} and figure~\ref{fig:uplane}.
    Top right: after the de-ghosting algorithm. 
	Bottom: Original charge measurement. Y-axis bin width (time) is ticks (2 microseconds), and Z-axis value represents the number of ionization electrons scaled by a factor 1/500 (comparable to ADC counts from raw waveforms). The nonfunctional wires are marked in gray.
  }
  \label{fig:vplane}
\end{figure}

\begin{figure}[!htbp]    
  \centering
  \begin{overpic}[width=1.0\figwidth]{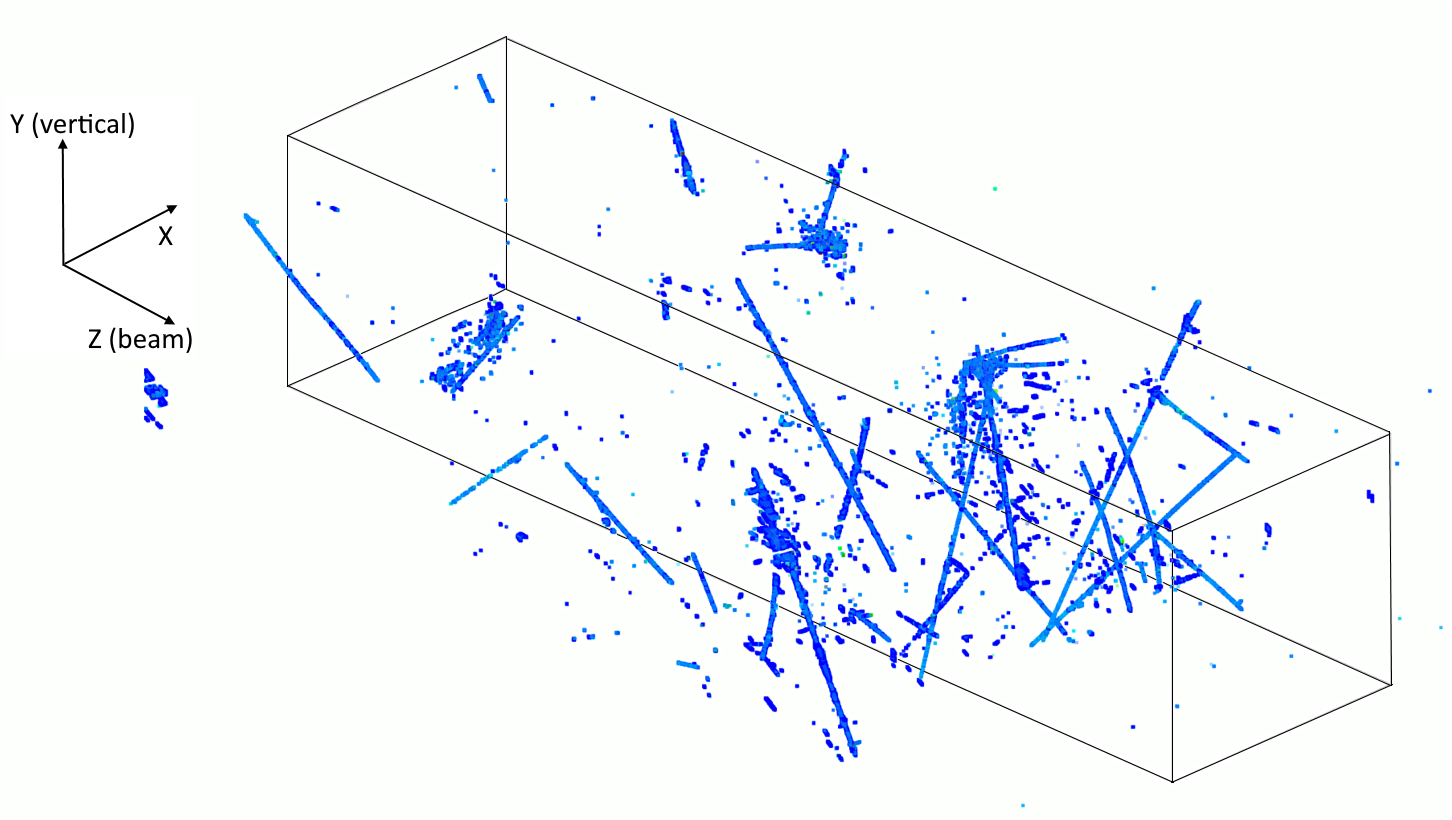}
      \put(25,10){\textbf{\small MicroBooNE Data}}
  \end{overpic}
  \begin{overpic}[width=1.0\figwidth]{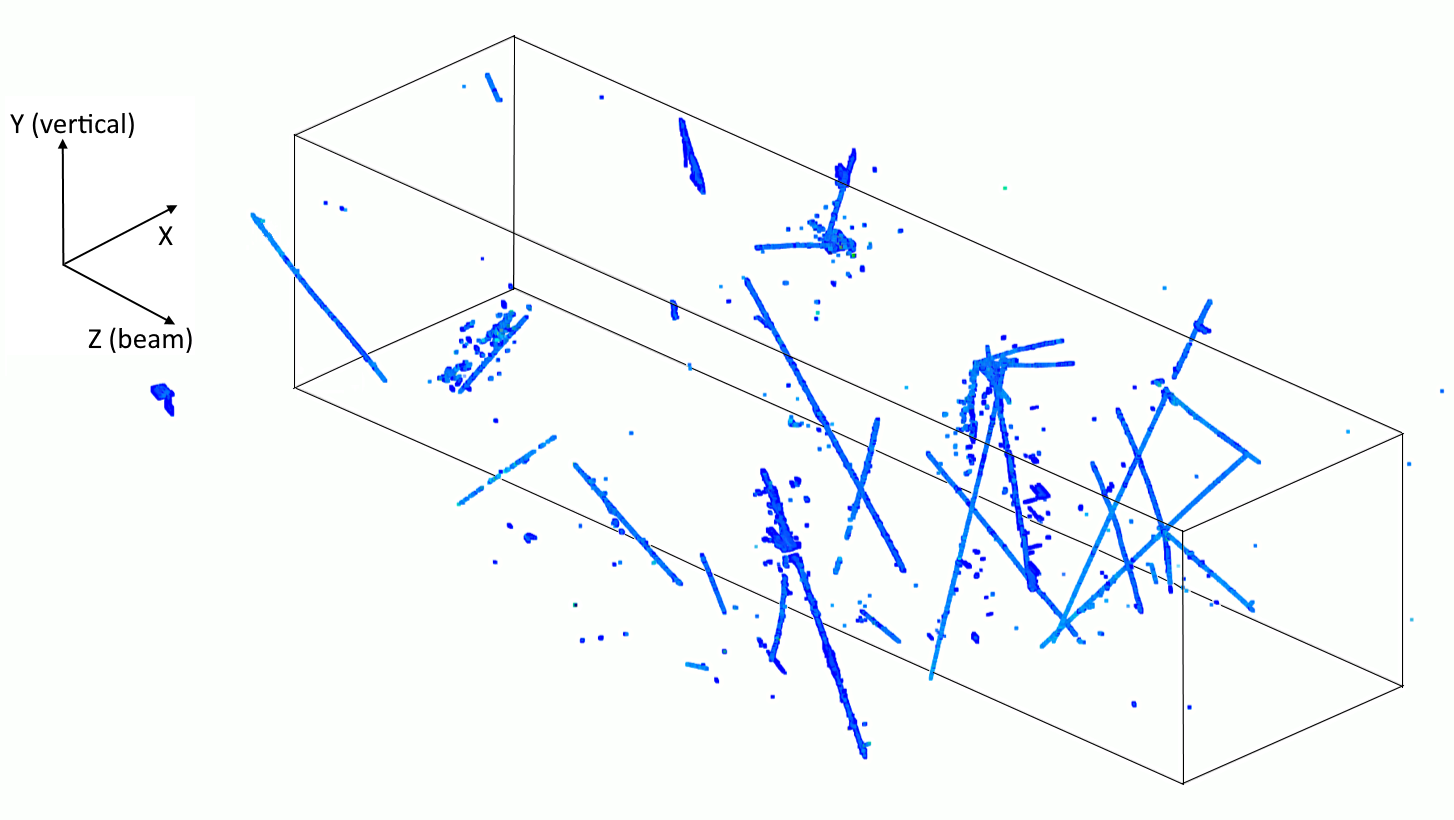}
      \put(25,10){\textbf{\small MicroBooNE Data}}
  \end{overpic}
    \caption{ Comparison of the 3D imaging results from MicroBooNE data (event 41075, run 3493) without (top) and with (bottom) the de-ghosting algorithm.
    Ghosts are significantly reduced after the de-ghosting. The solid black box represents the LArTPC active volume with an X-position (converted from the readout time) relative to the neutrino interaction time. The color scale indicates the charge density.
  }
  \label{fig:deghost_com}
\end{figure}

Below is an example of a MicroBooNE data event to illustrate the identification of ghosts.
Figure~\ref{fig:wplane}, figure~\ref{fig:uplane}, and figure~\ref{fig:vplane} show the 2D projections of the 3D image and the original charge measurements from the three wire planes: Y, U, and V, respectively.
In each figure, the top left is the result before de-ghosting and the top right is the result after de-ghosting,
and the bottom is the original charge measurement with the nonfunctional wires marked in gray.
The red circles in the three figures correspond to the same 3D volume in the TPC.
As can be seen, the ghosts in the Y plane's (collection plane) nonfunctional region overlap with the measurements in the U plane (induction plane), and those ghost proto-clusters are redundant since other proto-clusters can explain the same measurements in the U plane. 
In figure~\ref{fig:uplane}, the images in the red circle are nearly the same before and after the de-ghosting, and it hints that ghost tracks are redundant in terms of explaining the measured charge.
The ghosts in the V plane exhibit similar behavior, as shown in figure~\ref{fig:vplane}.
Note that after one round of the de-ghosting, another round of the charge solving is needed to reclaim the charge carried by the ghosts. 
The practical 3D imaging procedure is therefore iterative, and is summarized in section~\ref{sec:imaging_summary}.
Figure~\ref{fig:deghost_com} shows the imaging results with and without de-ghosting.

The occurrence of ghosts is aggravated by the inefficiency of the noise filtering~\cite{noise_filter_paper} and the signal processing~\cite{SP1_paper}, which may filter out some charges along the isochronous tracks as coherent noise, or fail to reconstruct the charges of prolonged tracks (a long signal along the drift direction) because of the bipolar cancellation of the induction plane signals\footnote{The recent advancement in TPC signal processing by leveraging the deep learning techniques\cite{Yu:2020wxu} is expected to reduce this signal inefficiency.}.
Consequently, one or two of the 2D wire-versus-time views of the charge measurements may have gaps along a track even on the functional wire planes.
This gap will lead to a separation in the 3D image since the charge measurements across the wire planes can no longer match. 
Consequently, the successfully reconstructed charges from the other wire planes corresponding to the gap could interplay with the charge measurements from other tracks and be erroneously explained by ghosts.
The removal of such ghosts requires a bridging of the gaps to connect the separated pieces of the track.
This will be further discussed in section~\ref{sec:clustering} in the context of 3D clustering, which results in the final TPC clusters based on the proto-clusters.

\subsection{Summary}~\label{sec:imaging_summary}
The actual procedure of the application of Wire-Cell 3D imaging in MicroBooNE is iterative,
containing multiple rounds of tiling, charge solving, and de-ghosting.
The number of iterations is based on an empirical evaluation based on data events used during the algorithm development. More iterations do not appear to significantly improve the results of the 3D imaging. 
A summary of procedures is shown in table~\ref{tab:imaging_flow}.
\begin{table*}[!htpb]
    \caption{Summary of the procedures of the Wire-Cell 3D imaging, including the 2-plane ($\ge$2 wire planes) tiling,
    the charge solving, and the de-ghosting.}
  \medskip
  \renewcommand{\arraystretch}{1.1} \centering
  \begin{tabular}{|l|l|}
    \hline
    Step & Description \\\hline
      1 & 2-plane tiling\\\hline
      2 & De-ghosting \\\hline
      3 & 1st round of charge solving \\\hline
      4 & 2nd round of charge solving with reweighting for connected blobs \\\hline
      5 & Repeat the steps 2, 3, 4 \\\hline
      6 & Repeat the steps 2, 3, 4 again \\\hline
  \end{tabular}\label{tab:imaging_flow}
\end{table*}

Since MicroBooNE is a near-surface detector with limited cosmic ray shielding, 20--30 cosmic-ray muons per event are input to the Wire-Cell imaging process in the full readout window of 4.8~ms. 
The time and memory consumption are practical issues to be addressed in the optimization and finalization of the algorithms.
Using $\sim$10k MicroBooNE data events, the average time and memory consumption (a single-threaded program) is estimated to be about 2~minutes and less than 2~GB on an Intel(R) Core(TM) i7-4790K CPU @ 4.00GHz. Most of the memory is used by the tiling to initialize and index the blobs from each time slice.
Most of the time is consumed by the charge solving and de-ghosting, which are critical to the quality of the 3D images.

The goal of the Wire-Cell imaging is to reconstruct the 3D image of the ionization electrons independently of the event topology and prior to the application of pattern recognition techniques (such as those presented in Ref.~\cite{uboone-pandora}).
The reconstructed 3D image is an input to the subsequent reconstruction, e.g.~the charge-light matching, to distinguish the in-beam neutrino candidate from the cosmic-ray backgrounds.
The 3D charge associated with each reconstructed space point is used in the prediction of PMT light signals. 

Isochronous tracks present a common problem in the LArTPC 3D imaging, as the wire readout ambiguity is drastically increased in the time slice containing them.
In Wire-Cell 3D imaging, this issue is mitigated by introducing tiling. On the other hand, the blobs of the isochronous tracks are significantly broadened, leading to a much worse 2D spatial resolution within the time slice, i.e. in the nominal Y-Z projection view, or the U-Z and V-Z wire plane views. An example can be found in figure~\ref{fig:coh_noise_removal}. 
Improvement of spatial resolution can be achieved via trajectory fitting in a later reconstruction stage. This is beyond the scope of this paper and will be presented in a future publication~\cite{Wire-Cell-Generic-PRD}. 

Because of the existence of  nonfunctional channels, a 2-plane ($\ge$2) tiling strategy is adopted to significantly enhance the image reconstruction efficiency at the cost of introducing more ghosts. 
Time, geometry, charge, sparsity, positivity, and proximity information is utilized to overcome the wire readout ambiguity and to remove the ghosts.
In addition to the de-ghosting steps performed during 3D imaging, another round of de-ghosting is performed in the clustering stage as discussed in section~\ref{sec:clustering_deghost}.
Quantitative evaluations of Wire-Cell 3D imaging in various cases are presented in section~\ref{sec:evaluation}.

\clearpage

\section{Matching Charge and Light}\label{sec:clustering_matching}

As introduced in section~\ref{sec:intro}, each triggered event in MicroBooNE contains a \SI{4.8}{\milli\second} TPC readout and a \SI{6.4}{\milli\second} PMT readout.
The Wire-Cell imaging reconstructs a 3D image of the TPC activities, which includes both cosmic-ray muons and a neutrino interaction if present.
The PMTs detect the scintillation light on a much shorter timescale than the drifting of the ionization electrons in the TPC, so it can be used to provide the interaction (start) time once it is paired with the corresponding charge signals.
The 32 PMTs' waveforms from a cosmic-ray muon or a neutrino interaction are processed to reconstruct a {\it flash}, which is a group of the PMT signals close in time (e.g.~within 100 ns). The detailed definition of a PMT flash can be found in section~\ref{sec:light}.
Typically, the cosmic-ray muon rate is 5.5 kHz in the TPC active volume, so there are 20-30 cosmic-ray muons within the 4.8 ms TPC readout window. 
Within the 6.4 ms PMT readout window, there are 40-50 PMT flashes which correspond not only to the activities inside the TPC but also those outside the TPC but within the LAr volume inside the cryostat.

As described in the previous section, the proto-cluster in the 3D imaging step is solely based on proximity, while a physical signal initiated by a primary particle's interactions could have disconnected pieces, such as from secondary neutral particles or because of imperfect signal processing or reconstruction. In order to accurately and robustly pair the TPC activities to the PMT flashes, an interaction 3D clustering is developed to group the proto-clusters further into a TPC {\it cluster}, which then represents signals initiated by an individual primary particle such as from a cosmic-ray muon or a neutrino interaction.

Given the TPC clusters and PMT flashes, a novel algorithm, {\it many-to-many charge-light matching}, is developed to match the clusters and the flashes simultaneously based on the predicted light signals generated by the 3D TPC clusters and the measured light signals from PMT flashes.
The TPC cluster(s) matched to an in-beam PMT flash is then regarded as a beam neutrino candidate.
All the remainders are rejected as cosmic-ray muons.
Compared to a previous single-to-single track-light-matching algorithm as described in ref.~\cite{Caratelli:2020dlg}, many-to-many charge-light matching enhances the cosmic rejection power and results in a cleaned-up 3D image of the neutrino activities.

The algorithms of the 3D clustering and the PMT light reconstruction are delineated in section~\ref{sec:clustering} and section~\ref{sec:light}, respectively.
The details of the many-to-many charge-light matching procedure are described in section~\ref{sec:matching_alg}.

\subsection{3D clustering}\label{sec:clustering}
Clustering as described in this section aims to group proto-clusters according to their physics origin into clusters.
This step is an initial separation of neutrino and cosmic activities, and is necessary to efficiently perform the subsequent many-to-many charge-light matching.

\begin{figure}[!thbp] 
  \centering
  \begin{overpic}[width=0.45\figwidth]{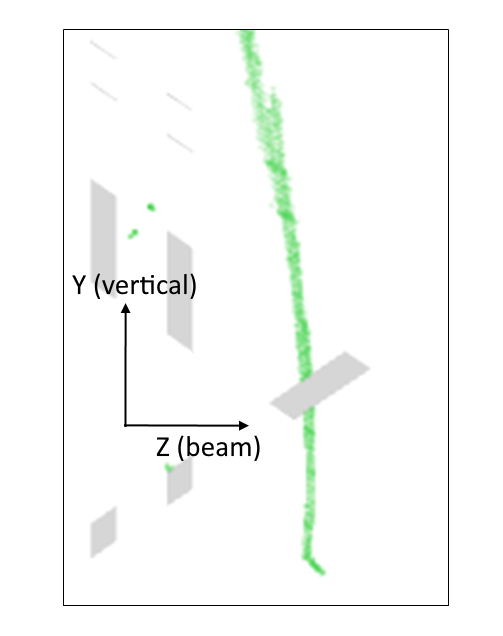}
    \put(10.5,97){\footnotesize MicroBooNE Data}
  \end{overpic}
    \caption{
      Zoomed in Y-Z view of a cosmic muon with a gap because of the nonfunctional regions. The nonfunctional regions are shown in gray.}
  \label{fig:gap_deadregion}
\end{figure}

\begin{figure}[!thbp]
  \centering
  \begin{overpic}[width=0.95\figwidth]{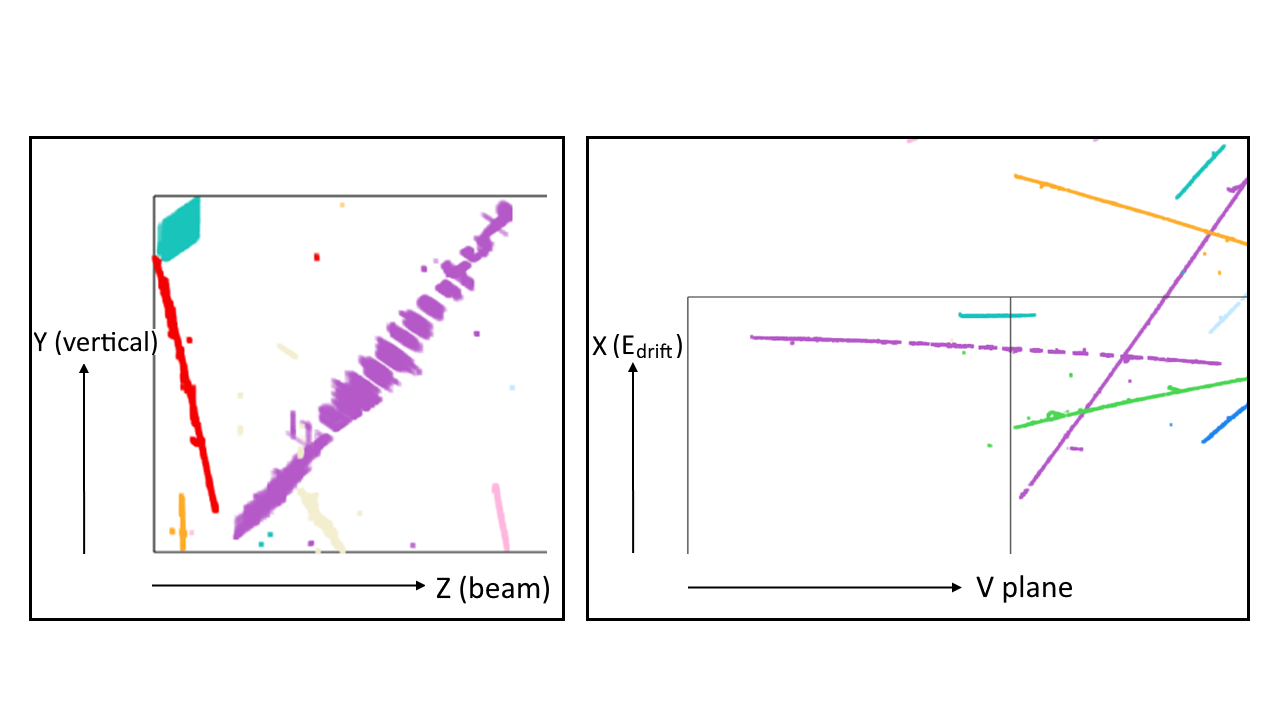}
    \put(2.5,46.5){\footnotesize MicroBooNE Data}
    \put(46,46.5){\footnotesize MicroBooNE Data}
  \end{overpic}
  \caption{Gaps along an isochronous track in different zoomed-in 2D views.
	(Left) Y-Z view of an isochronous track (magenta) from a MicroBooNE data event.
    (Right) X-V view of the same event. V plane direction represents the wire pitch direction of the V wire plane. 
    The black lines in the inner figures correspond to the boundaries of the 3D LArTPC active volume. Since the 3D boundaries are projected to the 2D visual shown, sometimes edges of the rectangular prism active volume appear in the center of the image.
    Cluster membership is indicated by uniform color within each plot. Some distant clusters could be marked in the same color because of a finite number of visibly distinctive colors available in the event display.
  }
  \label{fig:coh_noise_removal}
\end{figure}

\begin{figure}[!thbp] 
  \centering
  \begin{overpic}[width=0.95\figwidth]{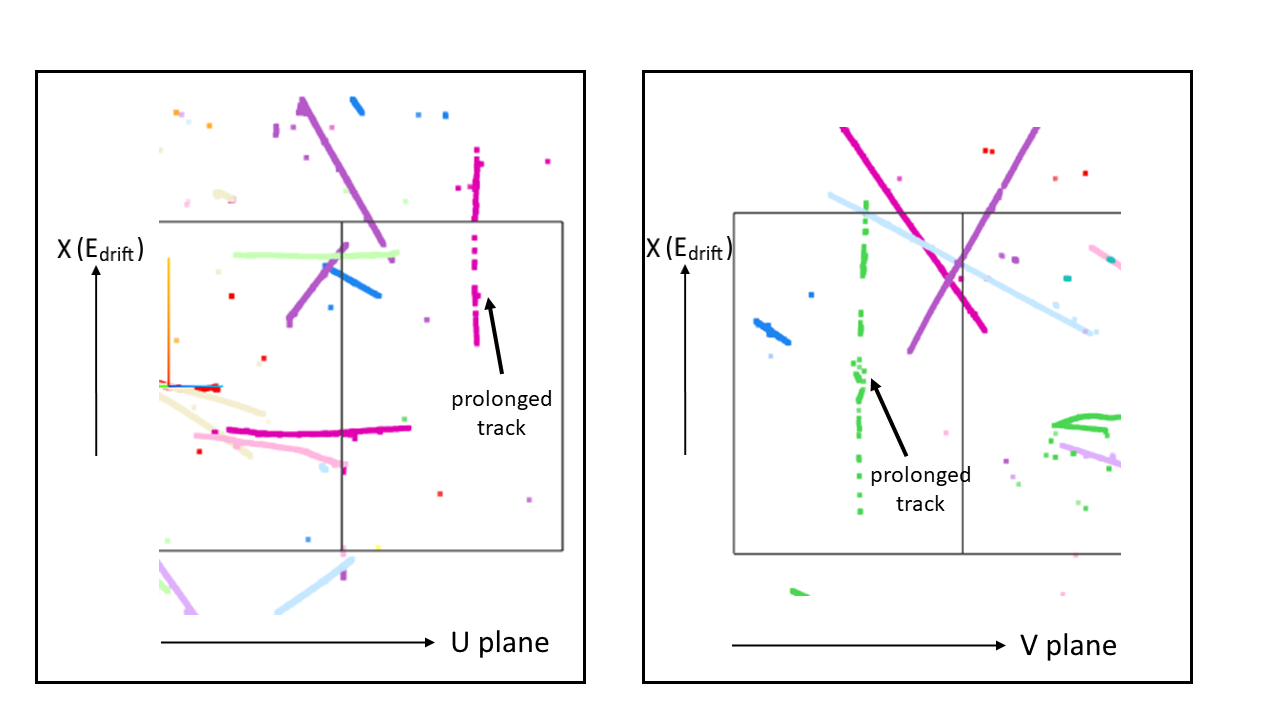}
    \put(3,51.5){\footnotesize MicroBooNE Data}
    \put(51,51.5){\footnotesize MicroBooNE Data}
    \end{overpic}
  \caption{Gaps along prolonged tracks in different zoomed-in 2D views.  
    (Left) X-U view for a prolonged cosmic muon track (magenta) from a MicroBooNE data event.
    (Right) X-V view for a prolonged cosmic muon track (green) from another MicroBooNE data event. 
    The black lines in the inner figures correspond to the boundaries of the 3D LArTPC active volume. Since the 3D boundaries are projected to the 2D visual shown, sometimes edges of the rectangular prism active volume appear in the center of the image.
    Cluster membership is indicated by uniform color within each plot. Some distant clusters could be marked in the same color because of a finite number of colors available in the event display.
  }
  \label{fig:gaps_prolong}
\end{figure}

Proto-clustering, which solely relies on proximity, has been carried out in the 3D imaging step (section~\ref{sec:deghosting}).
However, it doesn't meet the requirement of carrying out a high performance charge-light matching because of the following issues:

\textbf{Gaps}:~~The presence of gaps compromises the effectiveness of a proto-clustering based on proximity.
A gap mainly results from:
1) the $\sim$3\% nonfunctional regions, as shown in figure~\ref{fig:gap_deadregion};
2) incorrect removal of parts of the isochronous tracks (close to parallel to the wire planes) by the coherent noise filter, as shown in figure~\ref{fig:coh_noise_removal};
and 3) failures of the signal processing for parts of the prolonged tracks (a long signal along the drift direction) as shown in figure~\ref{fig:gaps_prolong},

\textbf{Coincidental overlap}:~~For LArTPCs operating near the surface (such as MicroBooNE), the
detector is bombarded by a large number of cosmic-ray muons. Although the cosmic-ray muons
generally pass through the detector at different time and locations, the 3D images from different TPC clusters, e.g.~two muons, can appear to be  connected when ionization electrons of different activities
arrive at the same location of the wire plane at the same time. 
This leads to an over-clustering of space points, causing mistakes in the charge-light matching.

\textbf{Residual ghosts}:~~The de-ghosting algorithm described in section~\ref{sec:deghosting}
is not completely sufficient because of the incomplete or improper proto-clustering as the two items explained above.

\textbf{Separated clusters from a neutrino interaction}:
Neutral particles from neutrino interactions with argon nuclei are very likely to travel some distance before
depositing their energy. The secondary charged particles from these neutral particles are therefore separated from the neutrino primary vertices.
For example, a $\pi^0$ is a potential final state particle of a neutrino interaction with an argon nucleus. 
It generally deposits its energy through two
electromagnetic (EM) showers from its decay $\gamma$'s. The two $\gamma$'s are in principle detached from the neutrino primary
vertex and other final state particles.
A dedicated algorithm is needed to group these separated particles from the primary interaction into a single cluster.

\subsubsection{Clustering in the presence of gaps}\label{sec:clustering_gap}
Clustering across gaps mainly relies on two sets of information: distance and directionality.
If two proto-clusters are close to each other along a line, the gap may be bridged and the two proto-clusters are grouped into a single cluster.
Many existing tools and algorithms operating on a point cloud (a collection of many 3D points) can be directly used, as a TPC cluster is a collection of the reconstructed 3D space points.
The distance between two clusters (point clouds) is defined as the minimal distance between a pair of space points, one from each respective point cloud. 
To calculate this distance rapidly, 
the k-d (k-dimensional) tree based algorithm as implemented in the ``naoflann'' package~\cite{blanco2014nanoflann} is employed.
Once the minimal distance and its direction are obtained, its direction is compared with the directions of the two proto-clusters. These are found using a voting scheme inspired by the Hough transformation~\cite{HoughT}.
The directional vector, parameterized by a polar angle and an azimuthal angle, is calculated for each point.
The most probable value in the 2D distribution of polar-versus-azimuthal is then taken as the primary direction of the point cloud.
Given the minimal distance vector and the two proto-cluster directions, the two proto-clusters are grouped (or not) based on their distance and consistency in directions.
In practice, a set of criteria are developed and optimized by analyzing hundreds of data events from various topologies.

\begin{figure}[!thbp]  
  \centering
  \begin{overpic}[width=0.95\figwidth]{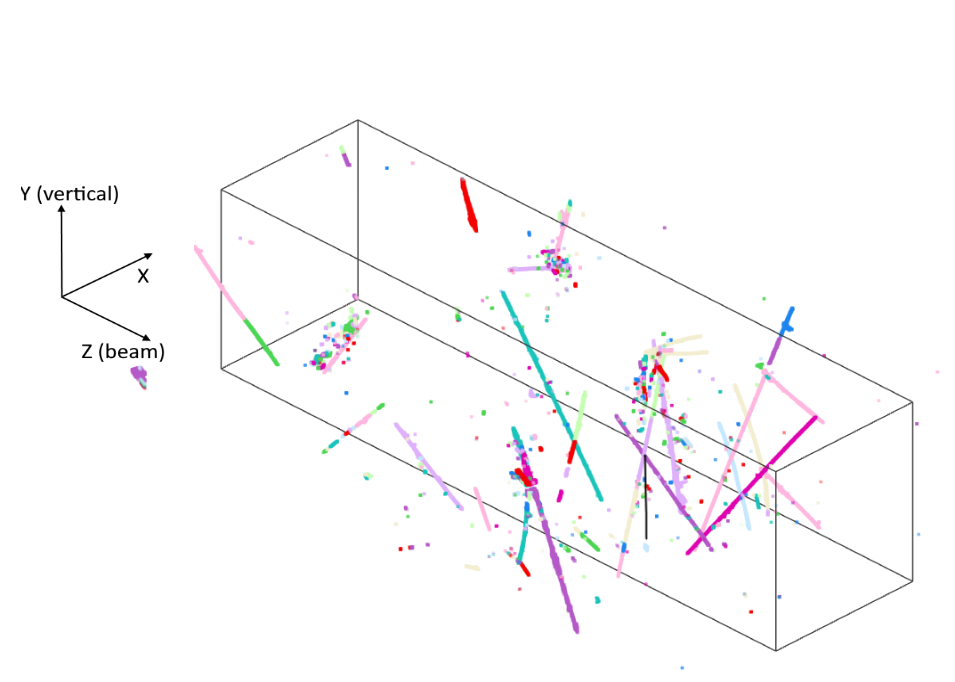}
    \put(10,10){\textbf{\small MicroBooNE Data}}
  \end{overpic}
  \begin{overpic}[width=0.95\figwidth]{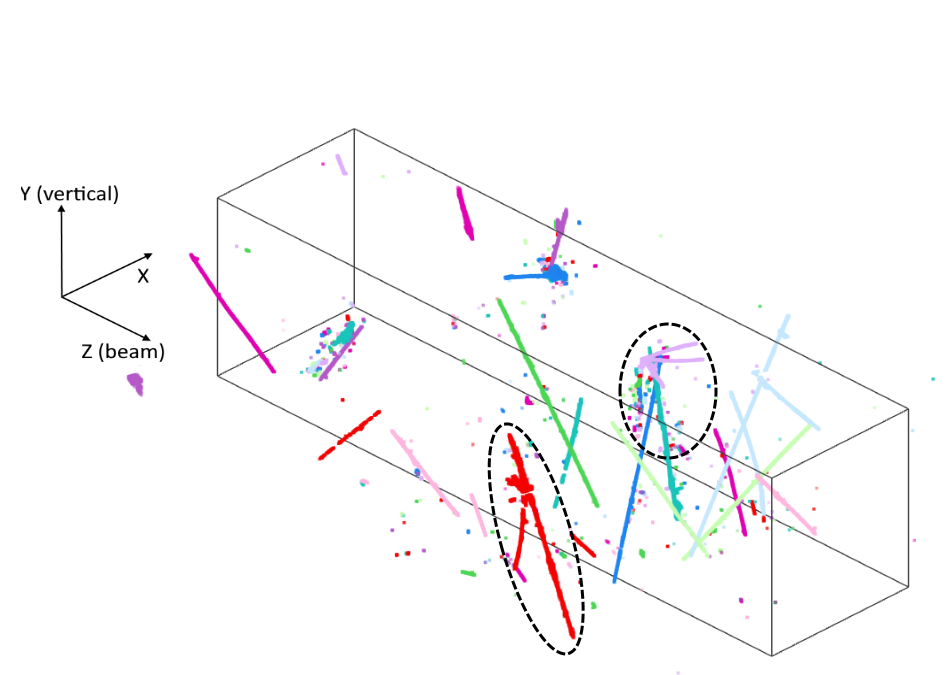}
    \put(10,10){\textbf{\small MicroBooNE Data}}
  \end{overpic}
    \caption{Demonstration of the effectiveness of the algorithm of bridging gaps.
    The solid black box represents the LArTPC active volume with an X-position (converted from the readout time) relative to the neutrino interaction time.
    Top: proto-clusters solely based on proximity. 
    Bottom: clusters after the application of the algorithm of bridging gaps. 
    The two circles indicate remaining clustering issues, e.g.~over-clustering of cosmic-ray muons and under-clustering of neutrino interactions.
    Cluster membership is indicated by uniform color.
  }
  \label{fig:full_gap}
\end{figure}

Figure~\ref{fig:full_gap} shows a comparison of the results before and after applying this clustering algorithm. Separate proto-clusters are successfully grouped into individual clusters.
Each cluster is marked by a different color. Some distant clusters are properly separated but they are in the same color because of a finite number of visibly distinctive colors available in the event display. In the bottom panel of figure~\ref{fig:full_gap}, one can find that there are still other clustering issues. For instance, there are two connected cosmic-ray muons and an incomplete neutrino cluster as indicated by the two black circles. These are dealt with using additional clustering algorithms as introduced in section~\ref{sec:clustering_sep} and section~\ref{sec:clustering_neutrino}. 
The resulting clusters are presented in figure~\ref{fig:neutrino1}.

\subsubsection{Separation of coincidental overlap clusters}\label{sec:clustering_sep}
In this section, we describe the algorithm to separate a ``coincidental overlap'' cluster,
and the steps are summarized in table~\ref{tab:sep_flow}.
\begin{table*}[!htp]
  \caption{Steps of the separation of a ``coincidental overlap'' cluster.}  
\medskip
\renewcommand{\arraystretch}{1.1} \centering 
\begin{tabular}{|c|c|}
  \hline
  Step & Key operations \\\hline
  1   & Identification of ``coincidental overlap'' cluster\\\hline
  2  & Find two end points of a primary track \\\hline
  3 & Form trajectory of this primary track  \\\hline
  4  & Collect space points of this primary track  \\\hline
  5   & Remove this primary track and repeat this procedure \\\hline
\end{tabular}\label{tab:sep_flow}
\end{table*}

The first step is to identify the ``coincidental overlap'' cluster.
Principal component analysis (PCA) is performed on each cluster after the bridging of gaps as described in section~\ref{sec:clustering_gap}. 
For a single-track-like cluster, only the primary component (axis) of the PCA has a significantly
larger eigenvalue in the data correlation matrix.
This is generally not true for a ``coincidental overlap'' cluster in which two or more tracks are crossing.
Once a candidate ``coincidental overlap'' cluster is identified, the sub-clusters representing different physical interactions are to be identified and separated one by one.

The separation of each sub-cluster starts with identifying the two end points of a primary track in this cluster. A primary track is the one that best matches one of the primary PCA axes, i.e.~the longest along this primary PCA axis.
Firstly, the quickhull~\cite{quickhull} algorithm
operates on the 3D space points of a coincidental overlap cluster to obtain the 3D convex hull, which is the smallest convex shape that contains all the space points.
The two end points of the current primary track must be contained or in close proximity with the convex hull's vertices.
Secondly, the nearby points around each convex hull's vertex are grouped together to form test clusters. 
The largest test clusters are used to discover the end points of the primary track, and this requires 1) a small distance to the PCA primary component;
2) a consistent direction of the test cluster with the PCA primary component. In general, such end points can always be found for a prominent cluster.
Once an end point is identified,
a Kalman-filter-based technique is used to crawl along this primary track until the other end point is determined.
Given the two end points, the trajectory of this primary track is obtained using a graph theory operation, the Dijkstra's shortest path~\cite{dijkstra}.
The connected component algorithm from graph theory is then used to collect the space points associated with this trajectory and form a sub-cluster.
After removing this sub-cluster from the current primary track, the remaining cluster is further examined and sub-clusters are removed until only one primary track is left.
Each removed sub-cluster is taken as an individual cluster in the end.

\begin{figure}[!thbp]   
  \centering
  \begin{overpic}[width=0.95\figwidth]{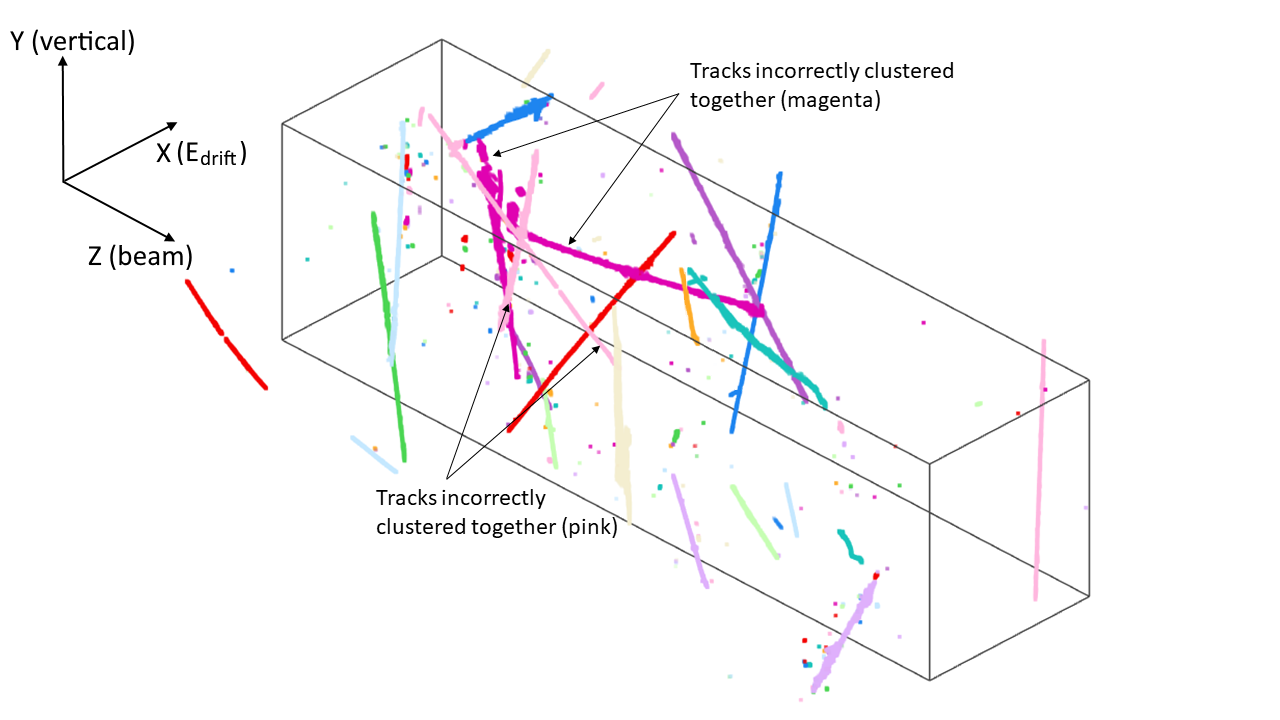}
    \put(10,10){\textbf{\small MicroBooNE Data}}
  \end{overpic}
  
  \begin{overpic}[width=0.95\figwidth]{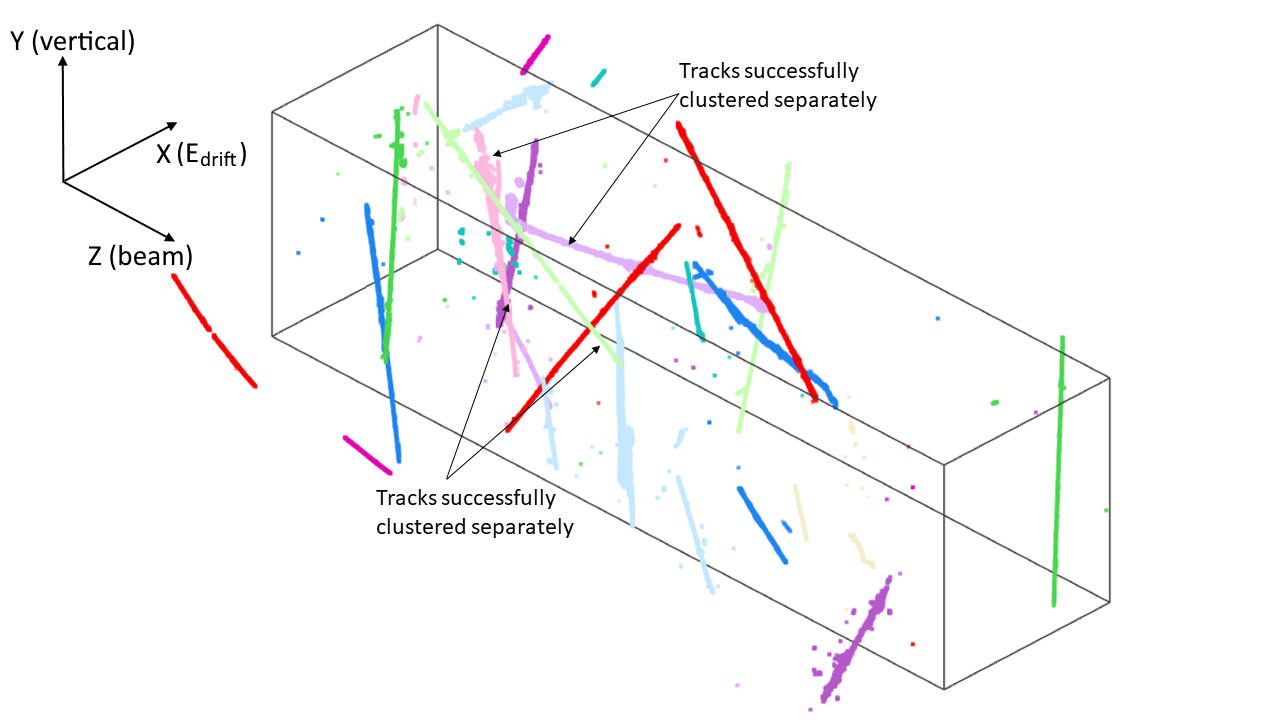}
    \put(10,10){\textbf{\small MicroBooNE Data}}
  \end{overpic}
  \caption{ Demonstration of the effectiveness of the clustering algorithm to separate a ``coincidental overlap'' cluster. The solid black box represents the LArTPC active volume with an X-position (converted from the readout time) relative to the neutrino interaction time. The top and bottom panels show the clusters before and after applying this algorithm. Cluster membership is indicated by uniform color.}
    \label{fig:sep2}
\end{figure}

\begin{figure}[!thbp] 
  \centering
  \begin{overpic}[width=0.95\figwidth]{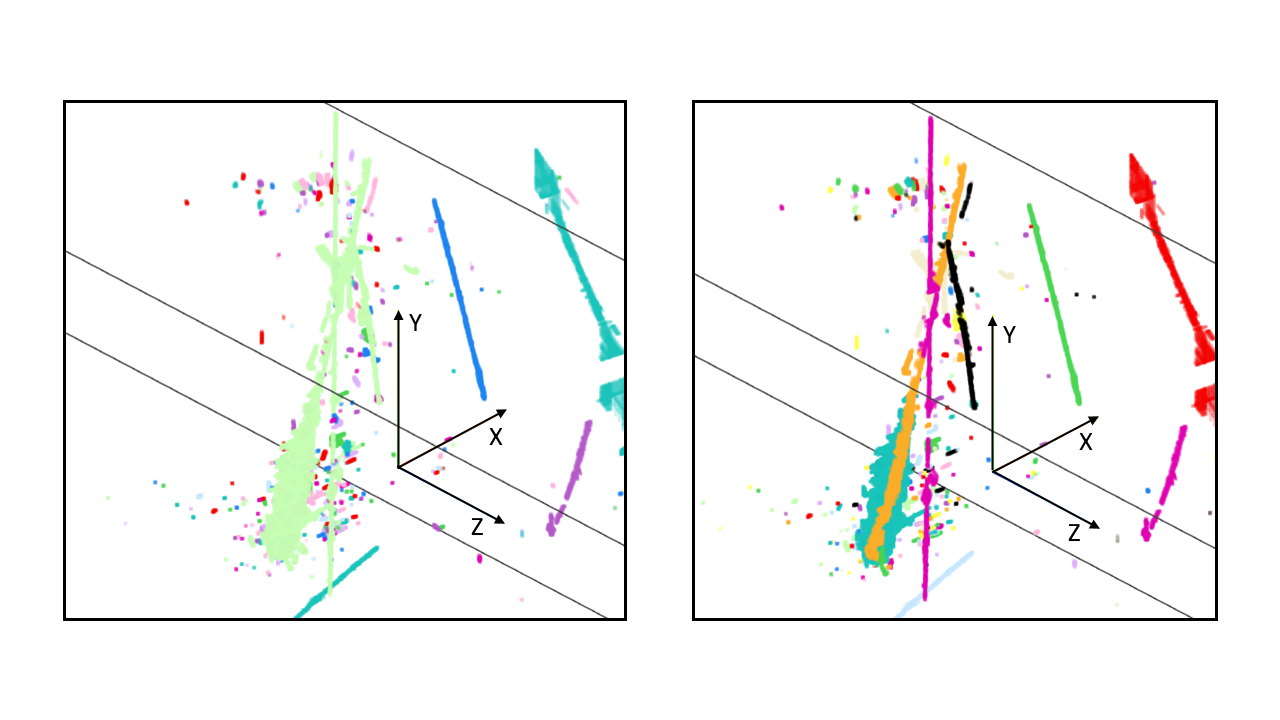}
    \put(5,49.25){\footnotesize MicroBooNE Data}
    \put(54,49.25){\footnotesize MicroBooNE Data}
  \end{overpic}
  \caption{ Demonstration of the effectiveness of the clustering algorithm to separate a ``coincidental overlap'' cluster. 
  Black lines in each subfigure correspond to the boundaries of the LArTPC active volume.
    The left and right panels show the clusters before and after applying this algorithm. The all-light-green 3D cluster in the left panel is broken into its components in the right panel.}
  \label{fig:sep3}
\end{figure}

Figure~\ref{fig:sep2} shows a comparison of the results before and after
applying the separation algorithm. Two ``coincidental overlap'' clusters show up in this event: one case has two cosmic-ray muons crossing 
each other, the other has a cosmic-ray muon grouped to a neutrino interaction.
Figure~\ref{fig:sep3} shows another example, where two cosmic-ray muons cross each other and one of the muons induces an EM shower.
After the separation step, part of the EM shower is improperly separated. This could be addressed by the many-to-many charge-light matching later, which can further group the clusters that as a whole match the same PMT flash.
Note that a cluster of a neutrino interaction with multiple final state particles might be incorrectly identified as a ``coincidental overlap''.
There is a protection against over-separating clusters because the neutrino final state particles are mostly forward-going along the beam direction while most cosmic-ray muons are pointing downward.
Additionally, a dedicated clustering algorithm to group the separate clusters from the same neutrino interaction is performed later as described in section~\ref{sec:clustering_neutrino}.

\subsubsection{Further de-ghosting}\label{sec:clustering_deghost}
As mentioned in section~\ref{sec:imaging}, a de-ghosting algorithm is applied in the 3D imaging stage to remove ghosts. This is done prior to clustering. 
This strategy is initially inefficient since a proto-cluster cannot appropriately represent complete TPC activities initiated by a primary particle's interaction. 
Given the improvements during the clustering stage as described above, the de-ghosting algorithm is run on the resulting clusters again to remove the residual ghosts.
\begin{figure}[!thbp]
  \centering
  \begin{overpic}[width=0.9\figwidth]{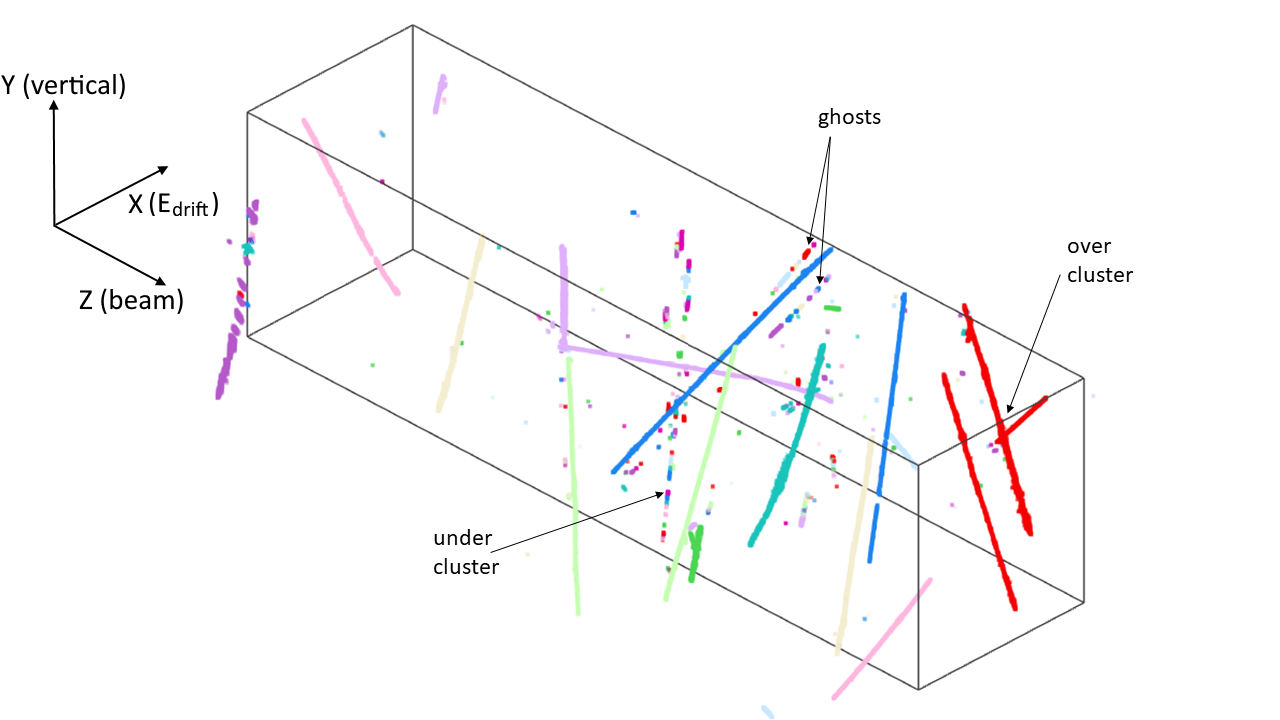}
    \put(10,8){\textbf{\small MicroBooNE Data}}
  \end{overpic}
  \begin{overpic}[width=0.9\figwidth]{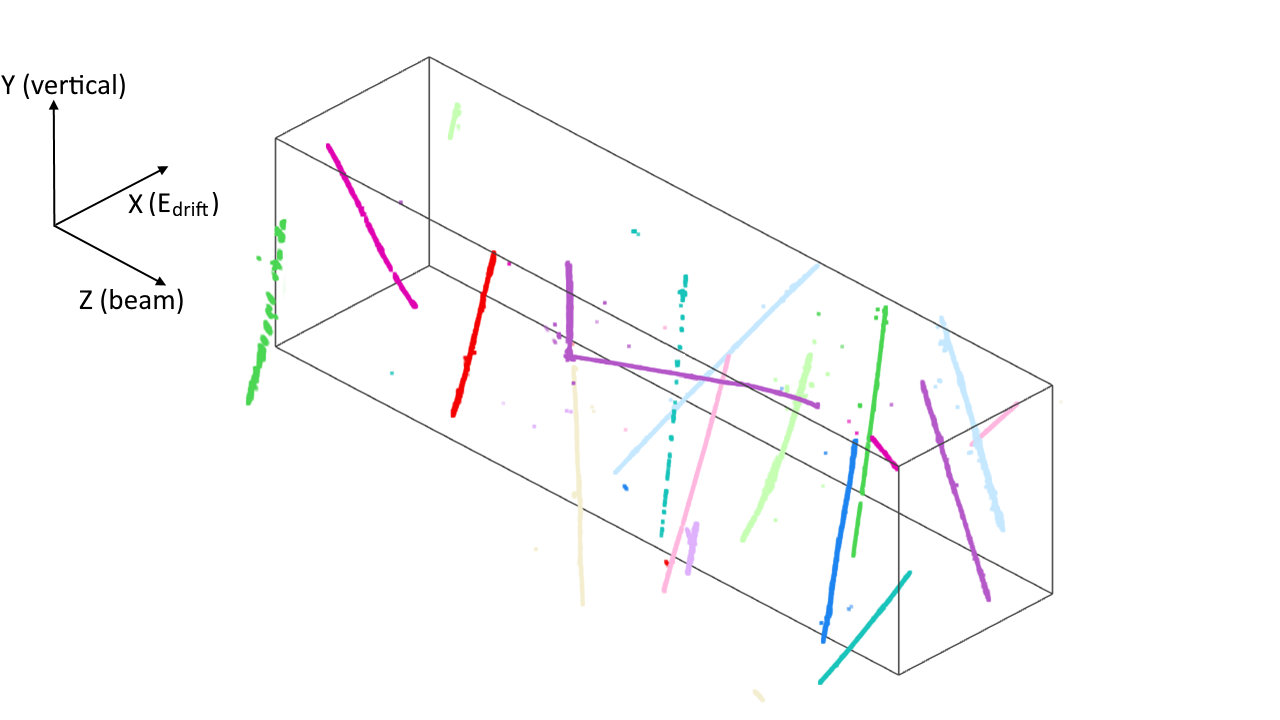}
    \put(10,10){\textbf{\small MicroBooNE Data}}
  \end{overpic}
    \caption{ Demonstration of the effectiveness of the de-ghosting algorithm with other advanced clustering algorithms applied. 
    The solid black box represents the LArTPC active volume with an X-position (converted from the readout time) relative to the neutrino interaction time.
    The top and bottom panels show the clusters before and after applying the de-ghosting algorithm after bridging gaps. Color indicates cluster membership.}
  \label{fig:prol}
\end{figure}

\begin{figure}[!thbp]   
  \centering
  \begin{overpic}[width=0.95\figwidth]{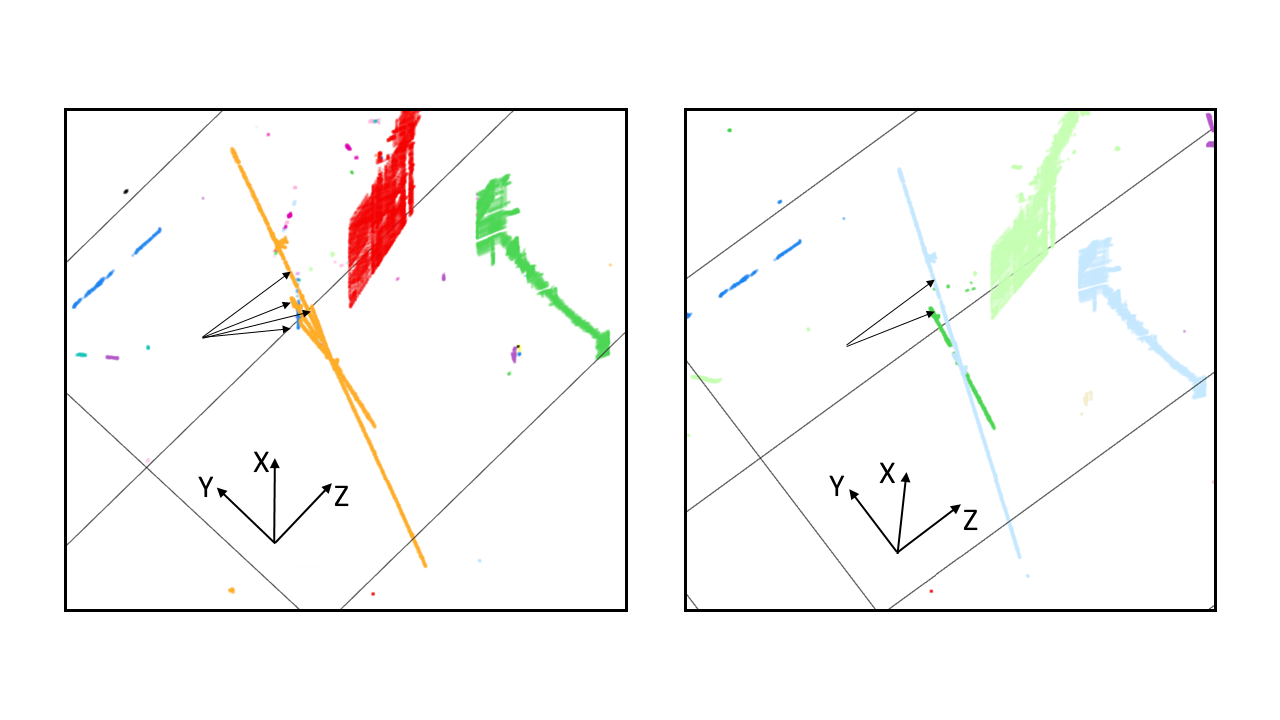}
    \put(6,48.5){\footnotesize MicroBooNE Data}
    \put(54,48.5){\footnotesize MicroBooNE Data}
  \end{overpic}
  \caption{ Demonstration of effectiveness of the de-ghosting algorithm with other advanced clustering algorithms applied. 
  The black lines inside each subfigure correspond to the boundaries of the LArTPC active volume.
    The left and right panels show the clusters before and after
    applying the de-ghosting algorithm following the separation of the ``coincidental overlap'' cluster. Color indicates cluster membership. The large clusters with much worse spatial resolution in Y-Z view correspond to big blobs of isochronous tracks as discussed in section~\ref{sec:imaging_summary}.
    }
  \label{fig:deghost1}
\end{figure}

We present some instructive examples of de-ghosting after clustering has been performed.
As shown in the top panel of figure~\ref{fig:prol}, there are some ghosts due to gaps along the prolonged tracks.
These ghosts cannot be removed during the 3D imaging since they are the only explanation of the charge measurements in functional wire planes.
With a bridging of the gaps, the original proto-clusters are grouped into a larger cluster, which as a whole can explain the charge measurements in all three wire planes. 
The ghosts related to gaps in this prolonged track can thus be removed.
Another example as shown in figure~\ref{fig:deghost1} has a four-track cluster, in which two tracks are ghosts.
This cluster is present in the region where there is a nonfunctional wire plane.
Since all of these four tracks including the ghosts are connected, the two ghost tracks survive the de-ghosting procedure in the 3D imaging stage.
After the application of the algorithm to separate the ``coincidental overlap'' cluster, the two ghost tracks are identified and removed individually.

\begin{figure}[!thbp]  
  \centering
  \begin{overpic}[width=0.9\figwidth]{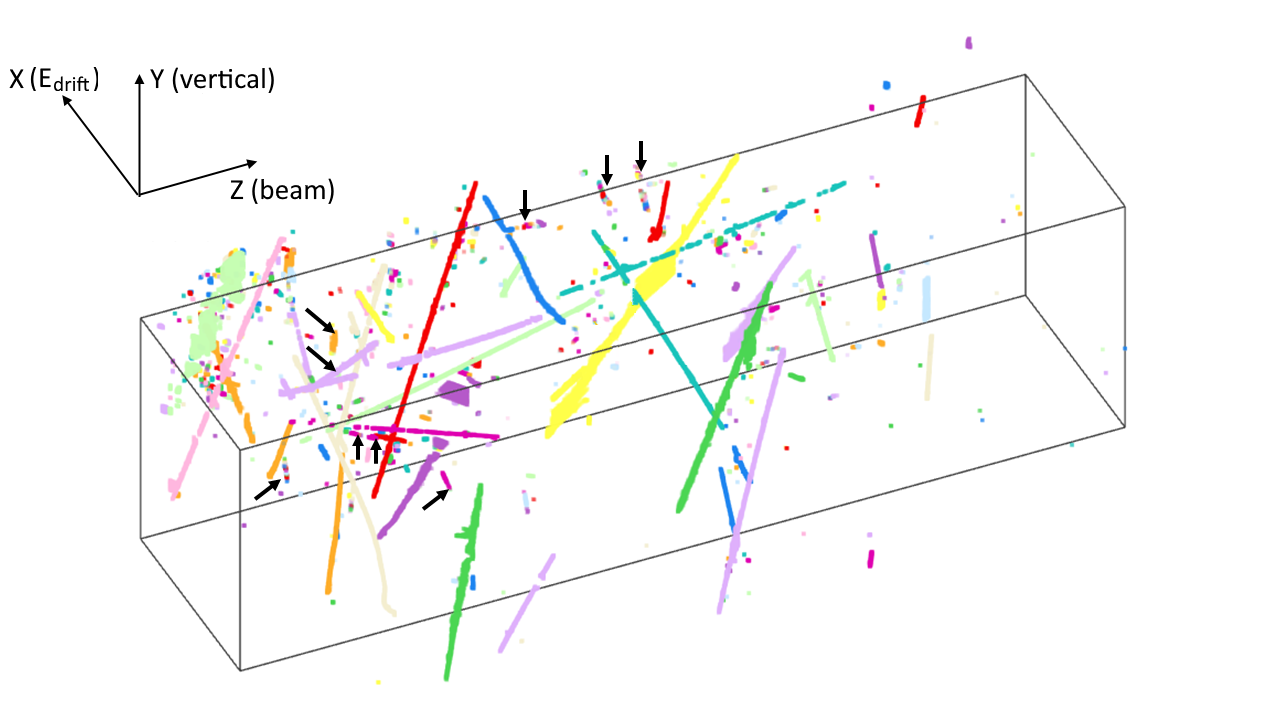}
    \put(60,8){\textbf{\small MicroBooNE Data}}
  \end{overpic}
    \begin{overpic}[width=0.9\figwidth]{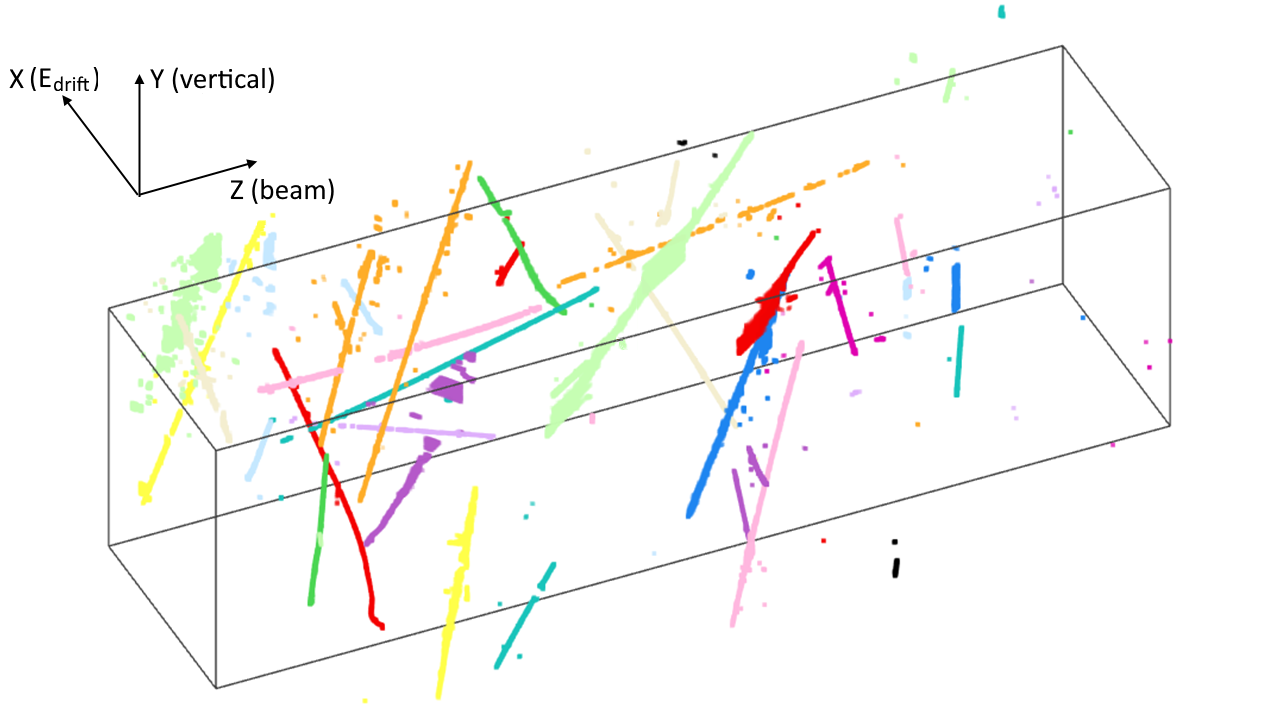}
      \put(60,8){\textbf{\small MicroBooNE Data}}
    \end{overpic}
      \caption{ Demonstration of the effectiveness of the de-ghosting algorithm with other advanced clustering algorithms applied. 
      The solid black box represents the LArTPC active volume with an X-position (converted from the readout time) relative to the neutrino interaction time.
    The dot-like clusters and superfluous clusters off the main trajectories are generally ghost tracks.
    The top and bottom panels show the clusters before and after applying the de-ghosting algorithm. The example ghosts in the top panel are indicated by the black arrows.
    This is a challenging case where multiple tracks go through a region where one wire plane (U plane) is largely nonfunctional. Color represents cluster membership.
    }
  \label{fig:deghost2}
\end{figure}

Figure~\ref{fig:deghost2} shows an example of a complex event with a large number of residual ghosts after imaging. The ghosts are indicated by black arrows in the top panel of this figure.
Many tracks including prolonged tracks and isochronous tracks go through the region (area on the left of figure~\ref{fig:dead_region}) where U wires are mostly nonfunctional.
Ghosts with various lengths and positions are reconstructed.
After bridging the gaps and separating coincidental overlap clusters, the number of ghosts is significantly reduced by re-running the de-ghosting algorithm.

\subsubsection{Clustering for neutrino events}\label{sec:clustering_neutrino}
\begin{figure}[!thbp]     
  \centering
  \begin{overpic}[width=0.9\figwidth]{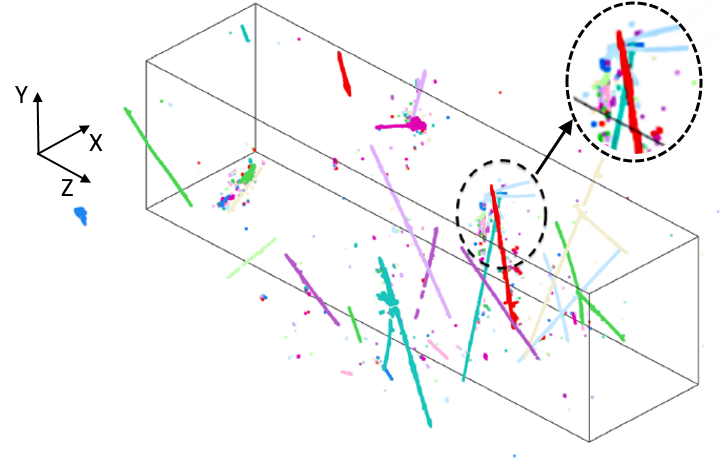}
    \put(10,10){\textbf{\small MicroBooNE Data}}
  \end{overpic}
  \begin{overpic}[width=0.9\figwidth]{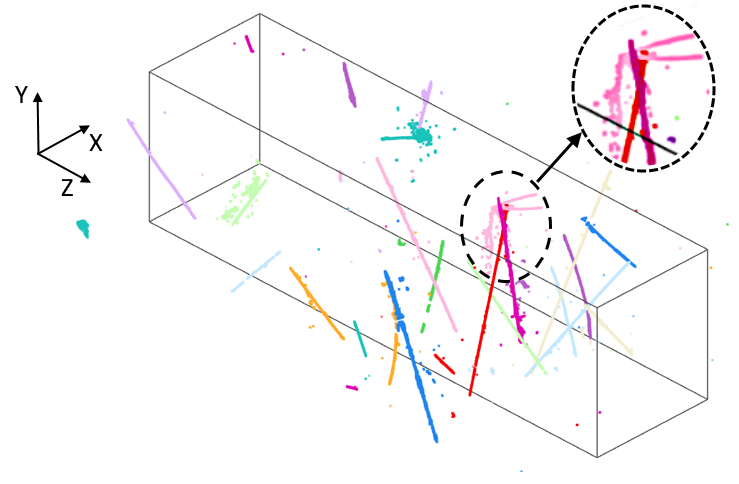}
    \put(10,10){\textbf{\small MicroBooNE Data}}
  \end{overpic}
    \caption{ Demonstration of the effectiveness of the clustering algorithm designed
    for neutrino interactions. 
    The solid black box represents the LArTPC active volume with an X-position (converted from the readout time) relative to the neutrino interaction time.
    Top and bottom panels show the clusters before
    and after applying the clustering algorithm. 
	The neutrino interaction (the light pink cluster) is in the black dashed circle with multiple particles emitted and two electromagnetic
    showers (two $\gamma$'s from a $\pi^0$ decay). 
    }
  \label{fig:neutrino1}
\end{figure}

In this section, we describe a dedicated clustering algorithm to group separate clusters from the same neutrino interaction. 
Generally, the neutral particles from neutrino interactions, such as a neutron or $\pi^0$,
can lead to clusters that are detached from the primary neutrino interaction vertex. 
These clusters are truly separated in 3D space and should be identified and grouped properly.
In order to do so, 
the major task is to find the common vertex based on the direction of each sub-cluster.
The operations to obtain the primary direction, find extreme points,
associate nearby points, and calculate the direction, are the same as those introduced in the previous sections.
The main steps are described below:
\begin{itemize}[nolistsep]
\item Only clusters within the drift window that correspond to the beam time are considered.
\item The direction of each sub-cluster is calculated. End points are examined to ensure that they do not belong to any isolated dot-like (less than 1-cm length) clusters, which are ignored because of their small size. 
\item Each cluster is extended with virtual space points along the track direction near each end point. 
\item The extended clusters are examined to find the ``intersection'' point with other clusters. This ``intersection'' is required to be formed by the extended part or the end points of the other clusters.
\end{itemize}
The ``intersection'' is not necessarily the primary neutrino vertex, as the separated clusters from the secondary interaction vertex are also expected to be grouped together.
An under-clustering issue may arise for neutrino interactions, but this is expected to be addressed by the charge-light matching step later when a many-to-many matching strategy is adopted.
Figure~\ref{fig:neutrino1} shows an example of a complex neutrino interaction.
Two $\gamma$'s from a $\pi^0$ decay and a detached charged particle are clustered properly.  

\subsection{PMT light signal reconstruction}\label{sec:light}
TPC clusters, which represent grouped TPC activities corresponding to either cosmic-ray muons or a neutrino interaction,
are formed by the clustering algorithms as described in the previous section.
Because of the asynchrony of the TPC readout system with the PMT readout system, TPC activities are mixed in the time sequence with an unknown interaction (start) time.
Scintillation light is produced and detected on a much shorter time scale by the spatially distributed PMTs.
An offline processing of the light signals from PMTs is thus important to 
perform the many-to-many charge-light matching to select the neutrino activities corresponding to the in-beam PMT signals that coincide with the beam spill.


As described in section~\ref{sec:intro}, 32 PMTs are used to detect the scintillation light in MicroBooNE.
In the PMT front-end motherboard (FEM), the PMT signal is separated by a splitter into high-gain (x10) and low-gain (x1) amplifiers,
allowing a wide dynamic range for a 64-MHz 12-bit ADC readout of the PMT pulses~\cite{Acciarri:2016smi}.
In the PMT readout system, there are two separate readout streams: {\it beam discriminator} and {\it cosmic discriminator}.
The beam discriminator starts 4 $\mu$s before the beam gate. It reads out 1500 consecutive samples ($\sim$23.4 $\mu$s) of the PMT waveforms.
The cosmic discriminator is a self-triggered PMT readout.
It reads out 40 consecutive samples ($\sim$0.6 $\mu$s) of the PMT waveforms,
which record the light information not only from beam-coincident activities but also activities out-of-time with the beam.

The PMT waveforms are processed offline to reconstruct the time and number of photoelectrons (PE) of a flash, which is a group of PMT signals close in time.
For the beam discriminator, 
a deconvolution using the Fast Fourier Transformation (FFT) is performed to unfold the electronics
responses from various RC circuits in the splitter and the shaper.
A flash is then formed if the PMT measurements satisfy the multiplicity requirement ($>$2 PMTs above a threshold of 1.5~PE) and the total integrated PE threshold ($>$6 total PE) in a 100 ns window.
A flash window lasts 7.3~$\mu$s in order to exclude noise and to include the contribution from the late scintillation light.
The scintillation light in liquid argon has a prompt and a slow component with decay times of about a few nanoseconds and 1.6 $\mu$s\footnote{The two lifetimes correspond to the molecular excimer states excited either in a singlet state or a triplet state.}, respectively.

\begin{figure}[thb]
  \centering
  \begin{overpic}[width=0.65\figwidth]{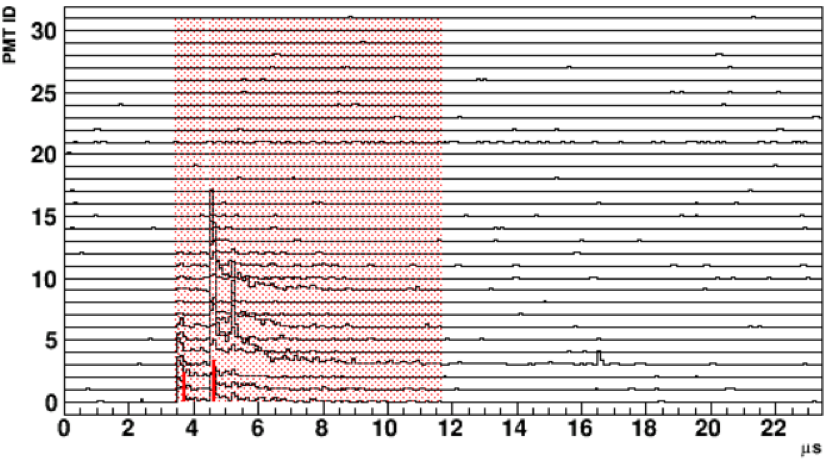}
    \put(77,56){\text{MicroBooNE}}
  \end{overpic}
  \caption{Illustration of two reconstructed flashes from beam discriminator PMT waveforms.
    The black curves are the deconvolved PE spectra for each PMT.
    The red lines represent the flash times and the red bands represent the flash windows.
	For the second flash at about 4.6$\mu$s, there is a Michel electron as indicated by the second peak (at about 5.3$\mu$s) of its PE spectra.
  }
  \label{fig:beam_sp}
\end{figure}

Within the flash window, the time bin with the maximal total PE from all PMTs marks the starting time of a flash.
The PE of each PMT in a flash is integrated over the entire flash window.
Though the average time between two adjacent flashes in MicroBooNE is $\sim$100 $\mu$s, a procedure is set to end the current flash window and start a new one if the new flash has a large starting PE, calculated as the total PE from all PMTs in the first 100 ns, and satisfies either of the two requirements: 
(1) the new flash is at least 1.6~$\mu$s later than the preceding one;
(2) a significantly different PMT hit pattern (number of PEs in each PMT) in the
first 100 ns of the new flash compared to the pattern in the last 100 ns of the preceding flash using a Kolmogorov-Smirnov test~\cite{KStest}. 
Figure~\ref{fig:beam_sp} shows an example of two adjacent reconstructed flashes from beam discriminator PMT waveforms.

For the cosmic discriminator, the readout window is shorter than the slow component of the scintillation light.
The light yield ratio of the slow to the prompt component is about 3:1 for the minimum ionizing particles.
The integrated PE of a cosmic discriminator is scaled by a factor of two to take into account
the slow component portion of the scintillation light not fully recorded by the readout window.
Because of the inefficiency of the cosmic discriminator,
the data from the cosmic discriminator is ignored when the beam discriminator data is present,
and the cosmic discriminator performance is calibrated by the beam discriminator data.

\begin{figure}[thpb!]
  \centering
  \begin{overpic}[width=0.65\figwidth]{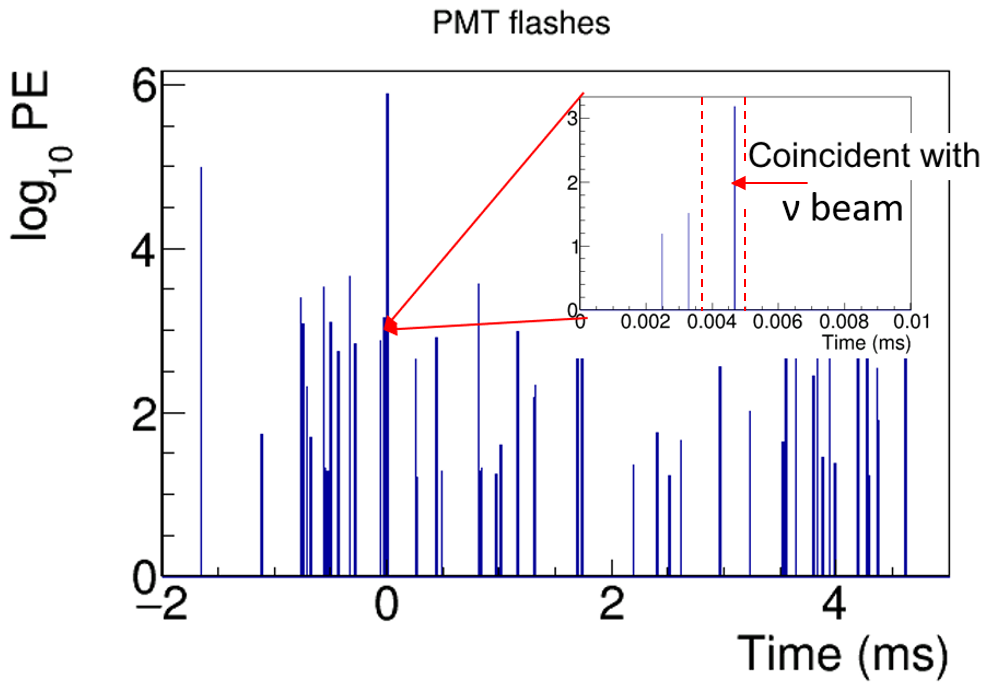}
    \put(76,66){\text{\small{MicroBooNE}}}
  \end{overpic}
  \caption{The reconstructed PEs of a flash as a function of flash time.
  The 6.4~ms PMT readout window is shown relative to the trigger time. The flashes from the beam discriminator
  (23.6~$\mu$s long) are shown as inset. The flash in coincidence with the BNB beam 
    spill (between dashed red lines) is indicated.  In general, there are 40--50 reconstructed PMT flashes in each BNB event.}
  \label{fig:light_PE}  
\end{figure}
Figure~\ref{fig:light_PE} shows the reconstructed PEs and time for each PMT flash from a data event. 
The flash corresponding to the neutrino interaction is shown in the inset figure between the dashed red lines that indicate the beam spill window.
One can see that about 50 flashes are reconstructed in this event and it is challenging to match the TPC clusters to these many PMT flashes. On the other hand, if a robust charge-light matching is developed, each TPC cluster's starting time measured by the PMTs can be used to reject the overwhelming cosmic-ray muon background in the neutrino selection.

\subsection{Many-to-many charge-light matching}~\label{sec:matching_alg}
Now that the TPC charge activities have been reconstructed and grouped into physically distinct clusters in section~\ref{sec:imaging} and the PMT light measurements have been reconstructed into distinct flashes in section~\ref{sec:light}, the next step is to match the 20--30 TPC clusters to the 40--50 PMT flashes for each recorded event. This will allow each matched cluster to be assigned the precise starting time measured by the PMTs, and enable using the short BNB time window to reject the vast majority of cosmic-ray muons as neutrino candidates.

\begin{figure}[!thb]   
  \centering
    \begin{overpic}[width=0.65\figwidth]{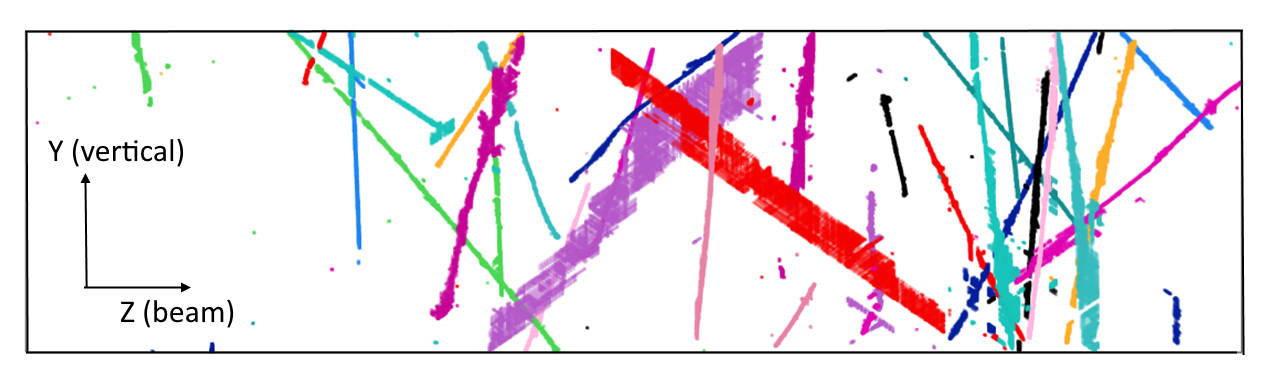}
        \put(5,30){\small\textbf{MicroBooNE Data}}
    \end{overpic}
  \includegraphics[width=0.95\figwidth]{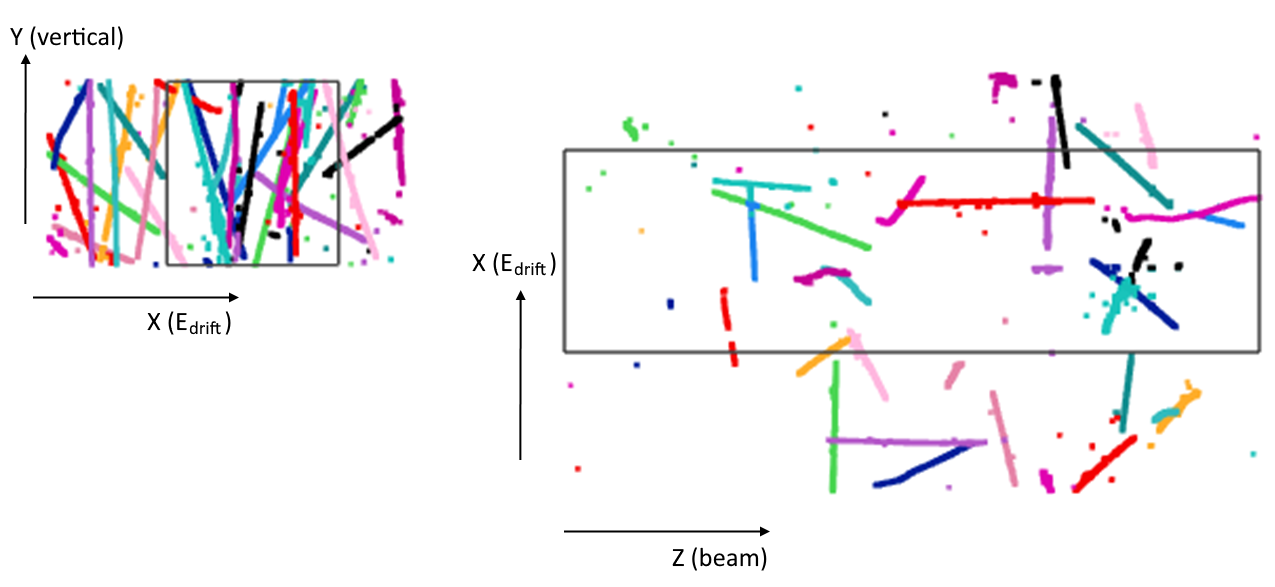}
  \caption{An example of all the TPC clusters from a MicroBooNE data event before charge-light matching.
    Different clusters are labeled in different colors, but each cluster is labeled in the same color in different 2D views. The borders represented by the black lines are the boundaries of the LArTPC active volume. 
    Top: front (Y-Z) view. 
    Bottom left: side (Y-X) view. 
    Bottom right: top (X-Z) view.
    The X-position of the black box corresponds to the starting time of the neutrino interaction, and the X-position shift of cosmic-ray muon clusters will be corrected after the charge-light matching. The entire readout time window, i.e. the X-axis range, is about 2 times the TPC width.
  }
  \label{fig:before_matching}
\end{figure}

As an example shown in figure~\ref{fig:before_matching}, there are many TPC clusters spanning the entire readout window with unknown electron drift start time. The X-position is assigned by a direct conversion from wire readout time relative to the trigger time.
More PMT flashes are generally recorded than the number of TPC clusters since PMTs sense not only the activity inside the TPC but also that outside the TPC where LAr is present within the cryostat.
On the other hand, a TPC cluster does not necessarily have a corresponding PMT flash since the light collection system (e.g.~the cosmic discriminators) has inefficiencies, especially for clusters either with low visible energy or near the cathode (far from the PMTs).
Also, as mentioned in section~\ref{sec:clustering}, the resulting clusters after the application of the clustering algorithm may still have an under-clustering issue, which is intended to be addressed in this matching stage by allowing several TPC clusters  to match to a single PMT flash.
In summary, there are two requirements in the matching algorithm: 
\begin{itemize}[nolistsep]
    \item[(A)] One TPC cluster can match to zero or at most one PMT flash.
    \item[(B)] One PMT flash can match to zero, one, or multiple TPC clusters. These multiple clusters that as a whole match the same PMT flash form a {\it cluster bundle}.
\end{itemize}
The ``match'' is defined as a good agreement between the predicted and measured light signals, considering the signal intensity of each individual PMT as well as the hit pattern of all 32 PMTs.
Assuming a TPC cluster to be associated with a PMT flash, a prediction of the PE distribution for the 32 PMTs can be made. 
The electron drift start time of the TPC cluster is shifted from the default BNB beam time to the measured time of the PMT flash. This enables a correction of the X-position of the TPC cluster. 
Then, the charge associated with each space point in the TPC cluster is used to predict the PMT light signals based on a photon library~\cite{PhotonLib-thesis}.
The TPC volume (2.56~m $\times$ 2.32~m $\times$ 10.36~m) is divided into 75 $\times$ 75 $\times$ 400 voxels.
Millions of optical photons of 128~nm wavelength from scintillation are generated and emitted with a 4$\pi$ angular distribution in each voxel, and the propagation of these photons is simulated with realistic optical photon processes of absorption and scattering in Geant4. The PMT acceptance of optical photons emitted at different locations in the TPC volume is calculated and recorded in the photon library.
With this photon library, the PEs from each of the PMTs for a given TPC cluster can be predicted by applying the PMT acceptance to the charge of each space point. 
An overall scaling factor is applied to take into account the calibrated scintillation light yield per unit deposited energy.

Interestingly, such a many-to-many matching problem is very similar to the charge solving problem as introduced in section~\ref{sec:solving}. 
There are more unknowns than knowns in this system, and the imaging equation of the first principle as shown in eq.~\eqref{eq:charge} can be used to relate the predicted light signals from all possible TPC clusters to the measured signals from PMT flashes.   
Hypothetical pairs of TPC clusters and PMT flashes are created and tested, in order to find the most compatible ones and eliminate the rest.
Again, the compressed sensing technique is utilized to perform this many-to-many matching by minimizing an $\ell_1$-regularized chi-square function.
In practice, a set of matching algorithms are developed to pre-select, fit the $\ell_1$-regularized chi-square, and re-examine the hypothetical TPC-PMT pairs.

\noindent\textbf{Pre-selection:} A pre-selection of the hypothetical TPC-PMT pairs is important to reduce the number of unknowns in the $\ell_1$-regularized chi-square fitting, allowing for a more robust minimization.
Two major tests, time range compatibility and PMT hit pattern compatibility, are performed to remove the incompatible TPC-PMT pairs.
For the time range compatibility, the TPC cluster is required to be fully contained within the maximum drift window corresponding to the PMT flash time\footnote{A precise cut can be applied since the space charge effects~\cite{SCE_paper} are insignificant along the drift direction.}.
For example, as shown in figure~\ref{fig:before_matching}, X-positions (along the drift) of the space points in any TPC cluster have an overall shift because of the unknown electron drift start time, but the in-beam activities must be contained in the nominal detector volume (black box) which is relative to the beam time.
For the PMT hit pattern compatibility, the pairs with highly incompatible predicted and measured light signals are ruled out.
A Kolmogorov-Smirnov test (K-S test) and a chi-square test, which inspect the hit pattern and the absolute normalization of the 32 PMTs' signals, respectively, are combined to discriminate the incompatible pairs.
Specifically, to enable a many-to-one TPC-PMT matching, the TPC clusters paired to the same PMT flash are jointly tested to maintain the many-to-one potential. 
The most compatible TPC-PMT pair is used as a basis and the other ones are added individually to check the change in compatibility. The pairs which significantly reduce the compatibility are ruled out.

\noindent\textbf{Chi-square fitting:} Given the passing candidate TPC-PMT pairs after the pre-selection, a chi-square function incorporating a $\ell_1$-regularization term is constructed to compare the predicted and measured light signals:
\begin{eqnarray}
\chi^2     & = & \sum_i \sum_j \chi^2_{ij} + \chi^2_{p1}  +\chi^2_{p2} + \chi^2_{p3},  \\
\chi^2_{ij} & = & \frac{(M_{ij} - \sum_k a_{ik} \cdot P_{ikj} - b_{i} \cdot M_{ij})^2}{\delta M_{ij}^2}, \\
\chi^2_{p1} & = & \sum_{i} \frac{(\sum_k a_{ik}-1)^2}{c_1^2}, \\ 
\chi^2_{p2} & = & \sum_{i} \frac{b_{i}^2}{c_2^2}, \\ 
\chi^2_{p3} & = & \lambda \cdot \sum_{i} \sum_k a_{ik}. 
\end{eqnarray}
For the input TPC-PMT pairs, the index $i$ runs through all PMT flashes, $j$ runs through all hit PMTs of each flash, and $k$ runs through all the TPC clusters.
$M_{ij}$ and $\delta M_{ij}$ represent the measured PE and its uncertainty of the $j$-th PMT in the $i$-th flash, respectively.
The uncertainties from light yield and charge measurements are conservatively assigned.
$P_{ikj}$ represents the predicted PE of the $j$-th PMT in the $i$-th flash from the $k$-th TPC cluster.
The $a_{ik}$'s, which represent the credibility of a correct match between the $k$-th TPC cluster and the $i$-th PMT flash pair, are the parameters of interest in the fit. All $a_{ik}$'s are constrained to be non-negative.
A well-matched TPC-PMT pair will have $a_{ik}$ close to 1, while a bad match will have $a_{ik}$ close to zero.
$\chi^2_{p1}$ applies the constraints that each TPC cluster should only be used once, i.e.~matched to at most one PMT flash.
The introduction of the $b_{i}$ term is to take into account the possibility that some of the PMT flashes may not be associated with any TPC clusters, in which case $b_{i}$ is close to 1, though the $\chi^2_{p2}$ term gives the constraint that $b_i$ is preferred to be close to 0.
The $\chi^2_{p3}$ term represents the application of the compressed sensing technique which prefers a best-fit solution where most of $a_{ij}$ terms are zero. $\lambda$ is the regularization strength.
$c_1$ and $c_2$ are two hyper-parameters to regularize the corresponding penalty terms, and the values are 0.01 and 0.025, respectively, tuned by real data.
After the fitting, the most incompatible TPC-PMT pairs with extremely small $a_{ik}$ values are eliminated from further consideration.
Naturally, PMT flashes that do not match any TPC clusters are eliminated as well.
The remaining TPC-PMT pairs go into the second round fitting to further approach the best solution, with the unnecessary $\chi^{2}_{p2}$ and other $b$ related terms removed.

\noindent\textbf{Re-examination:} After the two rounds of chi-square fitting, 
for each TPC cluster, the most probable TPC-PMT pair with the largest $a_{ik}$ is selected for further examination.
The hit pattern compatibility test as introduced in ``Pre-selection'' is performed.
Since many-to-one TPC-PMT matching is allowed in this procedure, the biggest TPC cluster that pairs to a PMT flash is defined as the principal component.
Then, another TPC cluster that is paired to the same PMT flash is added to the hit pattern compatibility test. If the test result becomes worse, this cluster is removed.
Otherwise, it is added to the many-to-one TPC-PMT pairs, i.e.~the cluster bundle.
After the re-examination of all selected TPC-PMT pairs, the unmatched TPC clusters will be tested against the unmatched PMT flashes to check if any possible pairings are missed.

\begin{figure}[!thbp]
  \centering  
  \includegraphics[width=0.38\figwidth]{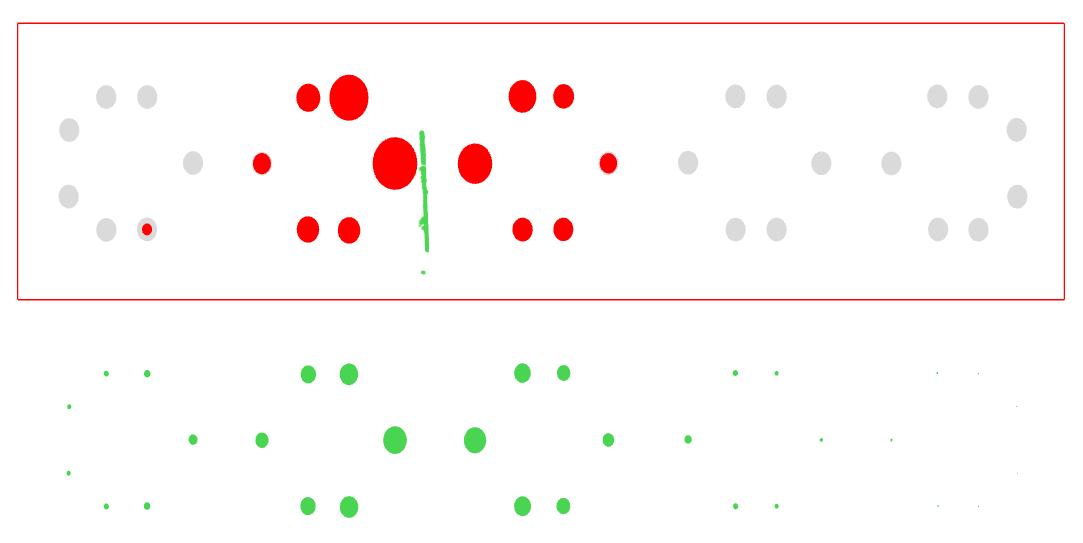}
  \begin{overpic}[width=0.45\figwidth]{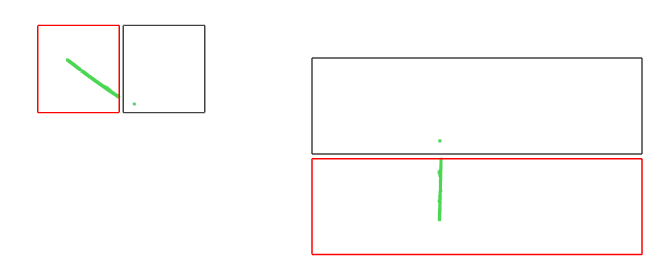}
    \put(57,42){\text{MicroBooNE Data}}
  \end{overpic}
  \includegraphics[width=0.38\figwidth]{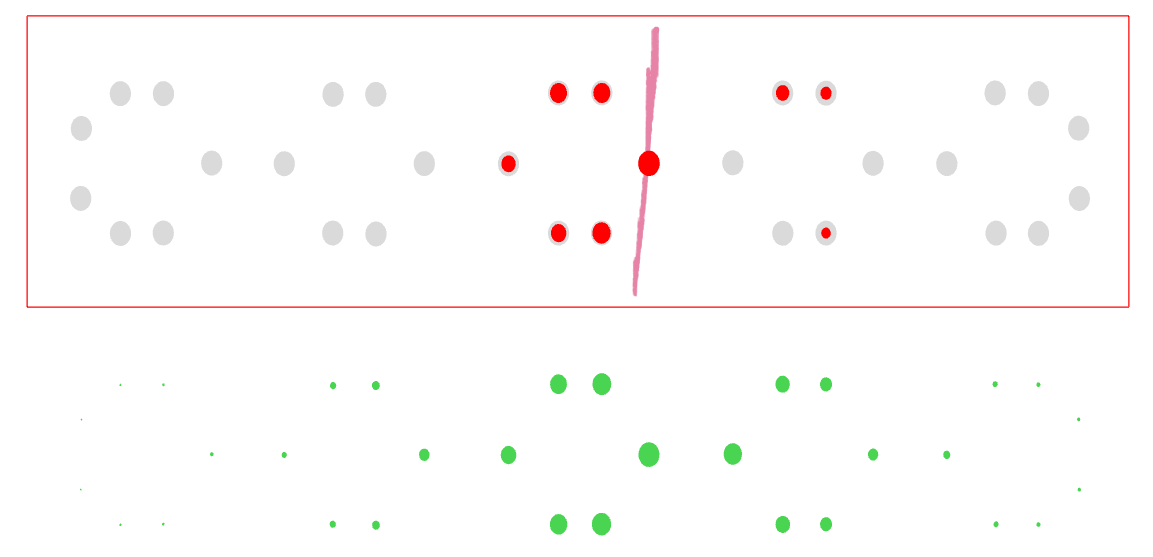}
  \includegraphics[width=0.45\figwidth]{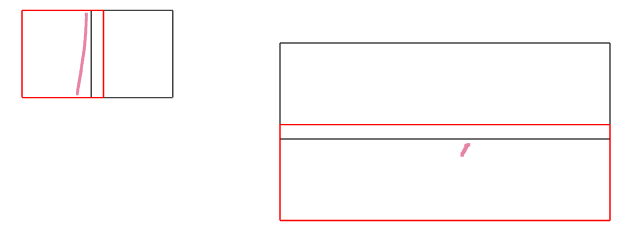}
  \includegraphics[width=0.38\figwidth]{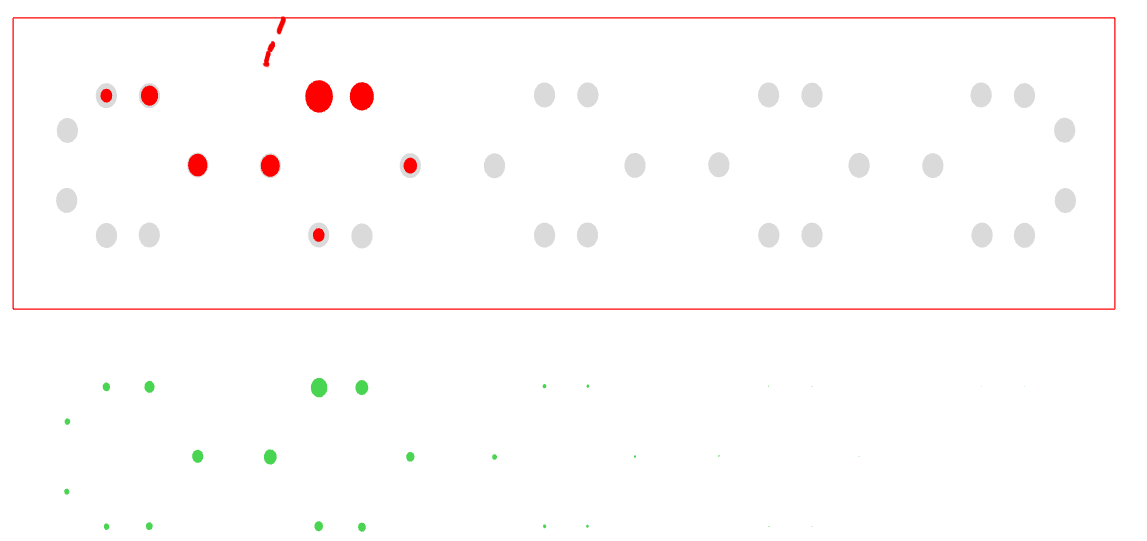}
  \includegraphics[width=0.45\figwidth]{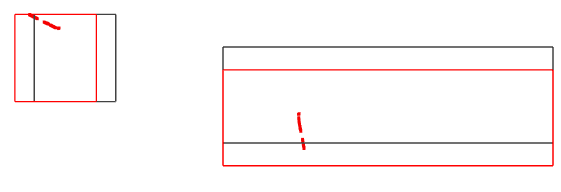}
  \includegraphics[width=0.38\figwidth]{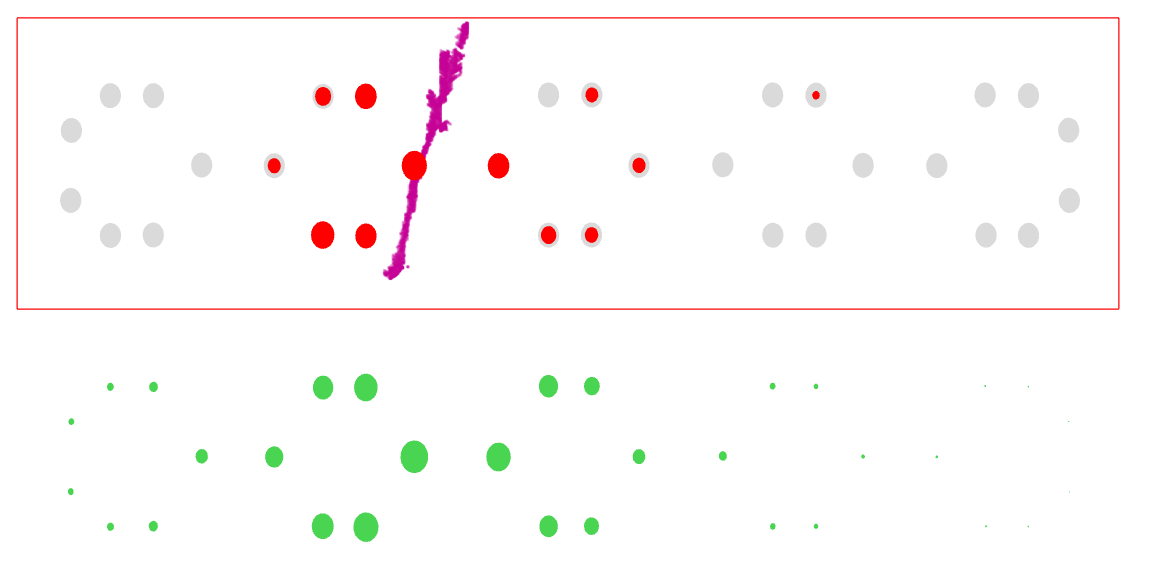}
  \includegraphics[width=0.45\figwidth]{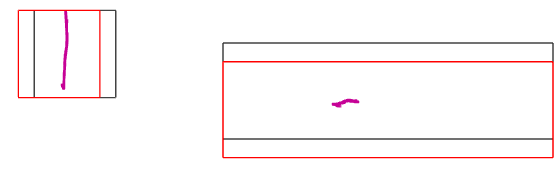}
  \includegraphics[width=0.38\figwidth]{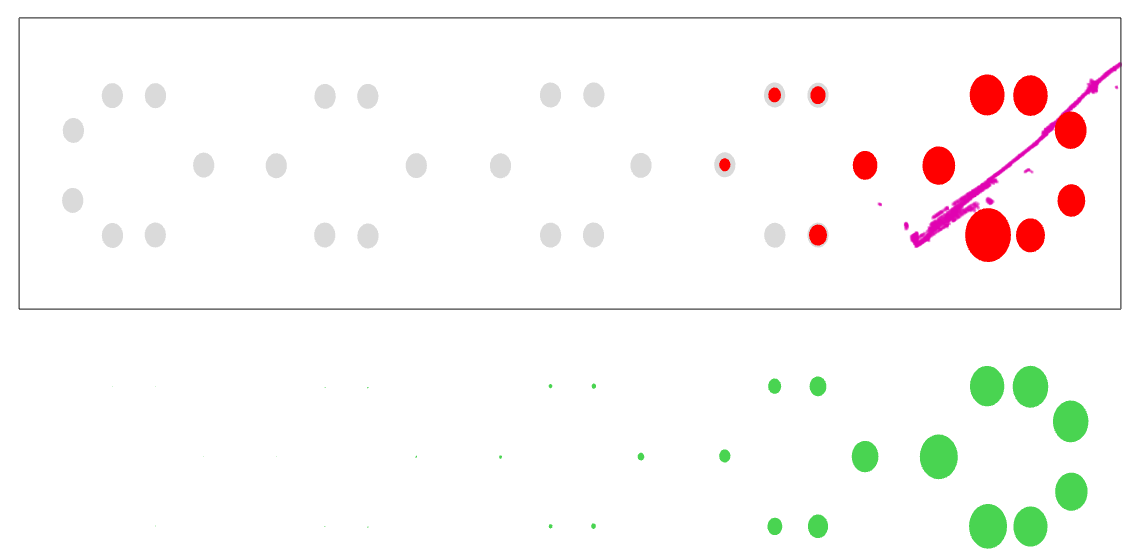}
  \includegraphics[width=0.45\figwidth]{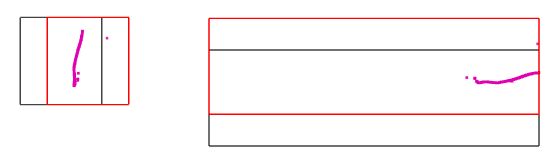}
  \includegraphics[width=0.38\figwidth]{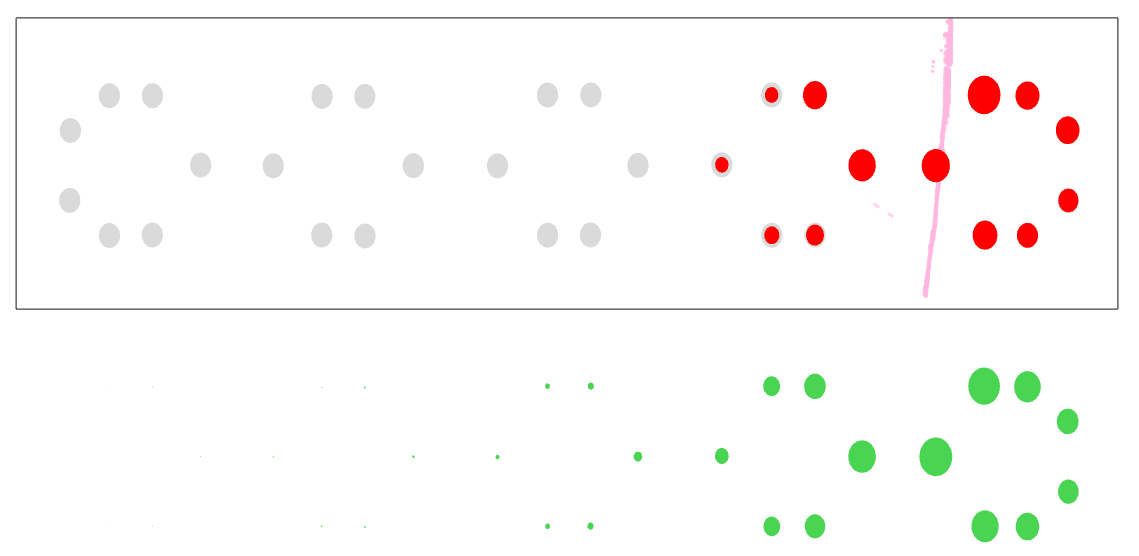}
  \includegraphics[width=0.45\figwidth]{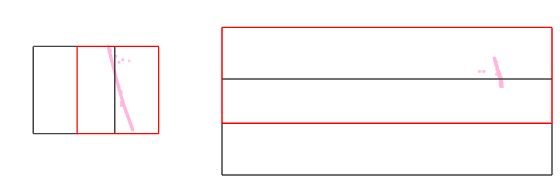}
  \includegraphics[width=0.38\figwidth]{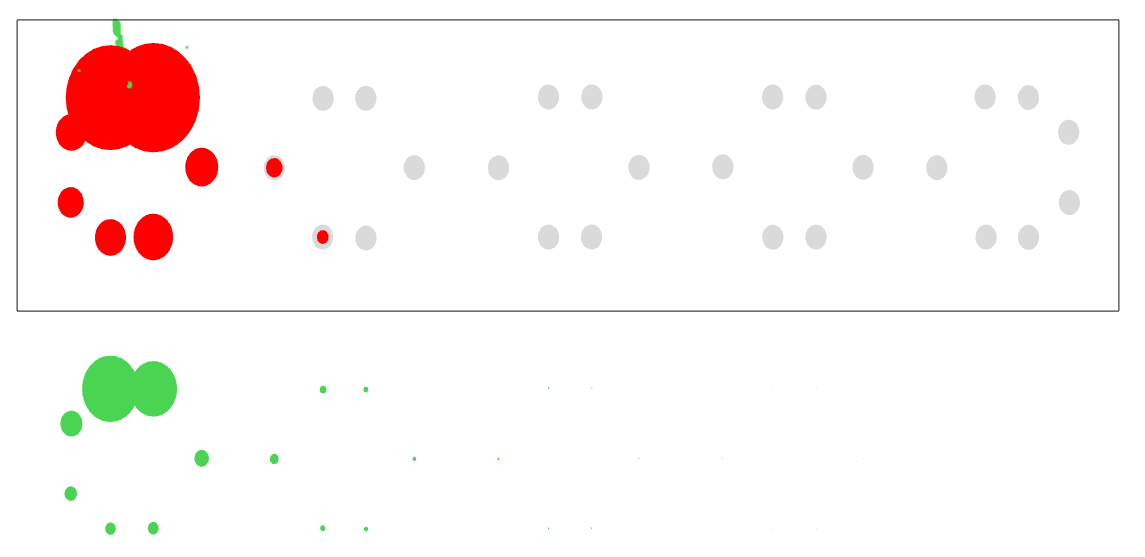}
  \includegraphics[width=0.45\figwidth]{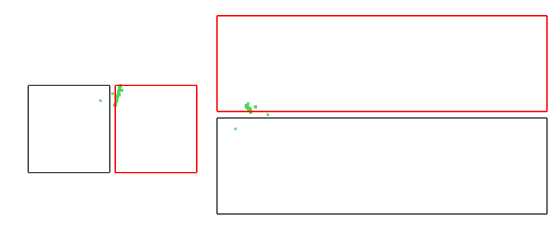}
  \caption{Selected 7 matched pairs out of the 31 pairs from a data event. 
    From left to right, they are the front (Y-Z), side (Y-X), and top (X-Z) views of the detector, respectively. The black or red boxes correspond to the LArTPC active volume. The gray solid circles in the front view represent PMTs in different locations. The red solid circles represent the measured PE in the PMTs. The green solid circles represent the predicted PE based on the matched TPC cluster(s). The area of the circle is proportional to the number of PEs. The black box has no X-position shift, and it corresponds to the starting time of the neutrino interaction. The red box corresponds to the time of the matched PMT flash, i.e. the starting time of the cosmic-ray muon, and the X-position shift is corrected.}
  \label{fig:matching_spec}
\end{figure}

\begin{figure}[!thb] 
	\centering
        \begin{overpic}[width=0.9\figwidth]{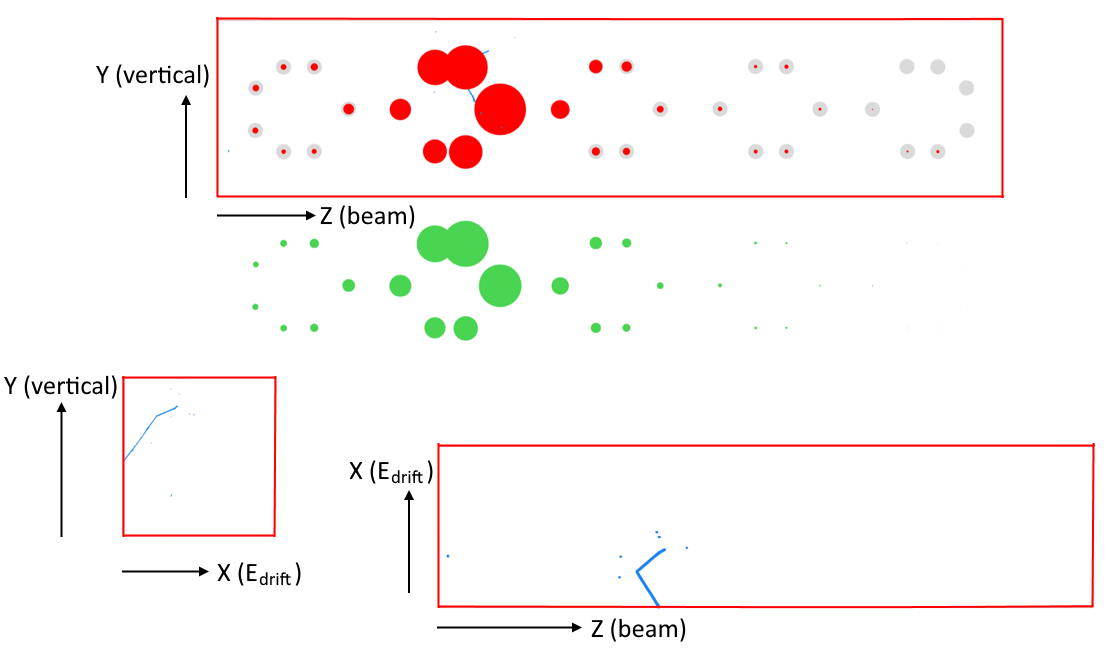}
          \put(66,58){\text{MicroBooNE Data}}
        \end{overpic}        
	\caption{A muon neutrino event is shown with its matched flash. The red boxes correspond to the LArTPC active volume. The gray solid circles in Y-Z view represent the PMTs in different locations. The red solid circles represent the measured PE in the PMTs. The green solid circles represent the predicted PE based on the TPC cluster(s). The area of the red or green circle is proportional to the number of PEs.}
	\label{fig:matching_6}
\end{figure}

\begin{figure}[!thb]
	\centering
        \begin{overpic}[width=0.9\figwidth]{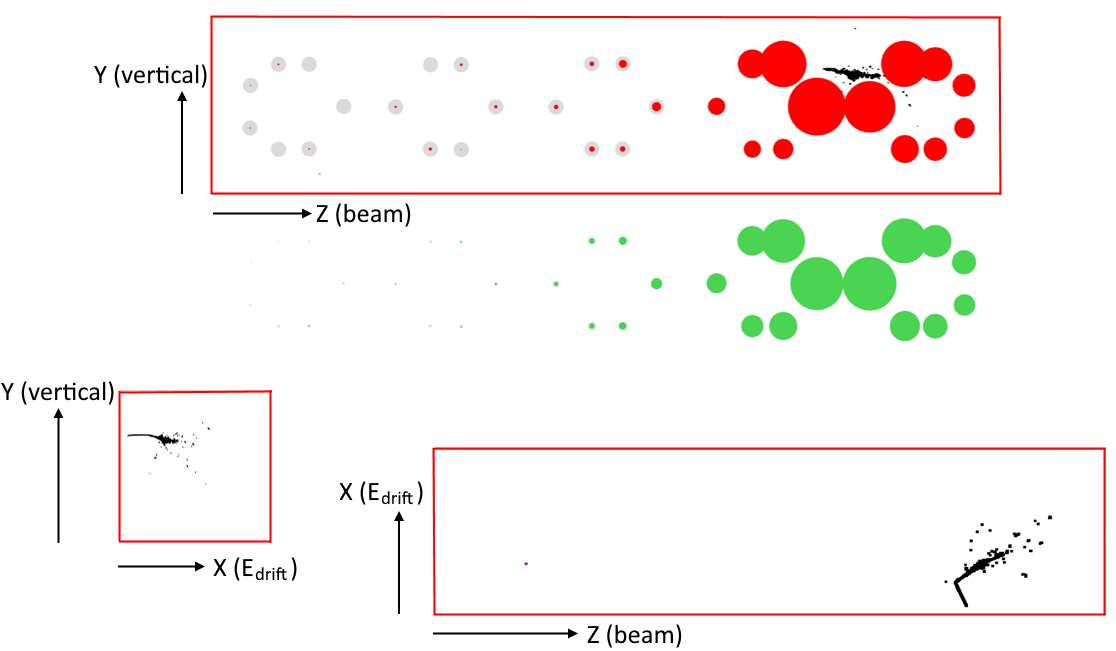}
          \put(66,58){\text{MicroBooNE Data}}
        \end{overpic}             
	\caption{An electron neutrino event is shown with its matched flash. The red boxes correspond to the LArTPC active volume. The gray solid circles in Y-Z view represent the PMTs in different locations. The red solid circles represent the measured PE in the PMTs. The green solid circles represent the predicted PE based on the TPC cluster(s). The area of the red or green circle is proportional to the number of PEs.}
	\label{fig:matching_7}
\end{figure}

Figure~\ref{fig:matching_spec} shows an example of 7 matched pairs out of a total of 31 matched pairs from one MicroBooNE data event.
After the many-to-many matching, the in-beam, flash-matched TPC clusters are taken to be neutrino interaction candidates, and the remainders are rejected as cosmic-ray background.
Figure~\ref{fig:matching_6} and figure~\ref{fig:matching_7} demonstrate successfully matching muon
and electron neutrino clusters to their respective in-beam flashes.
The performance of the matching algorithm is evident from these event displays and quantitative evaluations are provided in section~\ref{sec:evaluation}. 
On average, the charge-light matching consumes about 30 seconds per event with less than 1.5 GB memory on an Intel(R) Core(TM) i7-4790K CPU @ 4.00GHz. 

\clearpage
\section{Evaluation of the Wire-Cell 3D imaging and the charge-light matching}\label{sec:evaluation}
In this section, the quantitative evaluations of the Wire-Cell 3D imaging and the many-to-many charge-light matching are presented.
The performance of these three-dimensional approaches to reconstruct neutrino activities is demonstrated as well.
The intrinsic problem with 3D imaging stems from the wire readout ambiguity, and this is worsened by nonfunctional wires.
As a consequence, ghost tracks appear in the final 3D image and cannot be completely removed, despite the dedicated algorithms described in section~\ref{sec:imaging} and section~\ref{sec:clustering_matching}.
On the other hand, a true hit, which is a space point associated with true energy depositions, might be discarded in the charge solving and the de-ghosting steps.
Two major metrics are used to evaluate the quality of the 3D imaging result as follows:
\begin{itemize}[noitemsep, nolistsep, leftmargin=*]
    \item[] \textbf{Purity} of the 3D image -- the number of the reconstructed hits overlapping true TPC hits divided by the total number of the reconstructed hits.
    \item[] \textbf{Completeness} of the 3D image -- the number of the true hits overlapping the reconstructed hits divided by the total number of the true hits. The true hits are required to be within the TPC active volume and are weighted by their true deposited (visible) energy.
\end{itemize}
The 3D metrics are relevant to understand the performance of the subsequent Wire-Cell reconstruction. For example, the cosmic-ray background rejection and the pattern recognition are expected to operate on the 3D images in order to maximize the potential capability of LArTPCs.

Given the numerous cosmic-ray muons in the TPC, the 3D clustering and the many-to-many charge-light matching are applied to properly group the neutrino interaction and match it to the in-beam flash.
The clustering and charge-light matching may fail to select the neutrino interaction or suffer from both the over-clustering and under-clustering issues.
The correctness of the matching and the efficiency of selecting neutrino interactions after matching are evaluated as well. These two metrics are defined below and evaluated from simulation:
\begin{itemize}[noitemsep, nolistsep, leftmargin=*]
    \item[] \textbf{Correctness} of the charge-light matching -- the fraction of all in-beam neutrino candidates that true neutrino interactions. The incorrectly-matched candidates have no neutrino interactions but do have cosmic-ray muon activities with extremely low completeness values as defined above.
    \item[] \textbf{Efficiency} of selecting neutrino interactions -- the fraction of the events with neutrino interactions that have correct in-beam matches.
\end{itemize}

The development and optimization of the Wire-Cell 3D reconstruction techniques described in previous sections are based on $\sim$1500 data events.
The evaluations in this section are carried out using the MicroBooNE detector simulation.
The MicroBooNE simulation has incorporated a realistic detector response model which is in good agreement with data.
A data-driven noise model and long-range wire responses~\cite{noise_filter_paper, SP1_paper, SP2_paper} are implemented in addition to the capability to overlay real data from cosmic rays with a simulated neutrino interaction.
The MicroBooNE full detector simulation software {\it LArSoft}~\cite{larsoft} and {\it uboonecode}~\cite{uboonecode} are used to simulate the BNB neutrino charged current (CC) and neutral current (NC) interactions in 
the cryostat that contains the rectangularly shaped TPC active volume, as described in section~\ref{sec:intro}.
The GENIE neutrino generator~\cite{genie} and the Geant4 simulation toolkit~\cite{geant4_2003, geant4_2016} are incorporated into the MicroBooNE simulation software.

Three different Monte Carlo (MC) samples are used to perform the evaluations: 
\begin{enumerate}
    \item Ideal tracks -- lines of charge deposition corresponding to minimum ionizing particles (MIPs) to demonstrate the intrinsic performance of the 3D imaging and the impact from nonfunctional wires and the signal processing chain. See section~\ref{sec:eval_ideal}.
    \item Neutrino only -- full detector simulation of a neutrino interaction without cosmic-ray muons to demonstrate the performance of the 3D imaging on the complex topology of neutrino interaction final states. See section~\ref{sec:eval_nuonly}.
    \item Neutrino overlay -- full detector simulation of a neutrino interaction mixed with real cosmic-ray data to demonstrate the final performance after the Wire-Cell 3D imaging, clustering, and charge-light matching. This sample is used to show the correctness and the neutrino efficiency after the matching step. See section~\ref{sec:eval_nuoverlay}.
\end{enumerate}
By comparing the purity and the completeness results between sample B and sample C, the impact from cosmic-ray muons and the performance of clustering and charge-light matching on the neutrino interaction will be shown and discussed. 
In the neutrino-only or neutrino-overlay samples, the $\nu_{\mu}$ or $\nu_{e}$ energy spectra are from the BNB beam flux simulations. 
Only the neutrino interactions with their primary vertices in the TPC active volume are considered.
Neutrino interactions outside the active volume are largely or completely invisible because the ionization electrons outside the active volume cannot drift and be collected by the wire planes.
Evaluation of the performance on cosmic-ray only data is not specifically performed.
The coincident in-beam cosmic-ray activities is expected to be selected in this case, and they will be further rejected by dedicated cosmic-ray muon taggers in the later reconstruction chain~\cite{Wire-Cell-Generic-PRL, Wire-Cell-Generic-PRD}, which is out of the scope of this paper.

\subsection{Imaging performance of ideal tracks}\label{sec:eval_ideal}
About twenty one-meter-long ideal tracks (lines of charge depositions corresponding to MIPs) in each event are simulated in the MicroBooNE TPC.
The angular distribution is uniform in 4$\pi$.
The start position distribution is uniform in the TPC active volume.
The number of hit cells on the anode plane per unit time is close to the real data, mimicking the numerous cosmic-ray muons traversing the MicroBooNE detector.

Three scenarios of the simulation are constructed to study the performance of the 3D imaging as well as the impact from the nonfunctional wires and signal processing (SP):
\begin{itemize}[noitemsep, nolistsep, leftmargin=*]
    \item[] \textbf{Perfect SP}: The true charge deposition on each wire is only convoluted with the smearing effects from the diffusion during the charge drift and the software filters used in the signal processing. In this perfect signal processing procedure, there is no bias or failure of the charge extraction.
    \item[] \textbf{Dead + perfect SP}: Nonfunctional wires are added based on data observations and perfect signal processing is applied.
    \item[] \textbf{Dead + real SP}: Nonfunctional wires are added and realistic signal processing is applied. For a prolonged track which leaves a long signal in each individual wire readout, the realistic signal processing may fail to reconstruct the charge for the induction plane wires because of the bipolar signal cancellation. See ref.~\cite{SP1_paper} for more details. This results in gaps in the 2D wire-versus-time views of the charge measurement as mentioned in section~\ref{sec:imaging} and section~\ref{sec:clustering}.
\end{itemize}

\begin{figure}[thpb!]
  \centering
  \begin{overpic}[width=0.7\figwidth]{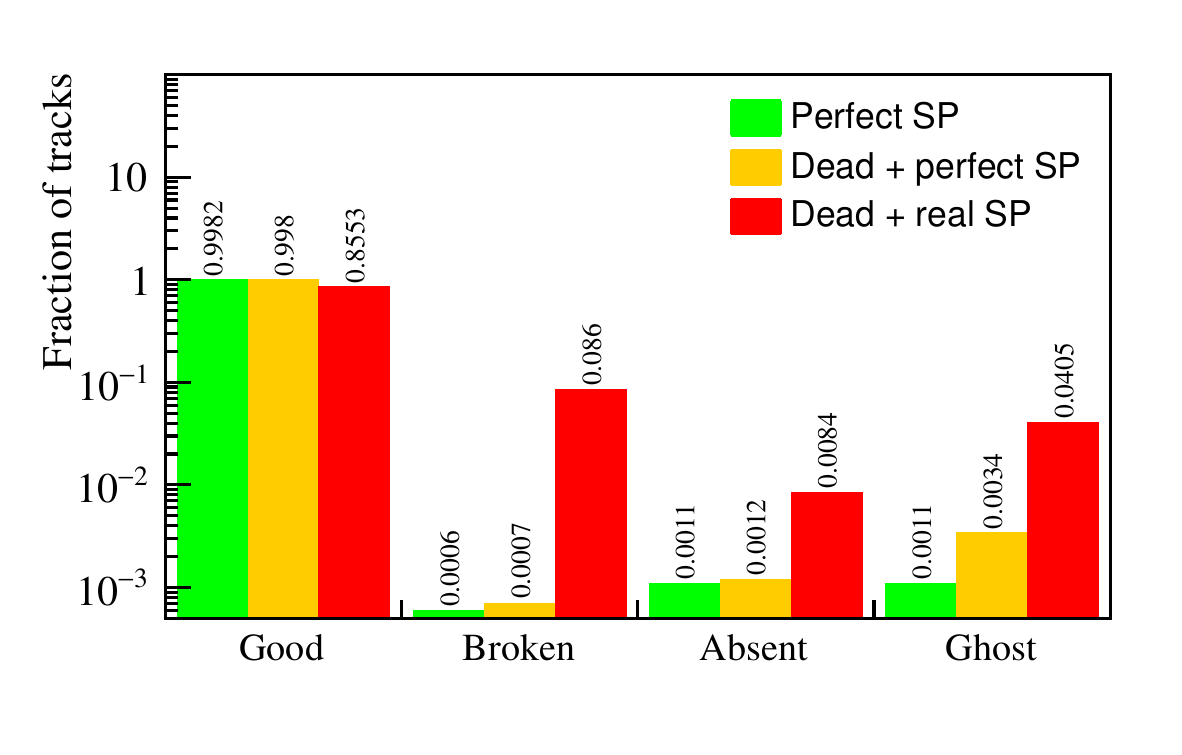}
    \put(56,59){\text{MicroBooNE Simulation}}
  \end{overpic}      
  \caption{The fraction of (good, broken, absent, ghost) reconstructed tracks from the Wire-Cell 3D imaging for different scenarios. 
    For good, broken, and ghost tracks, the fraction is weighted by their lengths and normalized to the total length of true tracks. See text for definitions of each category.} 
  \label{fig:eval_ideal_num_fraction}
\end{figure}

The results of reconstructed tracks by the 3D imaging are categorized into 4 types:
\begin{itemize}[noitemsep, nolistsep, leftmargin=*]
    \item[] {\bf Good} -- tracks are well reconstructed with at least 99\% completeness.
    \item[] {\bf Broken} -- tracks have gaps and are broken into separate segments.  
    \item[] {\bf Absent} -- tracks completely fail to be reconstructed. 
    \item[] {\bf Ghost} -- tracks have no overlap with any true track.
\end{itemize}
Based on thousands of simulated events, the fractions of each category of reconstructed tracks 
are shown in figure~\ref{fig:eval_ideal_num_fraction}.
For ``good'', ``broken'', and ``ghost'' tracks, the fraction is weighted by their lengths and normalized to the total length of true tracks. Therefore, the sum of the fractions for these three categories could be less than 100\% when there are gaps in the reconstructed tracks because the signal processing has inefficiency for events with a certain topology, e.g. for prolonged tracks. It could also be greater than 100\% because of the occurrence of ghost tracks in some places where there are no true charge depositions. Note that the sum of the fractions of ``good'', ``broken'', and ``ghost'' tracks is very close to 100\%, which indicates the ghost tracks explain the missing parts of the broken tracks.
The result of ``Dead + Perfect SP'' is very similar to the result of ``Perfect SP'' and almost all the tracks are well reconstructed.
This shows that the nonfunctional wire issue is properly addressed in the 3D imaging and a 97\% active volume efficiency has been achieved.
The impact of the nonfunctional wires on the quality of the 3D image will be further discussed in section~\ref{sec:eval_nuonly} and section~\ref{sec:eval_nuoverlay}. 
The fraction of the ghost tracks in the scenario of ``Dead + Perfect SP'' is three times larger than that of ``Perfect SP'' because of the presence of the nonfunctional wires, but it is still negligible.
In the scenario of ``Dead + Real SP'', there is a large increase of both broken tracks and ghost tracks.
The broken tracks come from the gaps, which as mentioned previously are attributed to the failure of signal processing for the prolonged tracks.
In this simulation of ideal tracks, there are a certain number of prolonged tracks since they are generated with a 4$\pi$ uniform angular distribution. 
The situation is better for the beam neutrino interactions, in which case the final state particles are mostly forward-going.

\begin{figure}[thpb!]
  \centering
  \begin{overpic}[width=0.7\figwidth]{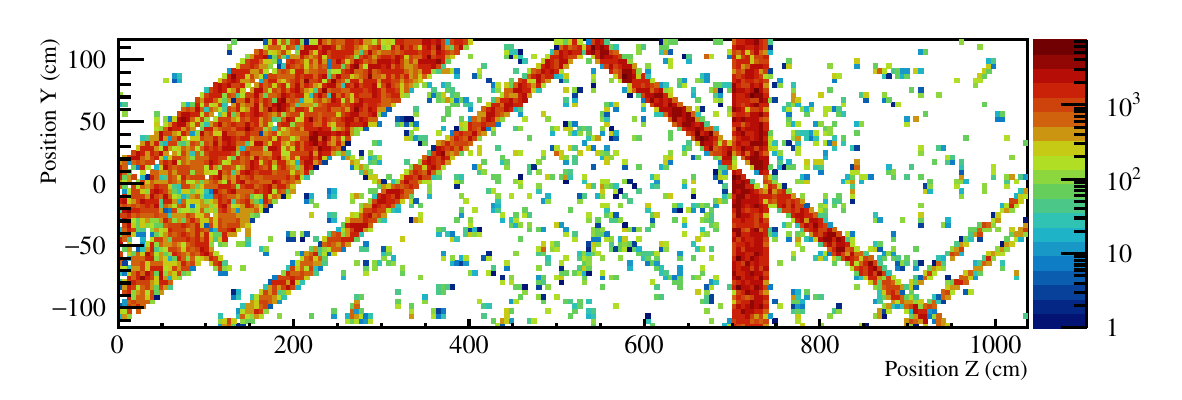}
    \put(57,32){\text{\small{MicroBooNE Simulation}}}
  \end{overpic}    
  \caption{The position (Y/vertical versus Z/beam) distribution of the ghost tracks in the scenario of ``Dead + real SP''. Color scale (Z-axis value) represents the count of space points in ghost tracks.
    The bands correspond to the nonfunctional regions as shown in figure~\ref{fig:dead_region}.}
  \label{fig:eval_ideal_ghost_points}
\end{figure}

With the realistic signal processing, more ghost tracks appear almost exclusively in the non-functional region as shown in figure~\ref{fig:eval_ideal_ghost_points}.
In one wire plane, the realistic signal processing, which may fail to extract the charges, could introduce a gap in the 3D image no matter the signal processing in the other two wire planes is successful or not. The measured charges originating from the TPC activities along this gap, if any in the other two wire planes, therefore tend to be explained by ghosts lying in a nonfunctional region where a 2-plane tiling is allowed.

\begin{figure}[thpb!]
  \centering
  \begin{overpic}[width=0.55\figwidth]{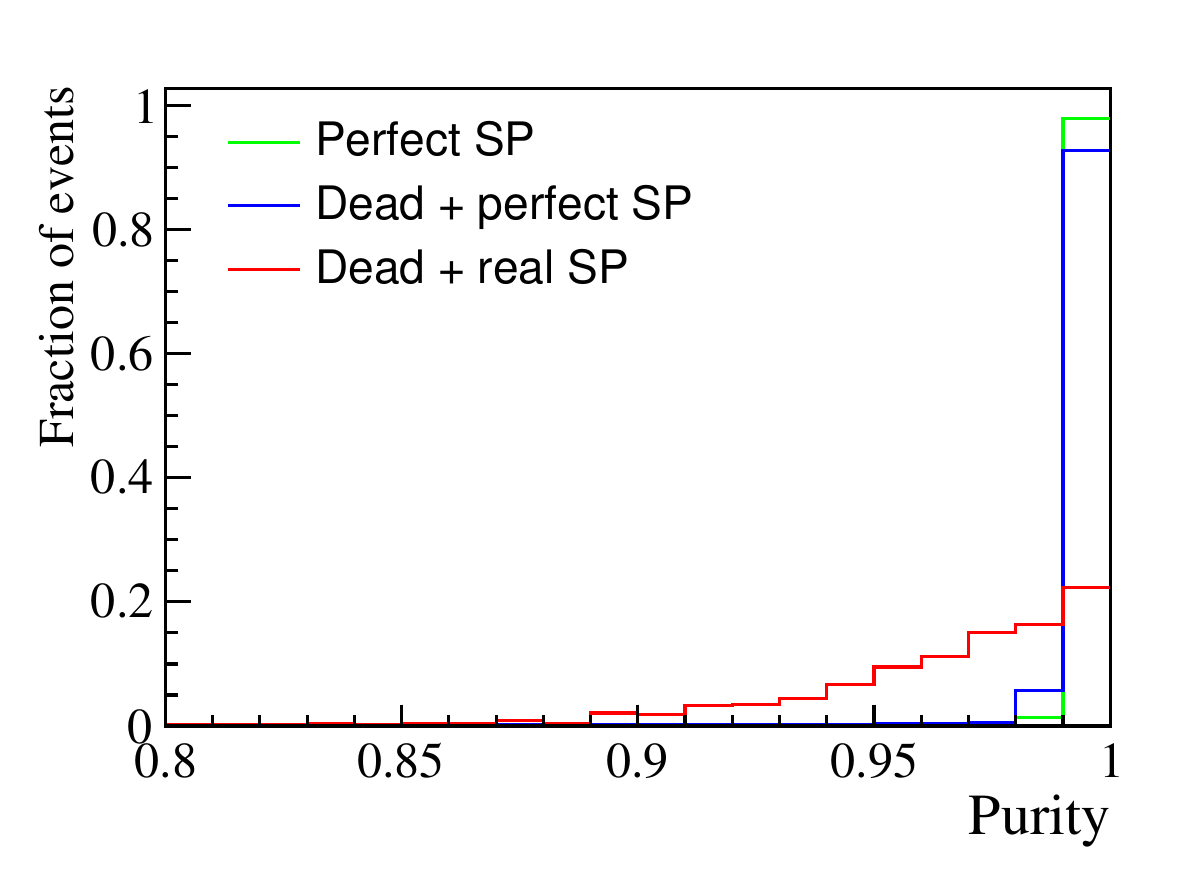}
    \put(49,70){\text{\small{MicroBooNE Simulation}}}
  \end{overpic}    
  \caption{Distribution of the purity of the reconstructed tracks from each event for different scenarios. The number of ghosts significantly increases with the presence of both nonfunctional wires and the real imperfect signal processing. The histograms are normalized separately for each category.}
  \label{fig:eval_ideal_purity}
\end{figure}

\begin{figure}[thpb!]
  \centering
  \begin{overpic}[width=0.55\figwidth]{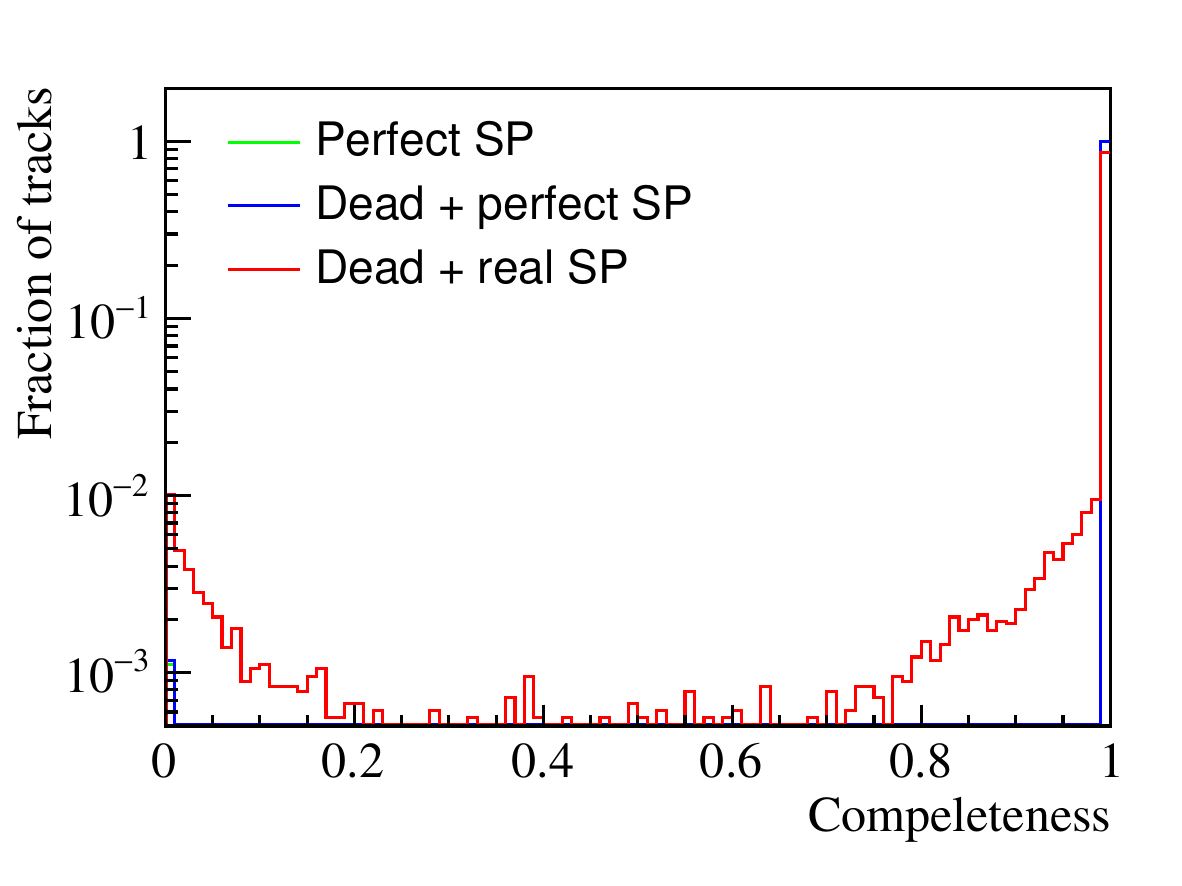}
    \put(49,70){\text{\small{MicroBooNE Simulation}}}
  \end{overpic}    
	\caption{Distribution of the completeness of each simulated track for different scenarios. The distributions are normalized for each category, respectively. The results of ``Perfect SP'' and ``Dead + perfect SP'' are basically the same in which case the green line is covered by the blue line.
    The inefficiency of the signal processing for prolonged tracks leads to very low completeness values.}
  \label{fig:eval_ideal_completeness}
\end{figure}

The purity for each event is calculated by dividing the total length of the non-ghost tracks by the total length of all the reconstructed tracks. 
The distribution of purity scores is presented in figure~\ref{fig:eval_ideal_purity}.
For ``Dead + real SP'', 96.4\% of the events have at least 90\% purity.
Figure~\ref{fig:eval_ideal_completeness} shows the distribution of the completeness for all simulated tracks. For the scenario of ``Dead + real SP'', 86.5\% (and 93.0\%) of the simulated tracks have at least 99\% (and 80\%) completeness. 
The low completeness values correspond to the prolonged tracks, especially those with directions close to normal to the wire planes.
This emphasizes again that good signal processing is important to retain the good quality of the 3D imaging result.

\subsection{Imaging performance of neutrino interactions}\label{sec:eval_nuonly}
Unlike the simulated ideal tracks in section~\ref{sec:eval_ideal}, the topology of a neutrino interaction's final state particles could be much more complicated than the single-track-like cosmic-ray muons.
Neutrino-only samples without cosmic-ray muons are used in this case.
In order to evaluate the performance of the 3D imaging, the clustering as well as the charge-light matching steps are bypassed and all the 3D space points reconstructed in the 3D imaging are taken as neutrino activities.
This is equivalent to performing perfect clustering and charge-light matching.
  
When a neutrino interacts with an argon nucleus, there are generally multiple final state particles.
On one hand, there is a very limited phase space for the final state particles to be in the prolonged or isochronous directions, in which case the 3D image may have gaps.
Note that a highly ionizing particle (HIP) may avoid such failures in the signal processing since it has a significantly higher signal-to-noise ratio.
On the other hand, with the complexity of the neutrino interactions, other failure modes may arise.
Some of the particles like neutrons, $\gamma$'s from pion decays, and primary or secondary electrons could yield low-energy (sub-MeV) depositions via nuclear recoil, Compton scattering, or Bremsstrahlung radiation, respectively.
These low energy depositions are likely to be suppressed because of the thresholding in the signal processing or removed in the 3D imaging as they resemble the dot-like ghosts.
As a result, the completeness distribution will be biased and smeared to lower values compared with that in figure~\ref{fig:eval_ideal_completeness}.
The thresholding in the signal processing is primarily to suppress fake signals from noise fluctuations.
A lower thresholding in the signal processing would create more fake charges, which can interplay with true charges and lead to ghost tracks in the nonfunctional region.

\begin{figure}[thpb!]   
  \centering
    \begin{overpic}[width=0.45\textwidth]{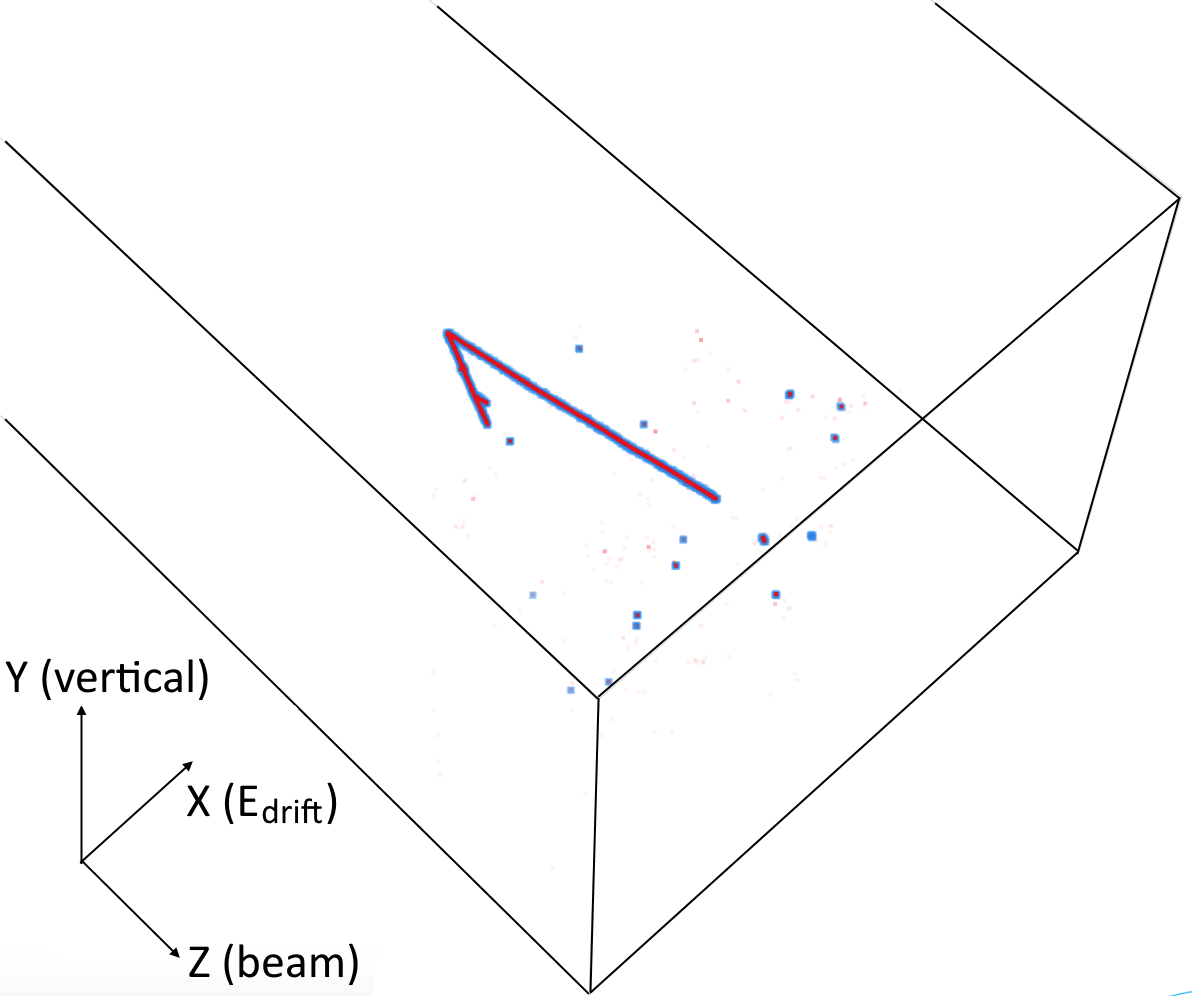}
        \put(10,90){\textbf{\small MicroBooNE Simulation}}
    \end{overpic}
    \begin{overpic}[width=0.45\textwidth]{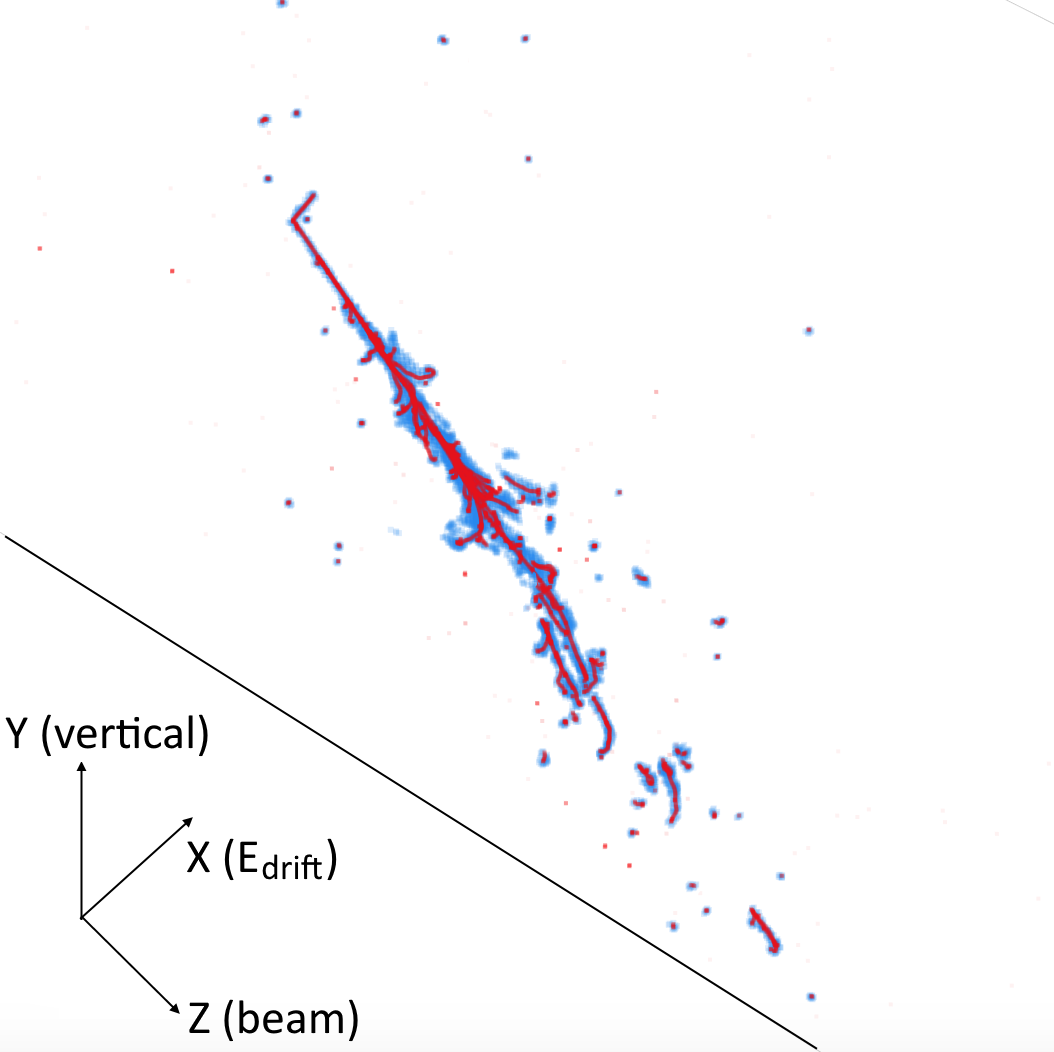}
        \put(10,90){\textbf{\small MicroBooNE Simulation}}
    \end{overpic}
    \caption{Left: 1$\mu$1$p$ $\nu_{\mu}$ CC interaction. Right: 1$e$1$p$ $\nu_{e}$ CC interaction. Blue: reconstructed 3D image. Red: truth trajectories.
    The voxel size and opacity are tuned for better illustration.
    }
  \label{fig:recovstruth_nuonly}
\end{figure}
\begin{figure}[thpb!]
  \centering
    \includegraphics[width=0.49\textwidth]{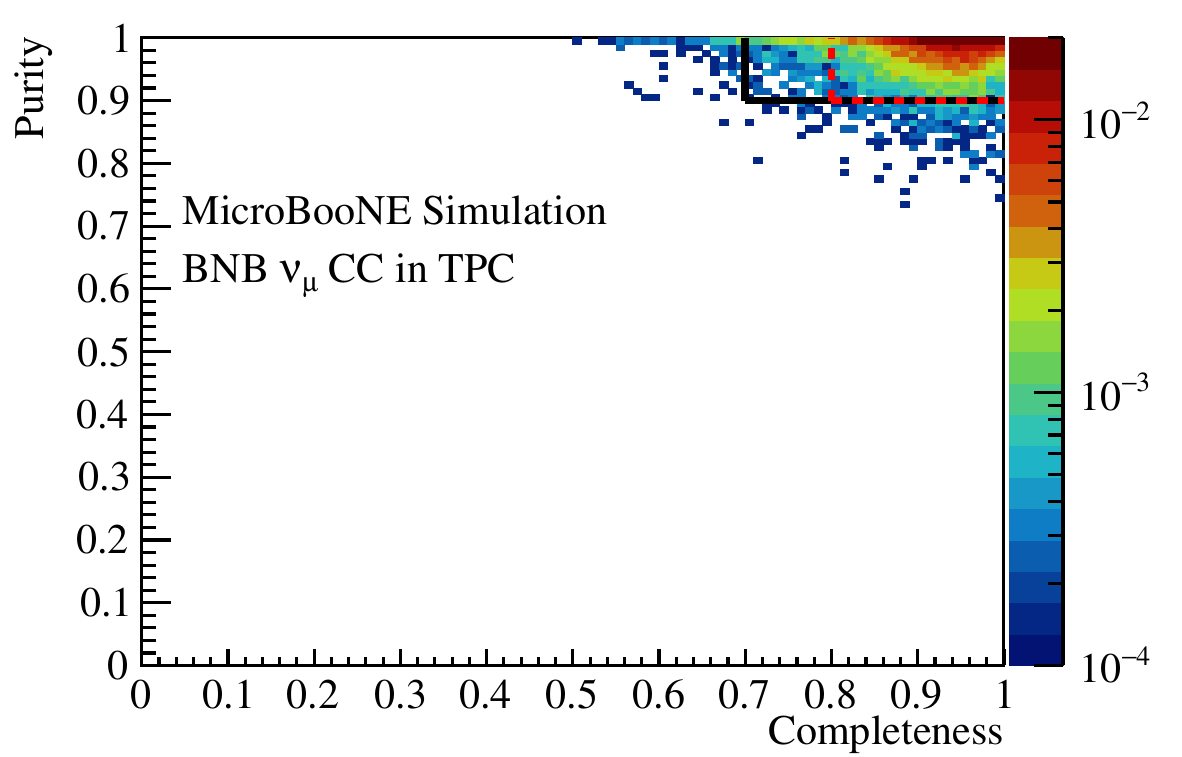}
    \includegraphics[width=0.49\textwidth]{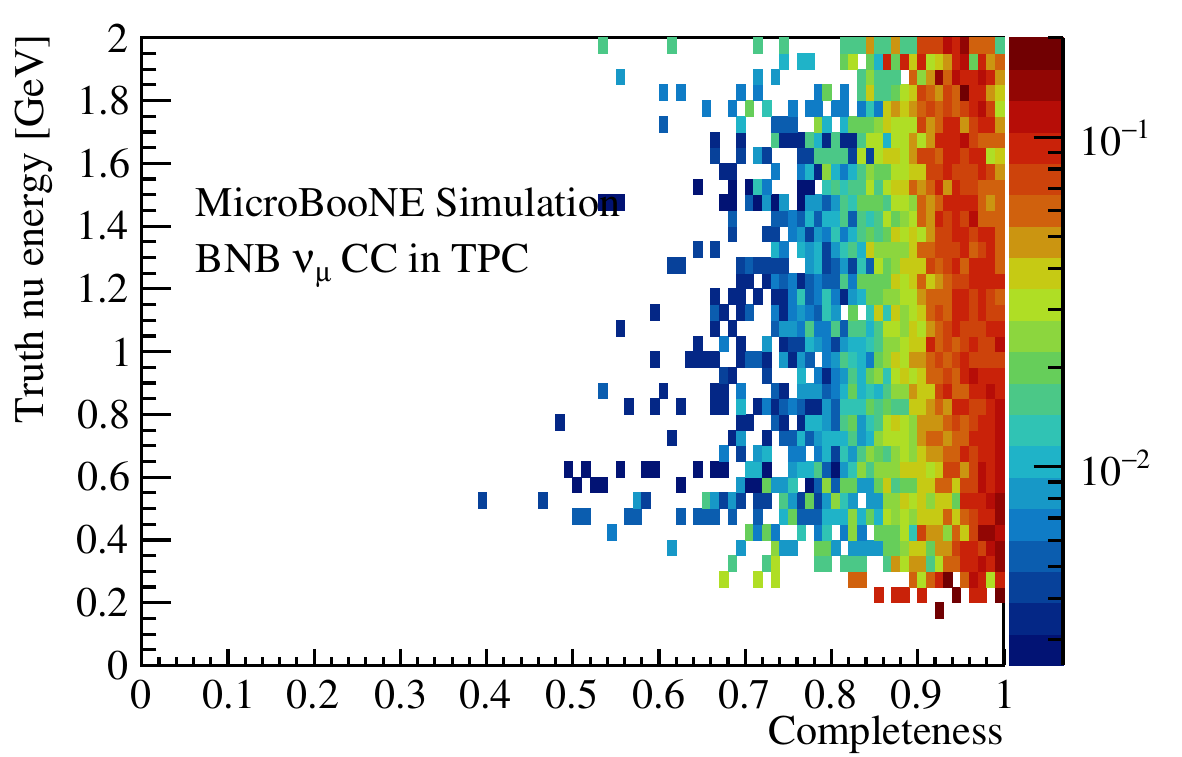}
    \includegraphics[width=0.49\textwidth]{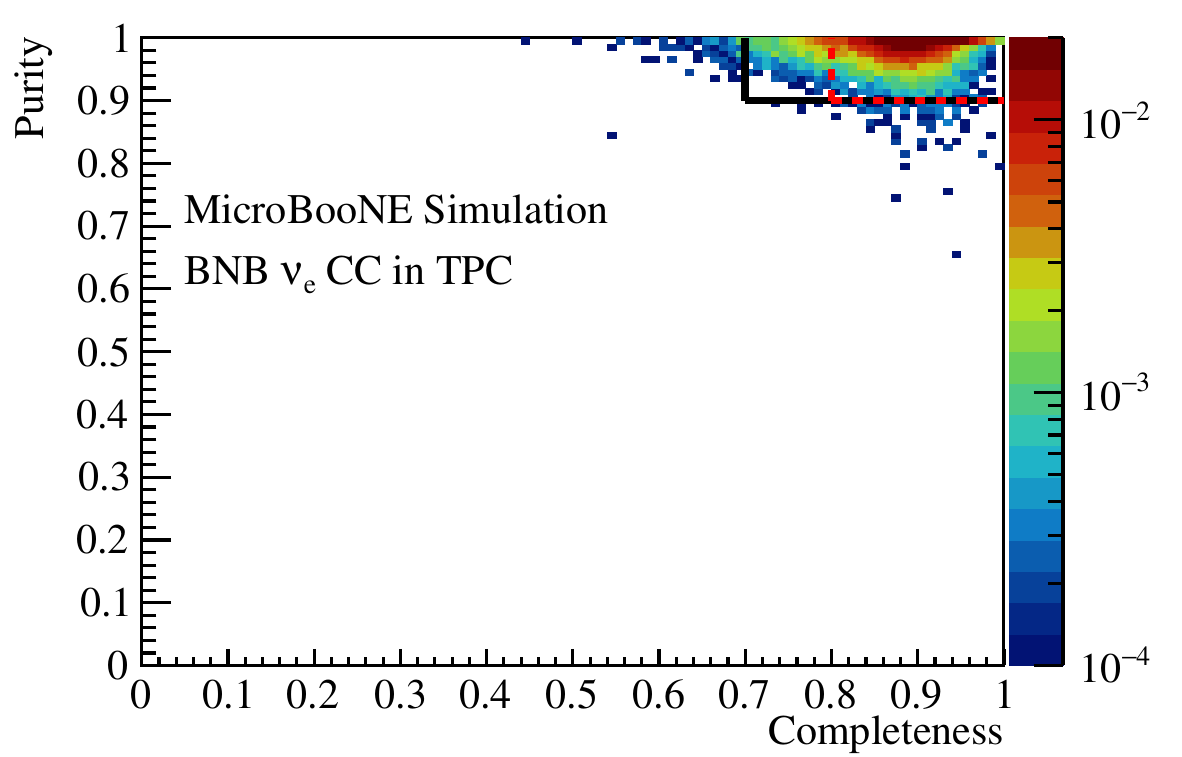}
    \includegraphics[width=0.49\textwidth]{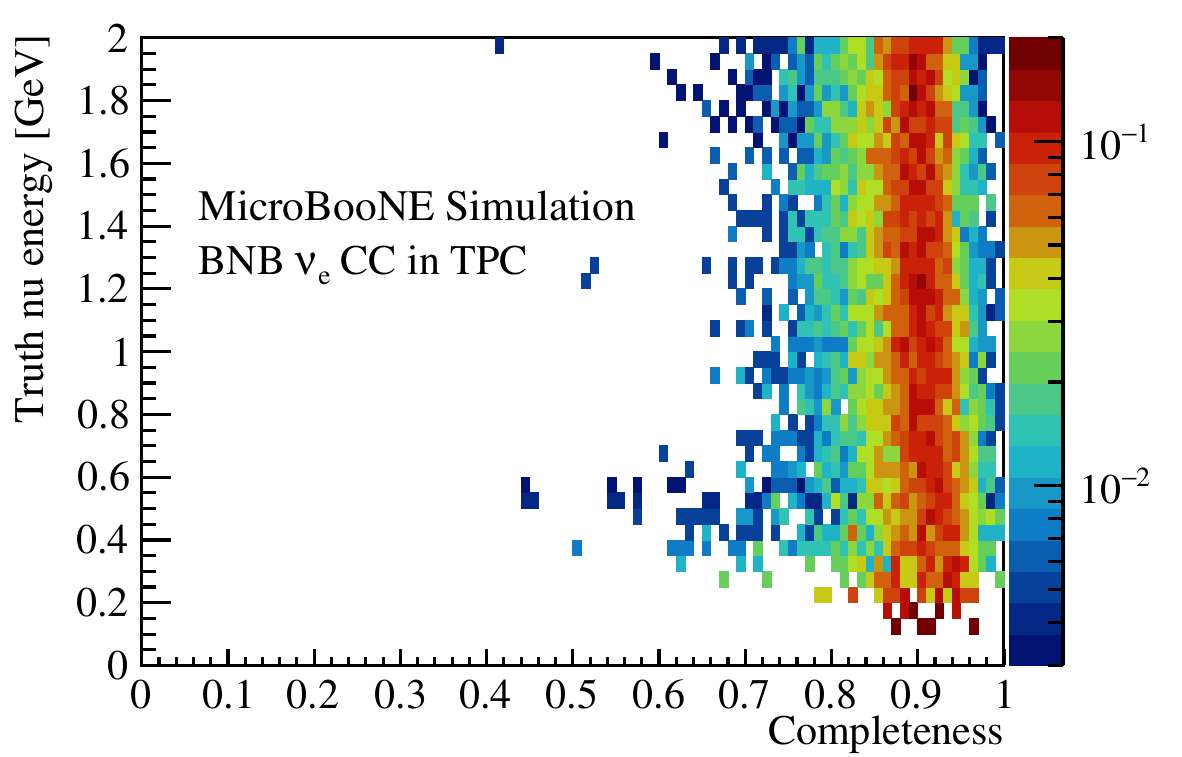}
    \includegraphics[width=0.49\textwidth]{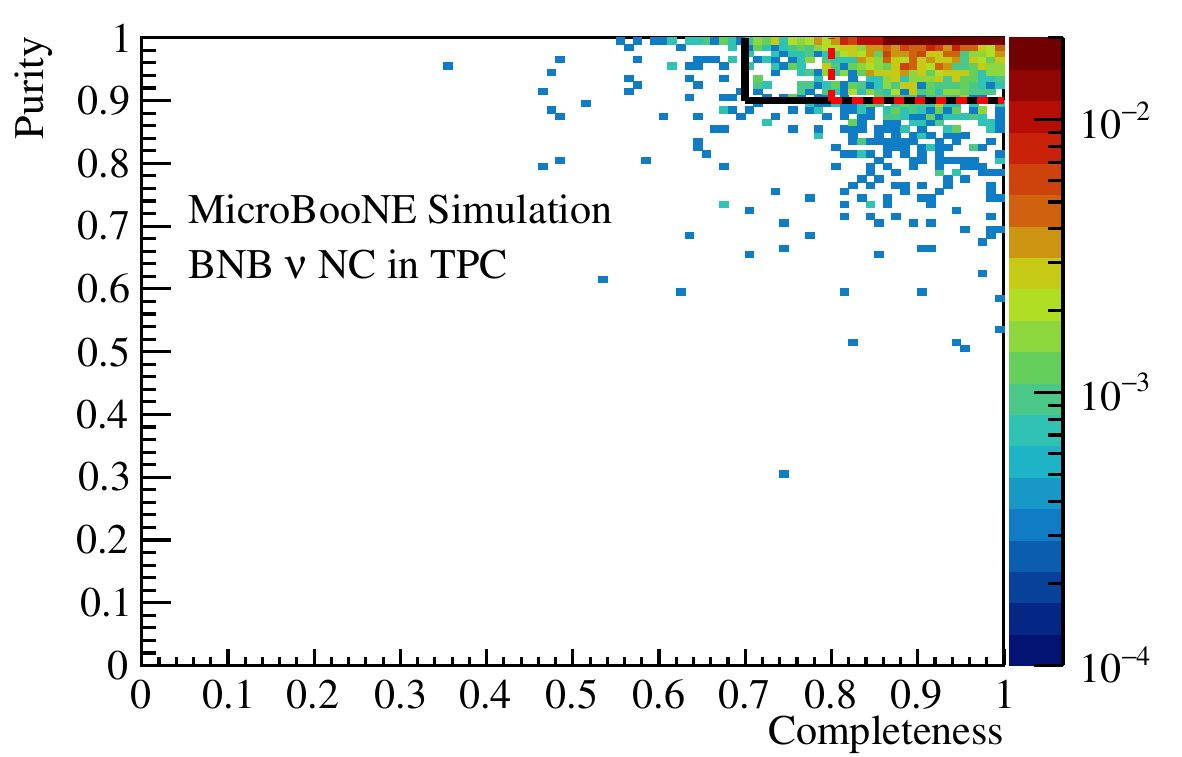}
    \includegraphics[width=0.49\textwidth]{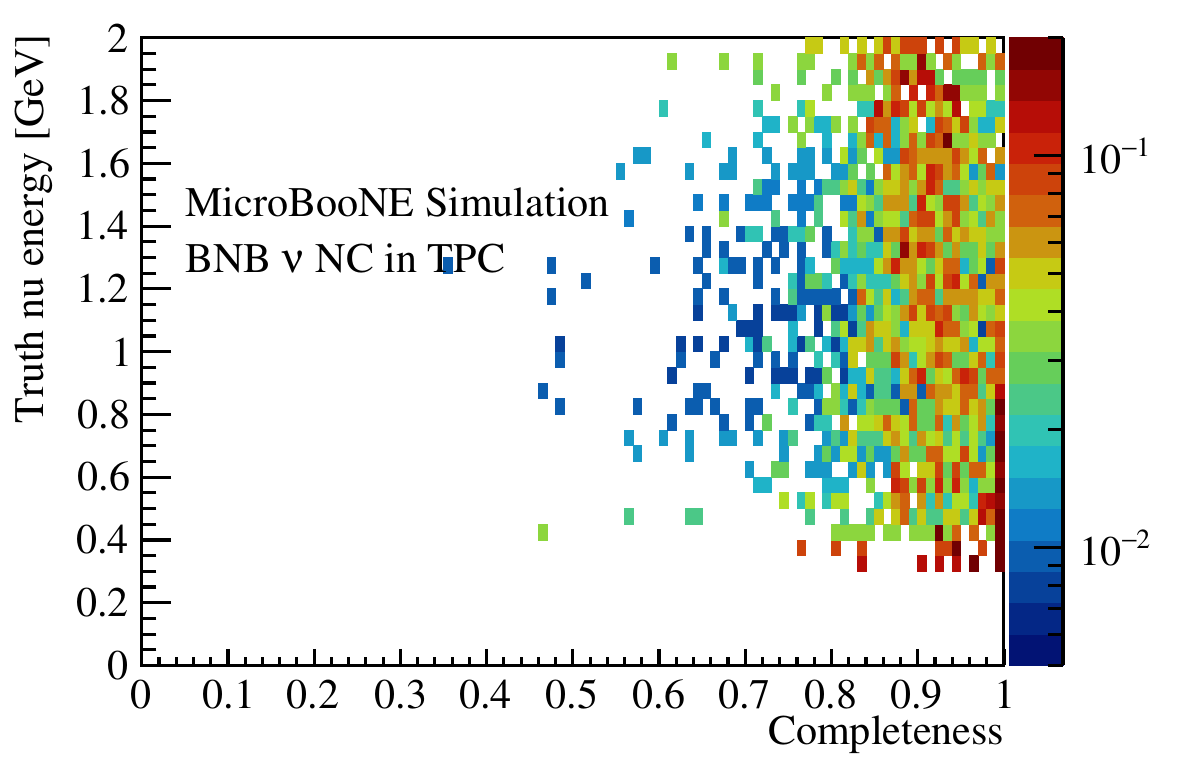}
    \caption{Two-dimensional distributions of the completeness and the purity of the 3D image for BNB $\nu_{\mu}$ CC, $\nu_{e}$ CC, and NC interactions in the TPC. 
    There are no cosmic-ray muons in this simulation.
    Left: purity vs. completeness for each neutrino interaction. The color scale (Z-axis value) represents the fraction of events.
    Right: true neutrino energy vs. completeness; the distribution is normalized for each row of the true neutrino energy bin. The color scale (Z-axis value) represents the fraction of events in each row.
    The integrated fraction of the events within the solid black and dashed red boxes can be found in table~\ref{tab:eval_nuonly}.
    }
  \label{fig:eval_bnb_only}
\end{figure}

 
Figure~\ref{fig:recovstruth_nuonly} shows two 2D snapshots of the 3D event displays. The left is a $\nu_{\mu}$ CC interaction producing a muon and a single proton (1$\mu$1$p$) in the final state. The right is a $\nu_{e}$ CC interaction producing an electron EM shower and a single proton (1$e$1$p$) in the final state. The red points represent the space points from Monte-Carlo truth and the blue ones represent the reconstructed space points in the 3D imaging.
The image of the reconstructed points are blurred because of the charge diffusion during the drift and the software filter smearing in the signal processing.
Generally speaking, the reconstructed 3D image has both good completeness and purity compared to the truth 3D image in these two examples; even the short tracks belonging to the EM shower and isolated energy depositions are reconstructed.
The quantitative evaluations of the purity and the completeness for BNB $\nu_{\mu}$ CC, $\nu_{e}$ CC, and NC interactions in the TPC are shown in figure~\ref{fig:eval_bnb_only}. The results are summarized in table~\ref{tab:eval_nuonly}.

\begin{table*}[tbhp!]
\centering
\begin{threeparttable}
    \caption{\label{tab:eval_nuonly}
    Fraction of the events that correspond to the completeness and purity values within the black or red boxes as shown in figure~\ref{fig:eval_bnb_only}. These numbers are the overall performance for the integrated BNB neutrino flux which has an average neutrino energy of about 800 MeV. See text for more discussions on the energy dependence.
    All neutrino interactions are simulated within the TPC active volume, without cosmic-ray muons.}
 \begin{tabular}{c|ccc}
 \hline
     Scenario: neutrino only  & BNB $\nu_{\mu}$ CC  & BNB $\nu_{e}$ CC &  BNB NC \\\hline
     Purity $>$90\% and &  &  &  \\
     Completeness $>80\%$   &  88.6\% & 89.2\% & 80.7\% \\
     Completeness $>70\%$   &  93.3\% & 96.7\% & 87.0\% \\\hline
 \end{tabular}
\end{threeparttable}
\end{table*}

The purity is high in the neutrino-only cases in which there are no cosmic-ray muons.
For neutrino energy less than 400 MeV, the purity performance, e.g. the fraction of events with greater than 90\% purity, is reduced by about 10\% compared to that in higher energy regions. This is due to the inefficiency of de-ghosting for low-energy events.
The lower purity for NC interactions mainly corresponds to the events with visible energy less than 100 MeV, in which case the 3D image consists of many dot-like or very short tracks.
Unlike figure~\ref{fig:eval_ideal_completeness}, figure~\ref{fig:eval_bnb_only} has no ultra-low completeness events because a final state of a single prolonged track can rarely happen for a neutrino interaction.
For $\nu_e$ CC interactions, since primary electrons lead to EM showers through significant Bremsstrahlung radiation, the peak completeness is biased down to $\sim$90\% because of the inefficiency for isolated low energy depositions in the shower. Such a bias is not critical to the track versus shower identification and can be corrected in the shower energy reconstruction.
NC interactions generally generate protons, neutrons, or pions. These particles could yield low energy depositions during their travel in the liquid argon as explained previously, introducing a much smeared completeness distribution.
The 100\% completeness peak for the low-energy neutrino NC interactions as seen in the bottom right panel of figure~\ref{fig:eval_bnb_only} mainly corresponds to elastic scattering with a single low-energy proton emitted. 

A dependence of the completeness on the true neutrino energy is indicated by the right panel of figure~\ref{fig:eval_bnb_only}. A high energy neutrino is more likely to produce multiple energetic hadrons, introducing distant or isolated low-energy depositions via nuclear recoils, de-excitation of argon nuclei, pion decays, etc. These low-energy TPC activities are more likely to be suppressed in the signal processing or 3D imaging as discussed previously.

\subsection{Final performance in realistic cases}\label{sec:eval_nuoverlay}
In this section, the neutrino-overlay samples are used to demonstrate the final performance of the Wire-Cell 3D imaging, clustering, and charge-light matching. 
Neutrino interactions are simulated and mixed with real cosmic-ray data.
The clustering, light signal reconstruction, and charge-light matching are applied on the 20--30 TPC clusters and 40--50 PMT flashes to select the in-beam neutrino activities.
The efficiency and correctness of the charge-light matching and the quality of the 3D images of the selected neutrino candidate clusters are keys to the downstream reconstructions.

\begin{figure}[thpb!]  
  \centering
    \begin{overpic}[width=0.8\textwidth]{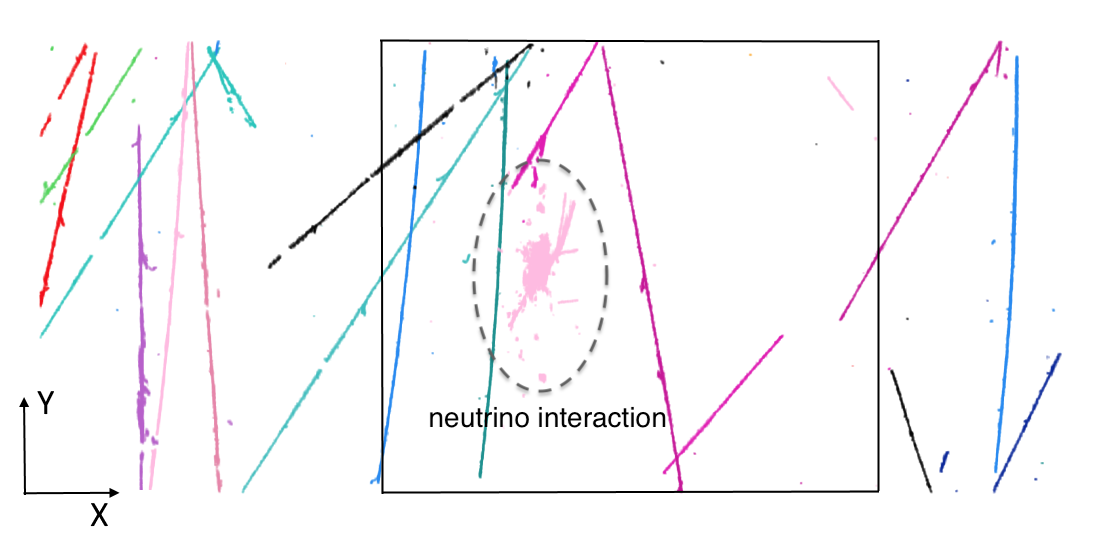}
        \put(10,48){\textbf{\small MicroBooNE Simulation}}
    \end{overpic}
    \begin{overpic}[width=0.8\textwidth]{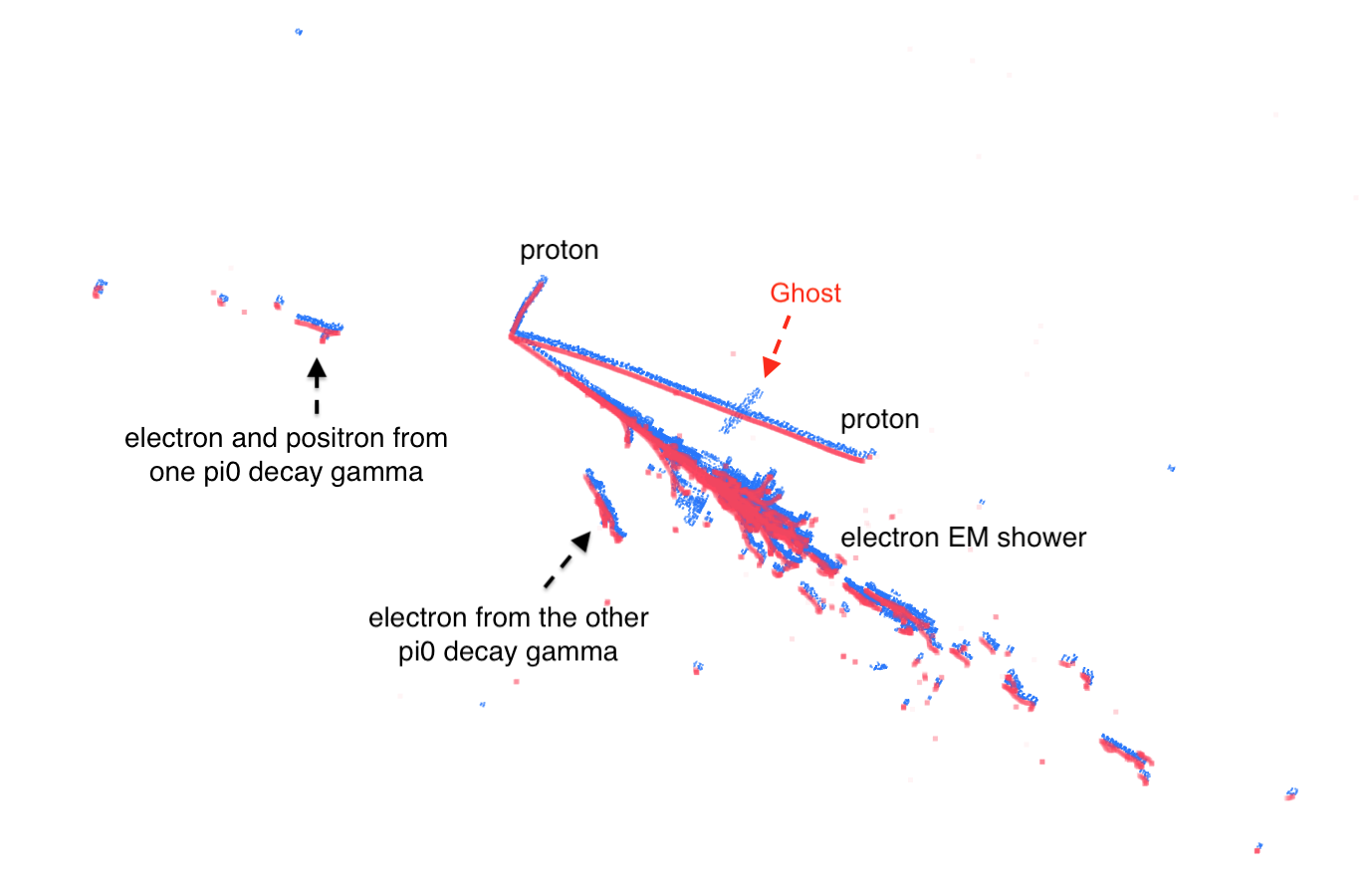}
        \put(10,48){\textbf{\small MicroBooNE Simulation}}
    \end{overpic}
    \caption{Event display of a 1$e$2$p$1$\pi^0$ $\nu_{e}$ CC interaction. 
    Top: side view of the full TPC readout; each cluster is labeled in one color. The black box corresponds to the LArTPC active volume with an X-position (converted from the readout time) relative to the neutrino interaction time.
    Bottom: the charge-light matching result -- the in-beam flash matched TPC activities; the blue points are the reconstructed 3D space points and the red ones are the true space points corresponding to the neutrino interaction.
    There is an offset of about 1 cm between the blue point and the red point to clearly show the event.
    The voxel size and opacity are also tuned for event display.
    }
  \label{fig:recovstruth_1e2p1pi}
\end{figure}

Figure~\ref{fig:recovstruth_1e2p1pi} shows an example of one of the most challenging cases.
The top panel shows the X-Y projection of all TPC activities including cosmic-ray muons and a neutrino interaction.
The bottom panel shows the reconstructed 3D image of the matched in-beam TPC activities and the truth trajectories of the neutrino interaction's final state particles.
In this example, there are two protons and an electron EM shower connected to the neutrino interaction vertex.
A $\pi^0$ is also created and decays into two $\gamma$'s. 
The two $\gamma$'s deposit energy through electrons from Compton scattering or pair production into electrons and positrons.
A proper clustering of the two detached $\gamma$'s is difficult considering the surrounding cosmic-ray muons.
In this example, there is also a ghost track which crosses one proton track in the 2D projection, but it is actually detached from the proton track in the 3D space. 
It resides in the nonfunctional wire region and originates from part of a cosmic-ray muon track.

Without any pattern recognition or topological reconstruction at this stage, the completeness is a more critical metric than the purity. There is little chance to fix the incompleteness issue in the downstream analysis chain once the charge is already lost. However, the purity can be further improved. For example, the ghost track in figure~\ref{fig:recovstruth_1e2p1pi} can be removed by checking the directionality, or by particle identification using dE/dx information, in which case this ghost track will be regarded as a cosmic-ray muon background.

\begin{figure}[thpb!]
  \centering
    \includegraphics[width=0.49\textwidth]{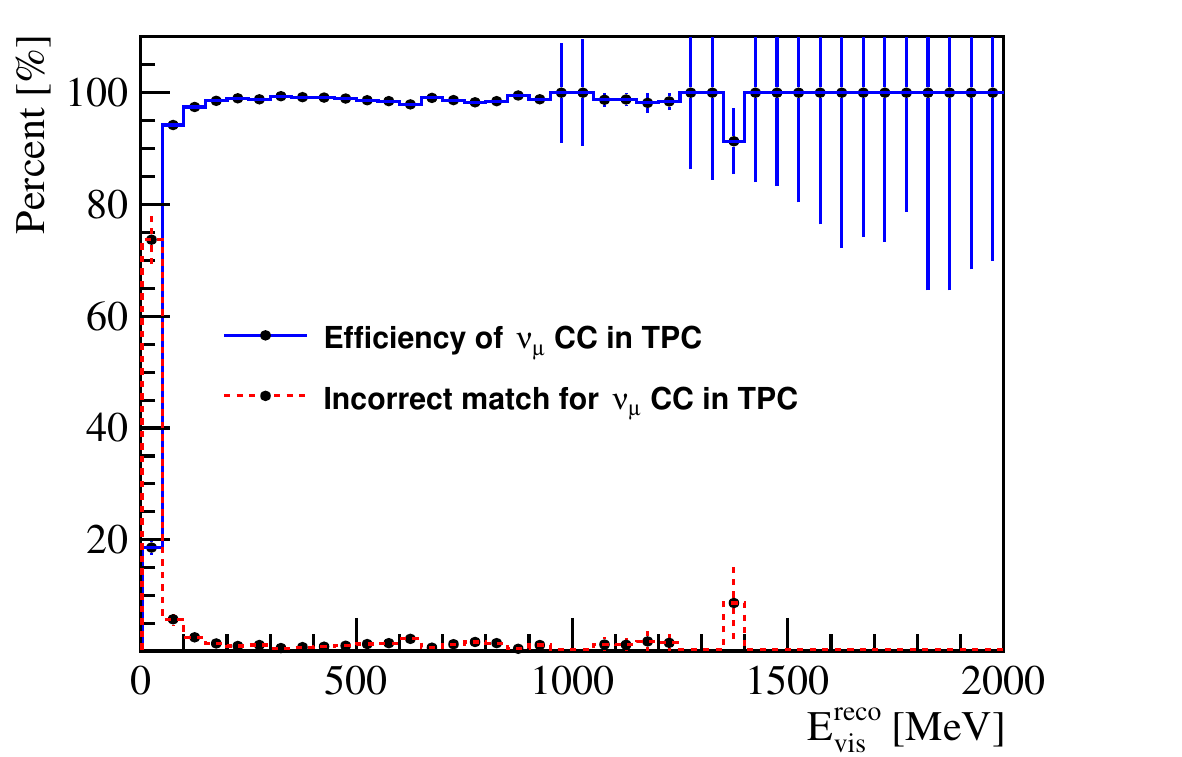}
    \includegraphics[width=0.49\textwidth]{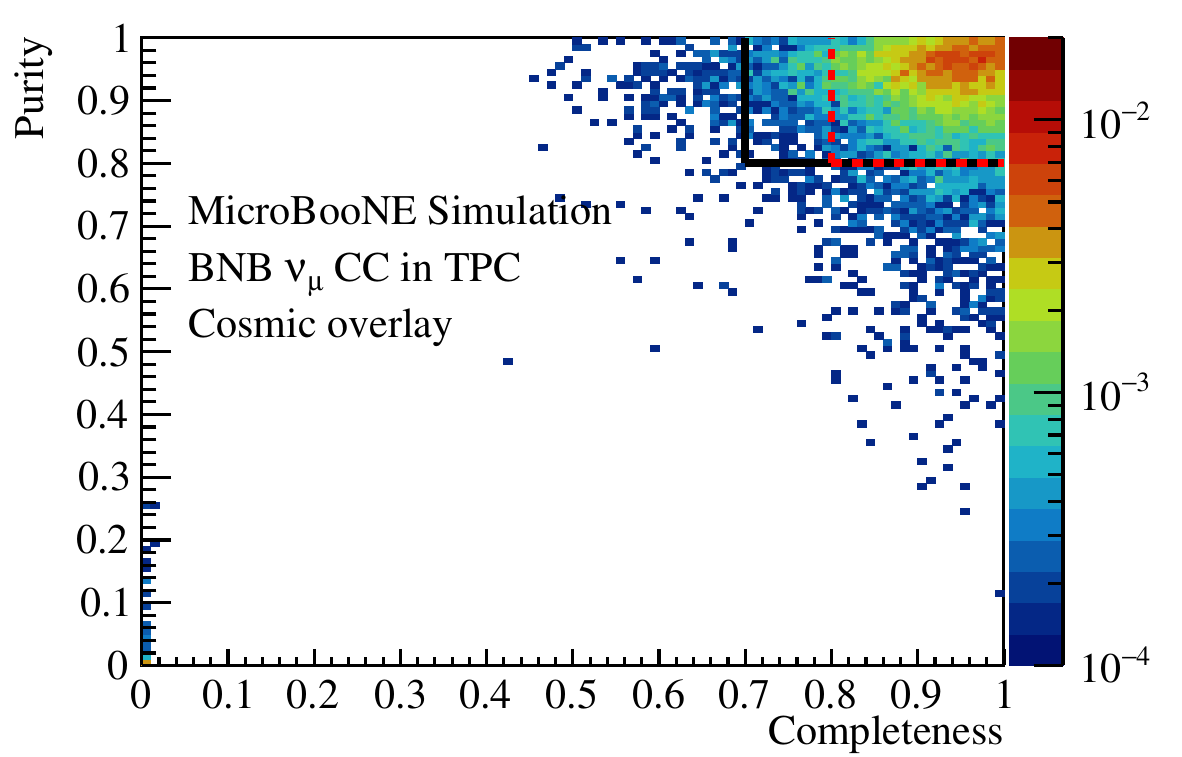}
    \includegraphics[width=0.49\textwidth]{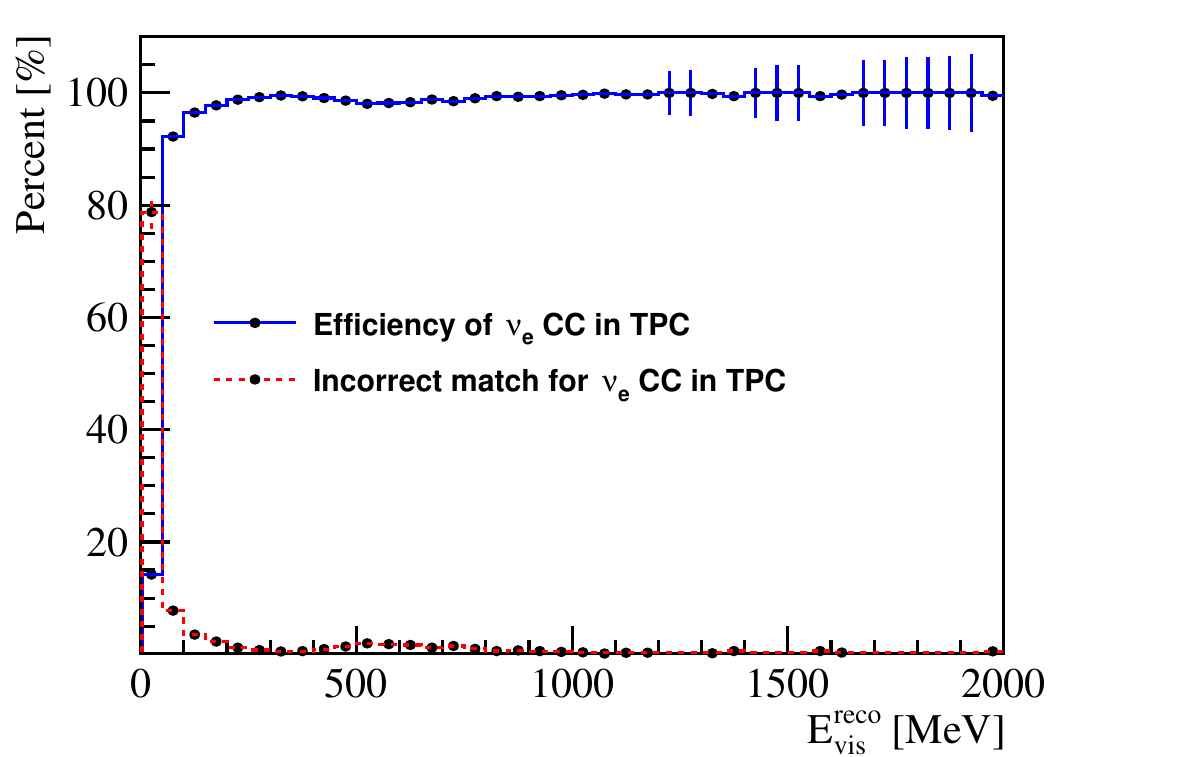}
    \includegraphics[width=0.49\textwidth]{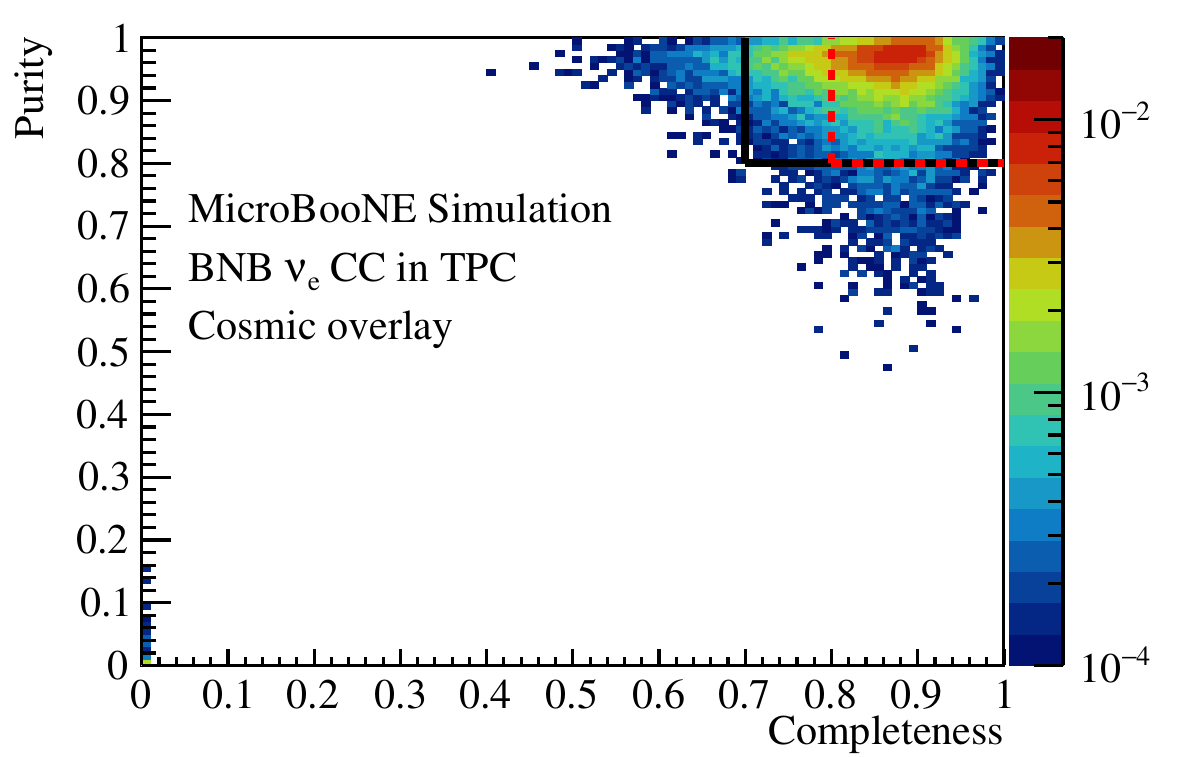}
    \includegraphics[width=0.49\textwidth]{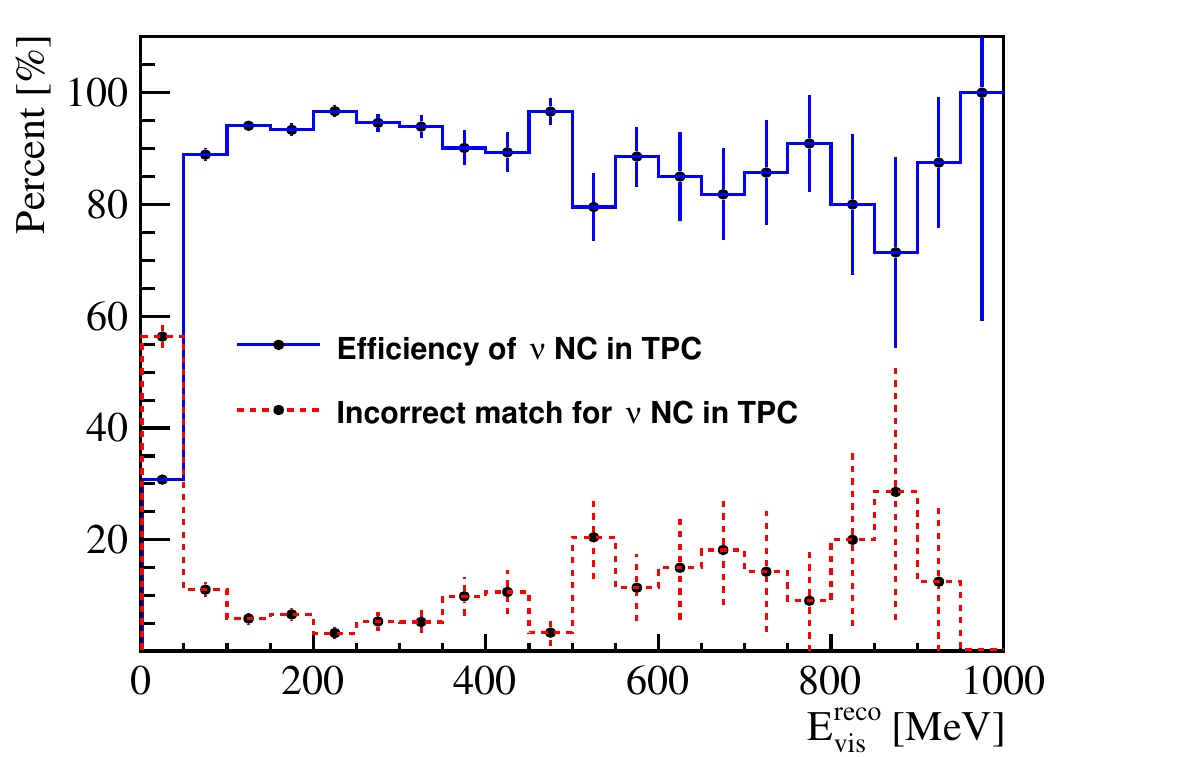}
    \includegraphics[width=0.49\textwidth]{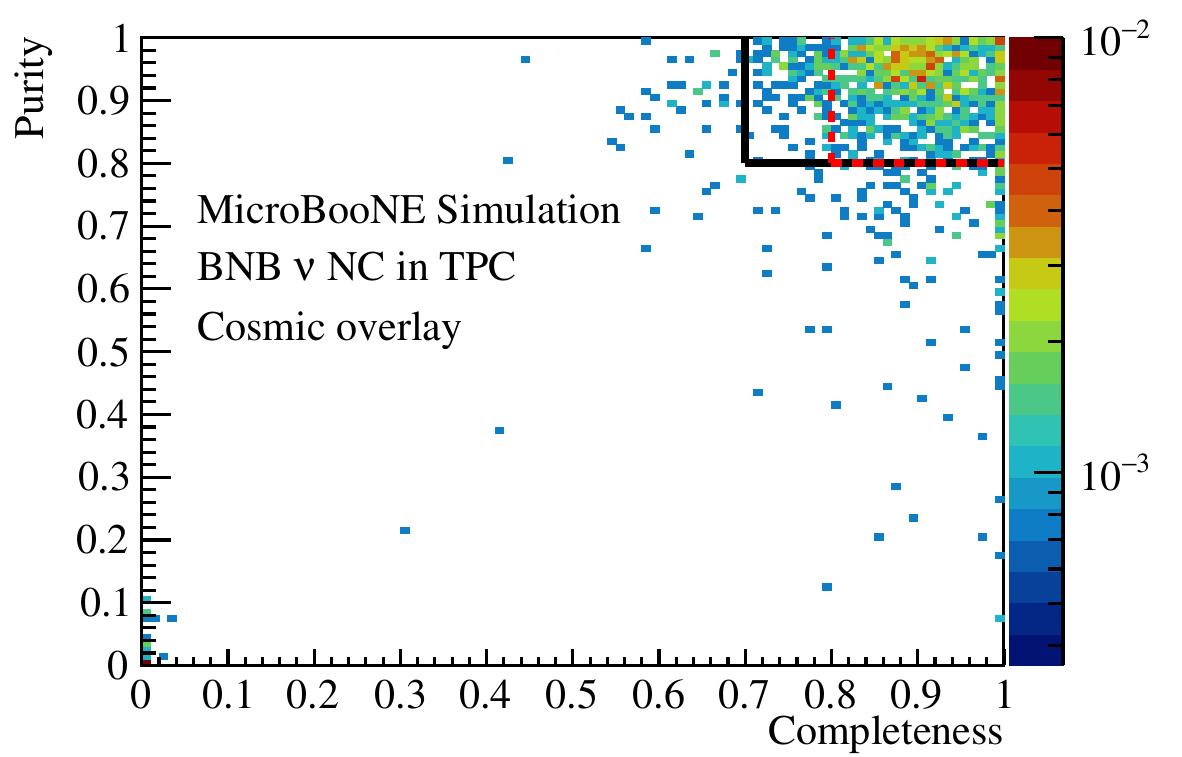}
    \caption{
    3D imaging and charge-light matching performance for BNB $\nu_{\mu}$ CC , $\nu_{e}$ CC, and NC interactions in the TPC.
    The neutrino interactions are simulated and overlaid with real data from cosmic rays. 
    The clustering and charge-light matching steps are applied to select the neutrino interaction.
    Left: efficiency and incorrectness of charge-light matching as a function of the simply reconstructed visible energy (a simple conversion from the reconstructed visible charge using a constant conversion factor); binomial statistics is used to calculate the efficiency uncertainty while Poisson statistics (large error bars in the plot) is used where the efficiency is 100\%, mainly for the low statistic bins.
    Right: purity vs. completeness for each selected neutrino interaction. The color scale (Z-axis value) represents the fraction of events.
    The integrated fraction of the events within the solid black and dashed red boxes can be found in table~\ref{tab:eval_nuoverlay}.
    }
  \label{fig:eval_bnb_overlay}
\end{figure}

Left panel of figure~\ref{fig:eval_bnb_overlay} shows the efficiency and incorrectness of the charge-light matching procedure.
The overall efficiency to select the neutrino CC interactions in the TPC active volume is 95\%, and the overall incorrectness values of charge-light matching are 4.6\%, 3.8\%, and 28.7\% for BNB $\nu_{\mu}$ CC, $\nu_{e}$ CC, and NC interactions, respectively.
The efficiency plus incorrectness is 100\% in this figure except for the first bin with low visible energy $<$50 MeV in which case some of events fail to match in-beam TPC activities to any PMT flash.
A neutrino interaction, close to the TPC boundary or with a significant number of neutral particles in the final states, tends to have a large portion of its charges escaping the active TPC volume, which then become invisible to the wire readout plane. However, the light signals originating from this neutrino interaction can still be collected if there is any charge deposition outside the TPC but still in the liquid argon volume.
Such inconsistent TPC activities and PMT signals may result in incorrect matches or no matches.

\begin{figure}[thpb!]
  \centering
    \includegraphics[width=0.48\textwidth]{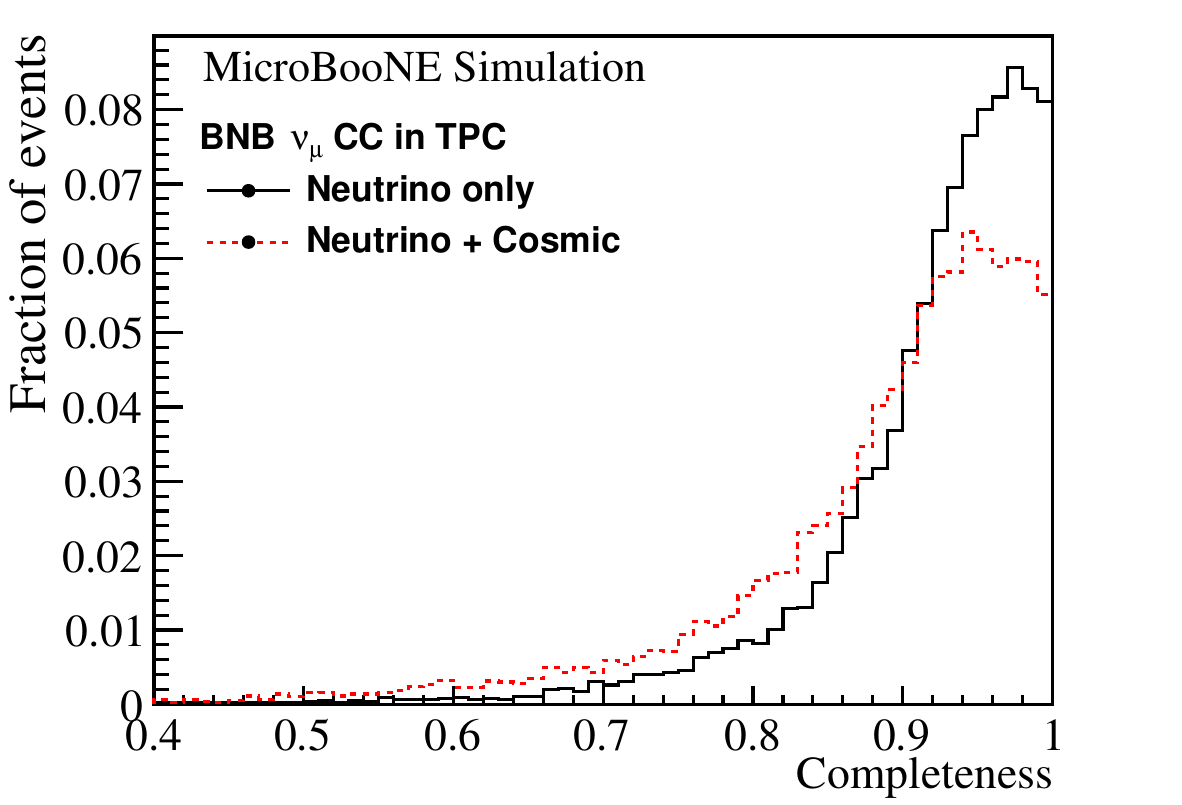}
    \includegraphics[width=0.48\textwidth]{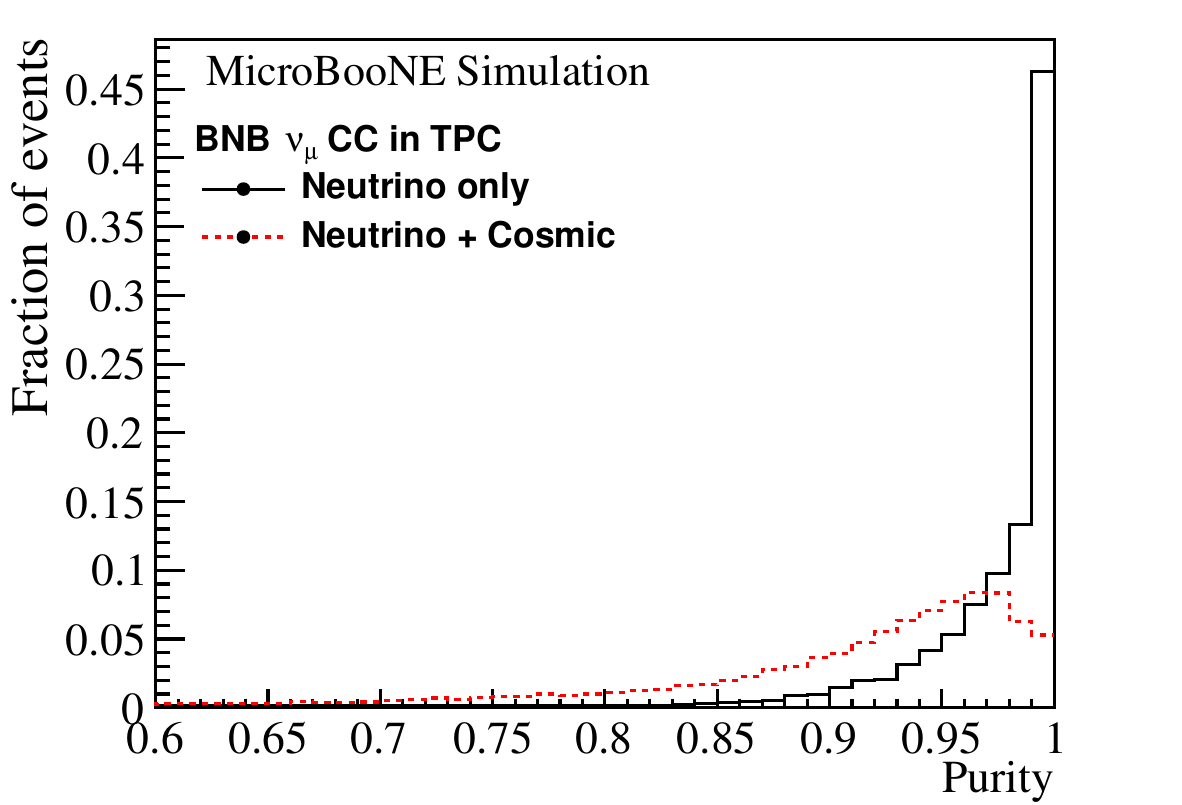}
    \includegraphics[width=0.48\textwidth]{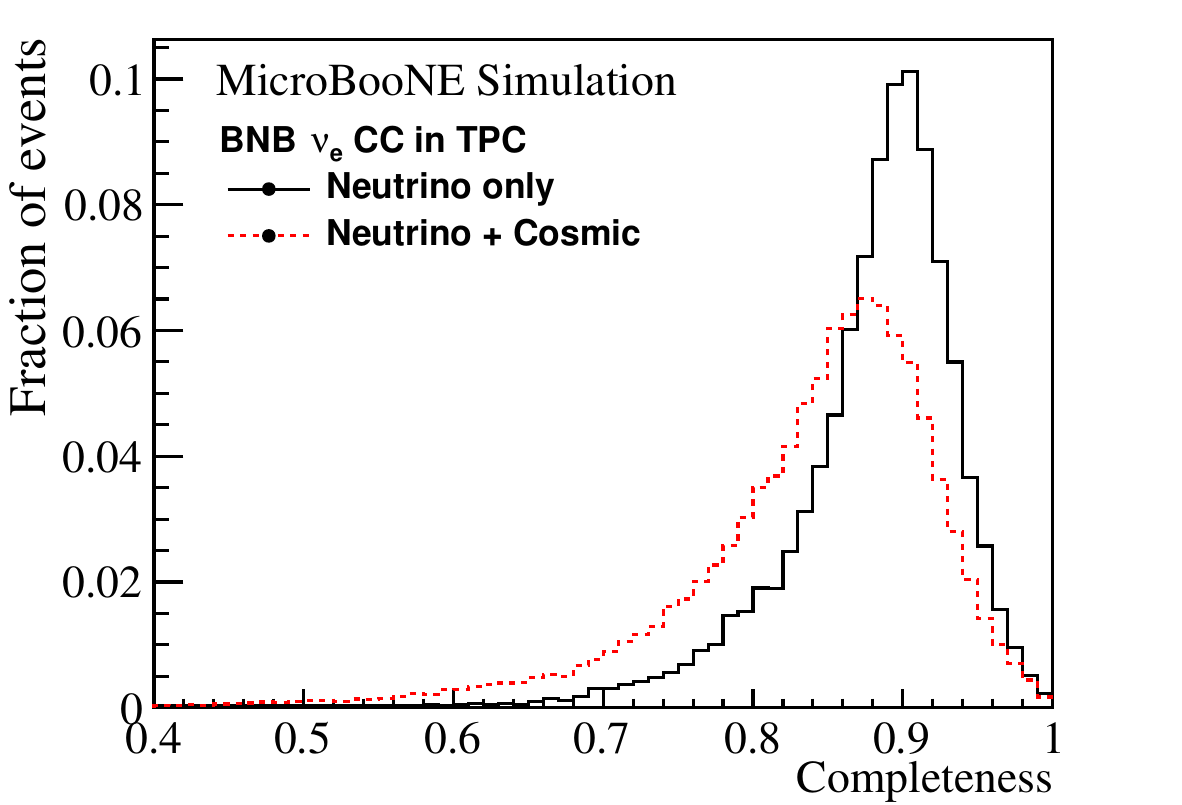}
    \includegraphics[width=0.48\textwidth]{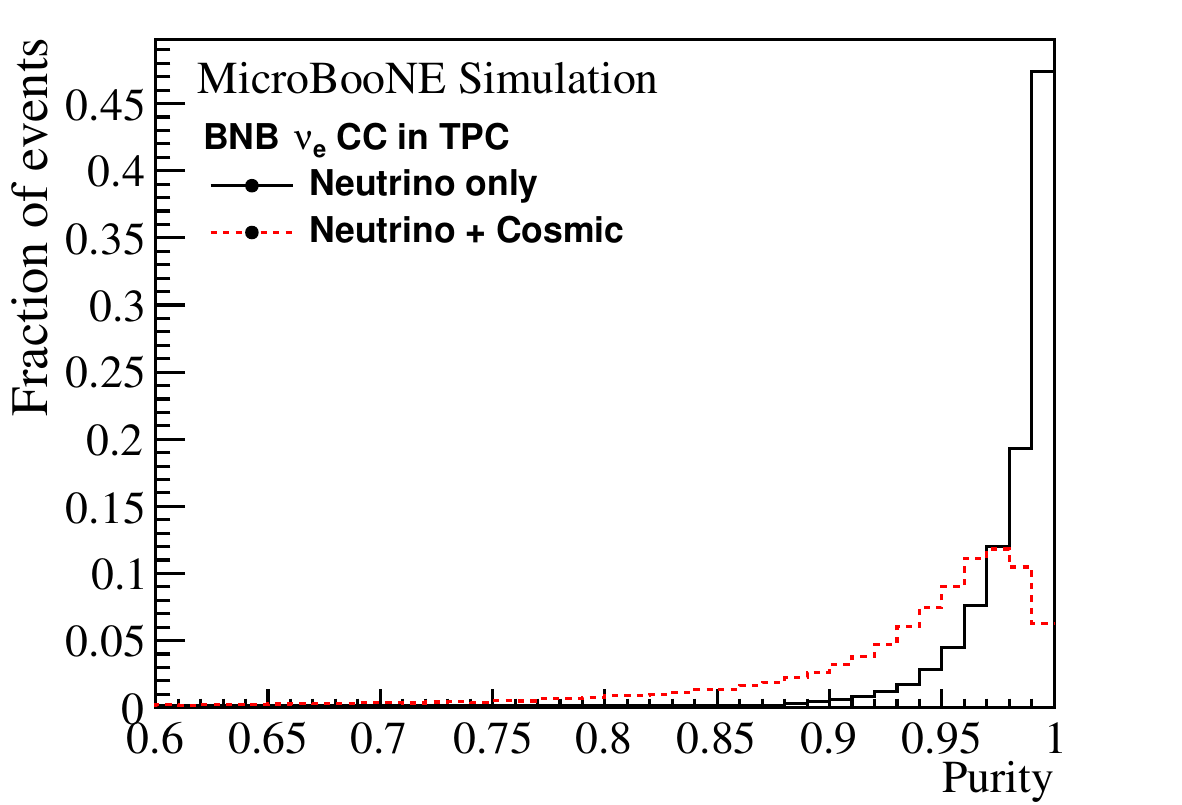}
    \includegraphics[width=0.48\textwidth]{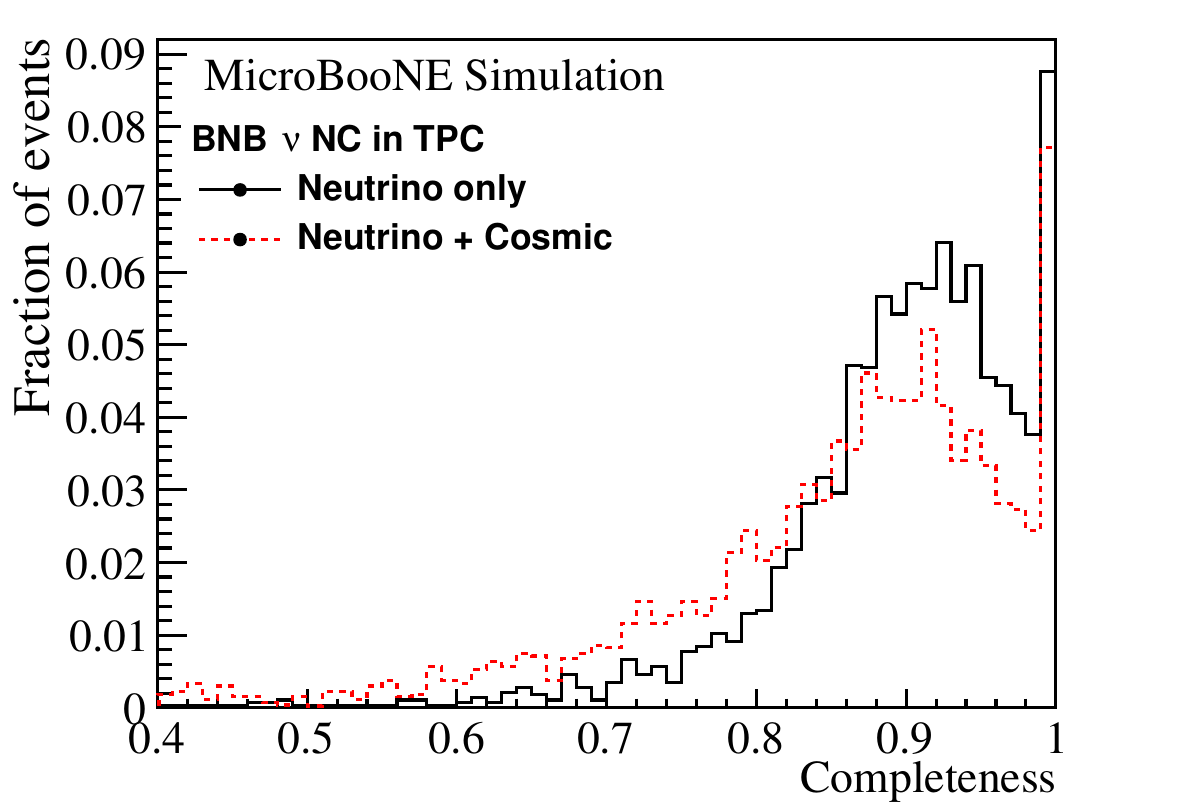}
    \includegraphics[width=0.48\textwidth]{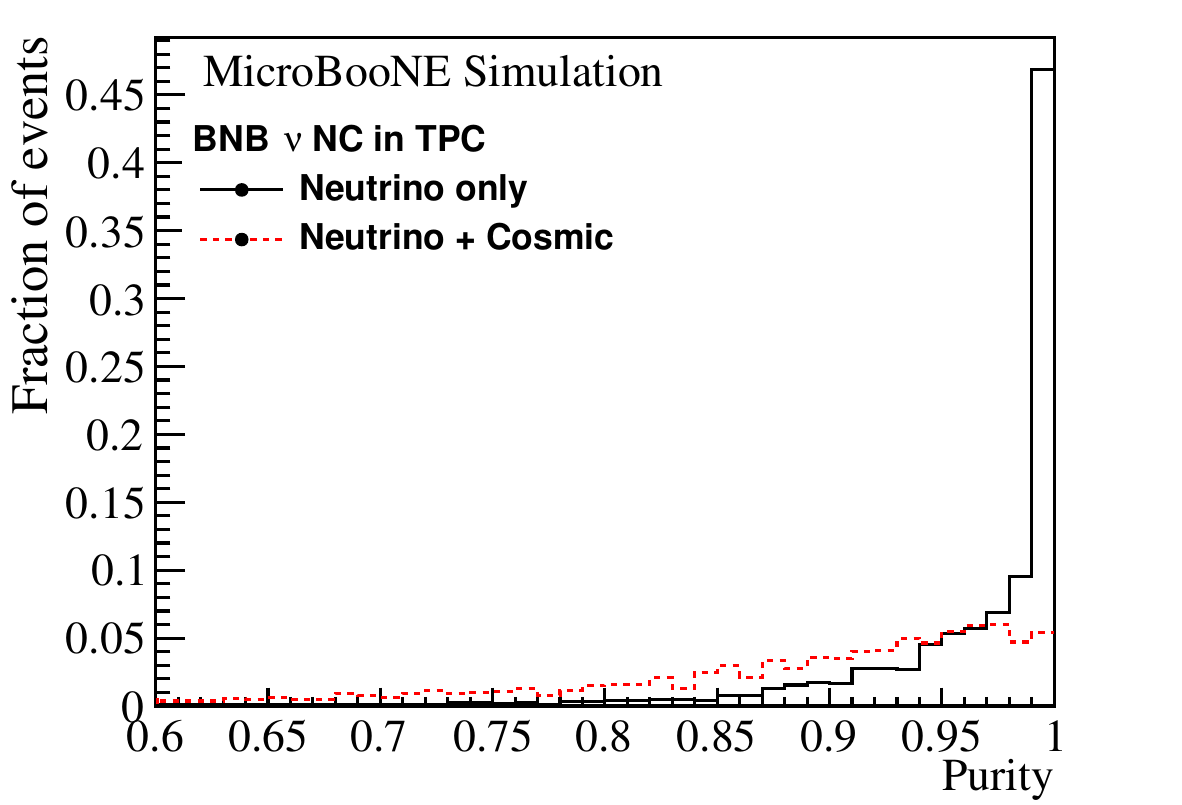}
    \caption{
    Independent comparisons of completeness and purity distributions for the two scenarios of ``neutrino-only'' and ``neutrino + cosmic''.
    The 3D clustering and charge-light matching steps are applied in the scenario of ``neutrino + cosmic'' to select the neutrino interaction.
    Top: BNB $\nu_{\mu}$ CC interactions in the TPC.   
    Middle: BNB $\nu_{e}$ CC interactions in the TPC. 
    Bottom: BNB $\nu$ NC interactions in the TPC; the ``$\sim$100\%'' completeness peak value mainly corresponds to the NC quasi-elastic scattering with a single low-energy (short) proton emitted and it can also be seen in the bottom right panel of figure~\ref{fig:eval_bnb_only} for low-energy neutrino NC interactions.
    See text for more details.
    }
  \label{fig:eval_compare}
\end{figure}

Right panel of figure~\ref{fig:eval_bnb_overlay} presents the completeness and purity of the selected TPC activities for BNB $\nu_{\mu}$ CC, $\nu_{e}$ CC, and NC interactions, respectively. The results of completeness and purity are summarized in table~\ref{tab:eval_nuoverlay}. The events with extremely low completeness and purity values as shown in the bottom left corner in each sub-figure of the right panel correspond to the incorrect charge-light matches as discussed previously.
\begin{table*}[tbhp!]
\centering
\begin{threeparttable}
    \caption{\label{tab:eval_nuoverlay}
    Fraction of the events that correspond to the completeness and purity values within the black or red boxes as shown in figure~\ref{fig:eval_bnb_overlay}. These numbers are the overall performance for the integrated BNB neutrino flux which has an average neutrino energy of about 800 MeV. 
    All neutrino interactions are simulated within the TPC active volume, with cosmic-ray data (beam-off) overlaid.}
 \begin{tabular}{c|cccc}
 \hline
     Scenario: neutrino + cosmic & BNB $\nu_{\mu}$ CC  & BNB $\nu_{e}$ CC &  BNB NC \\\hline
     Purity $>$80\% and  &  &  &  \\
     Completeness $>80\%$   &  73.0\% & 67.7\% & 56.0\% \\
     Completeness $>70\%$   & 80.2\% & 83.4\% & 66.5\% \\\hline
 \end{tabular}
\end{threeparttable}
\end{table*}
Comparing figure~\ref{fig:eval_bnb_only} and figure~\ref{fig:eval_bnb_overlay}, the degradation of the completeness and the purity can be attributed to the numerous cosmic-ray muons that traverse the detector.
Direct comparisons of the completeness and the purity are independently performed, and the distributions can be seen in figure~\ref{fig:eval_compare}.
The scenarios of ``neutrino-only'' and ``neutrino + cosmic'' correspond to figure~\ref{fig:eval_bnb_only} and figure~\ref{fig:eval_bnb_overlay}, respectively.
In the scenario of ``neutrino + cosmic'', the neutrino activities suffer not only an over-clustering issue with the cosmic-ray activities (or its related ghost tracks) but also an under-clustering issue since part of the detached activities from the neutrino primary cluster may be grouped to cosmic-ray muons.
These two issues introduce a smearing of both the completeness and purity distributions.
The typical values of the completeness and the purity for different scenarios and interaction types are summarized in table~\ref{tab:eval_compare}. 

\begin{table*}[tbhp!]
\centering
\begin{threeparttable}
    \caption{\label{tab:eval_compare}
    Summary of typical completeness and purity values corresponding to the distributions as shown in figure~\ref{fig:eval_compare}. Independent comparisons of completeness and purity are performed.
    The numbers are given as the fraction of the corresponding events. All neutrino interactions are simulated within the TPC active volume.}
 \begin{tabular}{c|c|ccc}
 \hline
     & Scenario & BNB $\nu_{\mu}$ CC  & BNB $\nu_{e}$ CC &  BNB NC \\\hline
     \multirow{2}{*}{\parbox{2.5 cm}{Completeness $>$80\% ($>$70\%)}} &  Neutrino + cosmic  &  84.5\% (93.4\%) & 74.9\% (92.5\%) & 73.1\% (87.9\%) \\
    &  Neutrino only &  92.8\% (97.8\%) & 90.8\% (98.5\%) & 90.2\% (97.4\%) \\\hline
     \multirow{2}{*}{Purity $>$80\%} & Neutrino + cosmic &  84.4\%  & 89.2\%  &  72.6\% \\
     &  Neutrino only &  99.4\%  & 99.8\%  &  97.1\% \\\hline
 \end{tabular}
\end{threeparttable}
\end{table*}
About 85\% of events have at least 80\% completeness for BNB $\nu_{\mu}$ CC interactions.
About 90\% of events have at least 70\% completeness for BNB $\nu_{e}$ CC or NC interactions.
The degradation of purity in the scenario of ``neutrino + cosmic'' is more severe than the degradation of completeness, as expected.
However, as explained previously, the completeness is more critical in the 3D imaging and charge-light matching stage, since the purity can be further improved in the later analysis chain.
In the optimization of the clustering and charge-light matching algorithms, the completeness is thus considered more favorably than the purity.
In reality, the final purity performance corresponding to the scenario of ``neutrino + cosmic'' is still very good and 80\%-90\% events have at least 80\% purity for the CC interactions.

In summary, the quantitative evaluations of the Wire-Cell 3D imaging, clustering, and charge-light matching have been presented in this section.
These techniques result in a high performance selection of the neutrino activities in the MicroBooNE LArTPC with a clean removal of the 20--30 cosmic-ray muons within a TPC readout.
The quality (completeness and purity) of the 3D images of the selected in-beam neutrino activities is very good considering the complexities from the wire readout ambiguity, nonfunctional wires, non-perfect signal processing, and numerous cosmic ray muons.

\clearpage
\section{Summary and Discussion}\label{sec:summary}
This article describes the principle and algorithms of the Wire-Cell 3D imaging, clustering, and many-to-many charge-light matching applied in the MicroBooNE LArTPC.
The 3D imaging tomographically reconstructs the 3D image of the ionization electrons using the fundamental information of charge, time, and geometry of the LArTPC detector. 
Other characteristics of the LArTPC physics activities such as sparsity, positivity, and proximity are utilized as additional constraints to improve the 3D imaging performance.    
The many-to-many charge-light matching with 3D clustering and light signal reconstruction is developed to pair the TPC clusters and PMT flashes to identify the neutrino interaction among numerous cosmic-ray muons.
Several realistic issues, e.g.~the nonfunctional wires, the gaps due to inefficient signal processing, detached neutrino activities, and coincidentally connected clusters, are properly addressed.
Using the MicroBooNE detector simulation, the realistic performance of the reconstruction techniques is evaluated.

In spite of the effort, there are some limitations in the 3D imaging as shown in the event displays in this paper.
For example, prolonged tracks, which often develop gaps in the signal processing stage, cannot be entirely fixed via the bridging of the gaps as implemented.
Similar issues occur for the isochronous tracks that often develop gaps because of the insufficient coherent noise filtering.
Subsequent pattern recognition techniques, e.g.~particle-level clustering and trajectory fitting, may further address this problem.
Isochronous tracks present another common problem for the LArTPC 3D imaging, as the wire readout ambiguity is drastically increased in the time slice containing them.
In Wire-Cell 3D imaging, this issue is mitigated by introducing tiling, however, the blobs of the isochronous track are significantly broadened, leading to a much worse spatial resolution in the Y-Z plane view.
This issue can be further mitigated via trajectory fitting in a later stage.

Generally speaking, the 3D event reconstruction techniques as presented in this paper are adequately accurate and efficient, and can successfully select neutrino interaction activities.
About 95\% of the neutrino CC interactions in the TPC active volume are selected, with a 30-fold reduction of non-beam-coincident cosmic-ray muons.
Good completeness and purity of the resulting 3D image of the selected neutrino activities have been achieved.
Greater than 80\% of the selected neutrino CC interactions have a reconstructed 3D image of at least 70\% completeness and 80\% purity.
These techniques will benefit the downstream pattern recognition and neutrino selection, 
and they are important steps towards realizing the full capability of single-phase LArTPCs.
In particular, the Wire-Cell based neutrino selection and analyses take full advantage of these tools to further reject cosmic muons and select neutrinos~\cite{Wire-Cell-Generic-PRD} and demonstrate a very promising high efficiency and high purity neutrino selection in LArTPCs.
Other analyses using techniques such as deep learning with convolutional neural networks~\cite{uboone-dl, uboone-dl2} and Pandora multi-algorithm pattern recognition~\cite{uboone-pandora} can also benefit from the outcome of the Wire-Cell 3D reconstruction tools, as it provides a clean and intact 3D image of the in-beam neutrino activities with the surrounding cosmic ray muons removed.

\clearpage
\acknowledgments

This document is prepared by the MicroBooNE collaboration using the resources of the Fermi National Accelerator Laboratory (Fermilab), a U.S. Department of Energy, Office of Science, HEP User Facility.
Fermilab is managed by Fermi Research Alliance, LLC (FRA), acting
under Contract No. DE-AC02-07CH11359.  MicroBooNE is supported by the following: the U.S. Department of Energy, Office of Science, Offices of High Energy Physics and Nuclear Physics; the U.S. National Science Foundation; the Swiss National Science Foundation; the Science and Technology Facilities Council (STFC), part of the United Kingdom Research and Innovation; and The Royal Society (United Kingdom). Additional support for the laser calibration system and cosmic ray tagger is provided by the Albert Einstein Center for Fundamental Physics, Bern, Switzerland.


\bibliographystyle{JHEP}
\bibliography{wire-cell-imaging}

\end{document}